\documentclass[twocolumn]{aa} 
\usepackage{amsmath}
\usepackage{hyperref}
\usepackage[makeroom]{cancel}
\usepackage[varg]{txfonts}
\usepackage{pifont}
\usepackage{epstopdf}
\renewcommand{\arraystretch}{1.2}
\usepackage{ulem}
\usepackage{tabularx}
\usepackage{float}
\usepackage{multirow,multicol}
\usepackage{enumitem}
\usepackage{graphicx}   
\usepackage{xcolor}
\usepackage{amstext}
\newcolumntype{R}{>{\raggedleft\arraybackslash}X}
\usepackage{bigdelim}
\usepackage{stackengine}
\usepackage{bbding}
\usepackage{layouts}

\begin{document} 

\abstract{The Rosetta spacecraft escorted Comet 67P/Churyumov-Gerasimenko {for} 2 years along its journey through the Solar System between 3.8 and 1.24~au. Thanks to the high resolution mass spectrometer on board Rosetta, the {detailed} ion composition within a coma has been accurately assessed in situ for the very first time.} {Previous cometary missions, such as {\it Giotto}, did not have the instrumental capabilities to identify the exact nature of the plasma in a coma because the mass resolution of the spectrometers onboard was too low to separate ion species with similar masses. In contrast, the Double Focusing Mass Spectrometer (DFMS), part of the Rosetta Orbiter Spectrometer for Ion and Neutral Analysis on board Rosetta (ROSINA), with its high mass resolution mode, outperformed all of them, revealing the diversity of cometary ions.} {We calibrated and analysed the set of spectra acquired by DFMS in ion mode from October 2014 to April 2016. In particular, we focused on the range from 13-39~u\textperiodcentered q$^{-1}$. The high mass resolution of DFMS allows for accurate identifications of ions with quasi-similar masses, separating $^{13}$C$^+$ from CH$^+$, for instance.} {We confirm the presence in situ of predicted cations at comets, such as CH$_m^+$ ($m=1-4$), H$_n$O$^+$ ($n=1-3$), O$^+$, Na$^+$, and several ionised and protonated molecules. Prior to Rosetta, only a fraction of them had been confirmed from Earth-based observations. In addition, we report for the first time the unambiguous presence of a molecular dication in the gas envelope of a Solar System body, namely CO$_2^{++}$.} {}
\keywords{Comets: individual: 67P/Churyumov-Gerasimenko; Plasmas; Molecular processes}

\title{ROSINA ion zoo at Comet 67P}

\author{A. Beth$^{1,2}$ \and K. Altwegg$^{3,4}$ \and H. Balsiger$^{3}$ \and J.-J. Berthelier$^{5}$ \and M.~R. Combi$^{6}$ \and J. De Keyser$^{7}$ \and B. Fiethe$^{8}$ \and S.~A. Fuselier$^{9,10}$ \and M.~Galand$^{1}$ \and T.~I. Gombosi$^{6}$ \and M. Rubin$^{3}$ \and T. Sémon$^{3}$ }

\institute{$^{1}$Department of Physics, Imperial College London, Prince Consort Road, London SW7 2AZ, United Kingdom\\
$^{2}$Department of Physics, Umeå University, 901 87 Umeå, Sweden\\
$^{3}$Physikalisches Institut, University of Bern, Sidlerstrasse 5, CH-3012 Bern, Switzerland\\
$^{4}$Center for Space and Habitability, University of Bern, Gesellschaftsstrasse 6, CH-3012 Bern, Switzerland\\
$^{5}$LATMOS, 4 Avenue de Neptune, F-94100 Saint-Maur, France\\
$^{6}$Department of Climate and Space Sciences and Engineering, University of Michigan, 2455 Hayward, Ann Arbor, MI 48109, USA\\
$^{7}$Royal Belgian Institute for Space Aeronomy (BIRA-IASB), Ringlaan 3, B-1180, Brussels, Belgium\\
$^{8}$Institute of Computer and Network Engineering (IDA), TU Braunschweig, Hans-Sommer-Straße 66, D-38106 Braunschweig,
Germany\\
$^{9}$Space Science Division, Southwest Research Institute, 6220 Culebra Road, San Antonio, TX 78228, USA\\
$^{10}$University of Texas at San Antonio, San Antonio, TX
}

\date{Received 24 September 2019; Accepted 7 July 2020}

\titlerunning{ROSINA ion zoo at Comet 67P}
\authorrunning{A. Beth}

\maketitle

\section{Introduction\label{sec1}}

Relatively small, with a nucleus size of a few kilometres to a few tens of kilometres, comets are only detectable once they are close enough to the Sun and display a bright tail. Compared to other planetary bodies and their atmosphere, the gas envelope of comets, the coma, behaves very differently. The coma results from the sublimation of ices near the nucleus' surface, which then undergoes an acceleration to several hundreds of m\textperiodcentered s$^{-1}$, continuously replenishing the coma. Mainly made of water, the coma contains a diversity of neutral species, such as CO$_2$, CO \citep{Krankowsky1986,Hassig2015}, and many others \citep[e.g.][]{Leroy2015} that have been detected in situ at 1P/Halley (hereinafter referred to as 1P) and 67P/Churyumov-Gerasimenko \citep[hereinafter referred to as 67P,][]{Churyumov1972}. Extreme ultraviolet (EUV) solar radiation penetrates and ionises the neutral gas envelope, giving birth to the cometary ionosphere. In addition to EUV, an additional source of ionisation is energetic electrons \citep{Cravens1987}. Depending on the local neutral number density, newborn cometary ions may undergo collisions with neutrals, yielding the production of cations which cannot result from direct ionisation of the neutrals. The diversity of ions is therefore richer than that of neutrals.


Cometary ions may be observed remotely at ultraviolet and visible wavelengths. Emissions in these wavelengths arise mainly from the resonant fluorescence of sunlight. These types of emissions from cometary molecular ions were first observed at Comet C/1907 L2 (Daniel) \citep{Deslandres1907,Evershed1907}. Although the emitting species was unknown at the time of the detection \citep{Larsson2012}, it was later identified as CO$^+$. The discovery of an ion tail that is always oriented anti-sunward led to the discovery of the solar wind \citep{Biermann1951,Parker1958}. Several cometary ions have since populated the list: N$_2^+$ and CH$^+$ \citep{Swings1942}, CO$_2^+$ and HO$^+$ \citep{Swings1950,Swings1956}, Ca$^+$ \citep{Preston1967}, H$_2$O$^+$ \citep{Herzberg1974}, CN$^+$ \citep{Lillie1976}, and H$_2$S$^+$ \citep{Cosmovici1984}. It is important to note that some ions are detected in cometary environments through observations in EUV and X-Rays \citep{Lisse1996}. Nevertheless, they are not cometary as we define here. The emission originates from the de-excitation of multiply-charged ions (e.g. O$^{6+}$), produced after a charge exchange between the high charge state solar wind ions (e.g. O$^{7+}$) and the neutral coma \citep{Lisse2004}. In addition, cometary ions may be detected at radio wavelengths. Emissions at these wavelengths are triggered by transitions between ro-vibrational or hyperfine levels \citep{Crovisier1991}. The first ion detected through radio-astronomy observations was HCO$^+$ \citep{Veal1997}, followed by H$_3$O$^+$ and CO$^+$ \citep{Lis1997}; all of them were at Comet C/1995 O1 (Hale-Bopp). Remote sensing detection of ionised constituents remains limited in terms of spatial resolution, species, and number density.


The most efficient way to probe the neutral and ion composition within a coma is in situ observations such as those performed by the European Space Agency's {\it Giotto} \citep{Reinhard1986} and Rosetta \citep{Glassmeier2007a} missions. At comet 1P, {\it Giotto} carried several instruments able to probe different mass and energy ranges of the cometary ions. The first measurements were performed by the {\it Giotto} Neutral Mass Spectrometer \citep[NMS,][]{Krankowsky1986b} when operated in the ion mode, which consisted of two analysers: a double-focusing mass spectrometer (M-analyser, for the range 1-37~u\textperiodcentered q$^{-1}$) and an electrostatic energy spectrometer (E-analyser, for the range 1-56~u\textperiodcentered q$^{-1}$ in ion mode). \citet{Krankowsky1986} reported the `quasi-unambiguous' identification of C$^+$, CH$^+$, O$^+$, HO$^+$, H$_2$O$^+$, H$_3$O$^+$, Na$^+$, C$_2^+$, S$^+$, $^{34}$S$^+$, and Fe$^{+}$. In addition, peaks were detected at all integer mass positions from 12 to 37~u\textperiodcentered q$^{-1}$, except at 22~u\textperiodcentered q$^{-1}$. A second {\it Giotto} instrument was the ion mass spectrometer \citep[IMS,][]{Balsiger1986b} consisting of two sensors: the high-energy-range spectrometer (HERS, dedicated to the study of ion composition and velocity outside the contact surface, covering the mass-per-charge range 1-35~u\textperiodcentered q$^{-1}$) and the high-intensity spectrometer (HIS, dedicated to measurements inside the contact surface, covering 12-57~u\textperiodcentered q$^{-1}$). \citet{Balsiger1986} reported the detection of C$^+$, CH$^+$, CH$_2^+$+N$^+$, CH$_3^+$, O$^+$, HO$^+$, H$_2$O$^+$, H$_3$O$^+$, CO$^+$, and S$^+$. The third instrument was the Positive Ion Cluster Composition Analyzer \citep[PICCA,][]{Korth1987} part of the Rème Plasma Analyser \citep[RPA,][]{Reme1987} which measured ions from 10~u\textperiodcentered q$^{-1}$ up to 203~u\textperiodcentered q$^{-1}$. As the E-analyzer in NMS, PICCA is not a true mass spectrometer as it separates ions in terms of energy-per-charge instead of mass-per-charge. However, as cometary ions are cold (i.e. their thermal speed is small compared with the ram speed of the spacecraft during the flyby), ions are collimated in the ram direction and their energy is roughly proportional to their mass allowing to deduce a mass spectrum with a limited mass resolution. \citet{Korth1986} reported the presence of ions belonging to the H$_2$O group (O$^+$, HO$^+$, H$_2$O$^+$, and H$_3$O$^+$), along with ions from CO$^+$, S$^+$, and CO$_2^+$ groups. In addition, spectra revealed periodic peaks in terms of u\textperiodcentered q$^{-1}$, around 64, 76, 94, and above 100~u\textperiodcentered q$^{-1}$. It was unclear if they were associated with ions from the Fe group, sulphur compounds, or hydrocarbons. \citet{Huebner1987} and \citet{Mitchell1987} suggested that these observations are consistent with the dissociation of polyoxymethylene (CH$_2$O)$_n$. Later on, \citet{Mitchell1992} showed that these peaks, repeating every $\sim15$~u\textperiodcentered q$^{-1}$, correlate with the number of combinations of C, H, O, and N, to form a molecular ion at a given u\textperiodcentered q$^{-1}$ \citep[see Fig.~2 in][]{Mitchell1992}. Similar patterns are observed at Titan \citep{Vuitton2007}, though the contribution of O-bearing molecules is negligible compared with that at comets. In addition, they observed a predominance of odd mass number ions. 

The term `unambiguous detection' or similar formulation should be taken with great care for {\it Giotto} data at 1P since the mass resolution of its instruments was about $\Delta m\sim1$~u. The combination and assemblage of the primary blocks, C, H, O, and N atoms, to build more complex molecules are limited at low masses \citep[typically below 25~u,][]{Mitchell1992}. At some specific values of u\textperiodcentered q$^{-1}$, there exists only one combination: C$^+$ (12~u\textperiodcentered q$^{-1}$), CH$^+$ (13~u\textperiodcentered q$^{-1}$), H$_3$O$^+$ (19~u\textperiodcentered q$^{-1}$), C$_2^+$ (24~u\textperiodcentered q$^{-1}$), C$_2$H$^+$ (25~u\textperiodcentered q$^{-1}$), if one disregards isotopes and isotopologues. There is even no candidate between 20~u\textperiodcentered q$^{-1}$ and 23~u\textperiodcentered q$^{-1}$. At other u\textperiodcentered q$^{-1}$ (in particular 18~u\textperiodcentered q$^{-1}$ which corresponds to H$_2$O$^+$ and NH$_4^+$), photo-chemical models are needed to infer the relative contribution of each ion, or, conversely, constrain the neutral composition \citep{Haider2005}.

At comet 67P, the Rosetta orbiter carried two instruments which performed a true mass analysis of the ambient ions: the Ion Composition Analyzer \citep[ICA,][]{Nilsson2007}, part of the Rosetta Plasma Consortium \citep[RPC,][]{Carr2007}, and the Double Focusing Mass Spectrometer (DFMS), part of the Rosetta Orbiter Spectrometer for Ion and Neutral Analysis \citep[ROSINA,][]{Balsiger2007}. Although we may compare RPC-ICA with its homologue RPA-PICCA, the former suffers from limitations to probe cold cometary ions. As Rosetta was moving slowly with respect to the ambient plasma, ions were not collimated along the spacecraft velocity and RPC-ICA has a wide field of view. In addition, its minimum energy acceptance is 4-5~eV, such that it only observed ion species after they were energised either as pick-up ions or accelerated by the spacecraft potential prior to entering RPC-ICA. Nevertheless, RPC-ICA was perfectly designed for probing the energetic solar wind ions, such as H$^+$, He$^{++}$, and He$^{+}$, unlike ROSINA-DFMS. ROSINA-DFMS (described in Section~\ref{sec2}) has the ability to probe neutrals as well as ions with two different mass resolutions either $m/\Delta m \approx 500$ in the `Low Resolution' LR mode or $m/\Delta m>3000$ in the `High Resolution' HR mode. DFMS is the most powerful spectrometer in terms of mass resolution ever flown on board a spacecraft so far. Previous analyses of DFMS ion spectra in high resolution revealed the unambiguous detection of H$_2$O$^+$, NH$_4^+$, and H$_3$O$^+$ \citep{Fuselier2016,Beth2016}. Low resolution spectra have been also analysed and highlighted the presence of other species either at large heliocentric distances \citep{Fuselier2015} or near perihelion \citep{Heritier2017a} with the support of photo-chemical modelling. 

In this paper, we present in situ detections of cometary ions at 67P over the range 13-39~u\textperiodcentered q$^{-1}$ in high resolution and 13-141~u\textperiodcentered q$^{-1}$ in low resolution. In HR, DFMS pinpointed the mass-per-charge ratio of impinging cometary ions with such a high accuracy that their composition and identity can be ascertained without any ambiguity. The DFMS spectrometer and data processing are presented in Section~\ref{sec2}, followed by a review of the mass spectra acquired during the period Oct 2014-Apr 2016 in Section~\ref{sec3}. Section~\ref{sec4} highlights the main results including the key different ion family behaviours (\ref{sec41}), the protonated molecules (\ref{sec42}), water isotopologues (\ref{sec43}), and dications (\ref{sec44}). Discussion and conclusions are presented in Section~\ref{sec5}.

\section{ROSINA-DFMS\label{sec2}}

\subsection{Ion Mode: principle, data processing, and limitations}

The description of the instrument and its capabilities have been provided in \citet{Balsiger2007} and \citet{Leroy2015}. In the ion mode, the ionised constituents are directly admitted in the ion optics from which they exit towards the detector. Surrounding the entrance of the instrument, a negatively-biased grid was designed to attract ions in case of a positive spacecraft potential as anticipated from simulations based on plasma conditions in the dense cometary ionosphere encountered by {\it{Giotto}}. However, the spacecraft potential of Rosetta was very negative during most of the escort phase \citep{Odelstad2017} and the grid was permanently set to a small negative potential, of -5~V. Once inside the instrument, ions are accelerated significantly by a large negative potential so that their energy in the ion optics is much higher than their energy at the entrance of the instrument: they undergo a first deflection in the electrostatic energy analyser which selects the ion energy before they exit through either the LR or the HR energy slit, the former being 6.5 times wider that the latter, into the magnetic analyser where they are deflected according to their mass and charge. Exiting the magnetic analyser, ions impinge on the detector which consists of a Micro Channel Plate (MCP) followed by a Linear Electron Detector Array (LEDA). Since the magnetic field intensity in the magnet varies with the temperature \citep[see][]{DeKeyser2019}, the impact position of a given ion on the detector will depend on the temperature as well. The LEDA is split into two identical rows (hereafter referred to as channel A and channel B) with 512 pixels, 25-$\mu$m wide and 8-mm long perpendicular to the mean axis of the row. When an ion hits the MCP, a cascade of electrons is produced and the total amount of negative charges collected by the LEDA, known as the MCP gain, depends on the voltage applied to the MCP. The gain is not uniform over the entire MCP area and, for each pixel, one can define a `pixel gain', which modulates the average MCP gain, and determine the actual number of electrons collected by the corresponding LEDA pixel.

The pixel gains vary during the Rosetta escort phase and have been regularly determined through dedicated in-flight calibrations. Pixel gains degraded during the mission \citep{Schroeder2019}, especially for pixels located close to the centre of each row (from pixel 200 to pixel 400), where H$_2$O$^+$ ions strike in both neutral and ion modes of DFMS. To partially compensate this degradation and the loss of sensitivity, on the 27$^{th}$ of January 2016, the post-acceleration was modified in order to move the central pixel $p_0$, such that the position on the detector of the selected mass of each spectrum was moved forwards, on pixels with a less degraded gain. Pixels at the very edge of the LEDA rows have a poor gain as well but they are not included in the analysis. 

DFMS has two basic modes of operation. In the `neutral' mode, the neutral species are ionised and fragmented through electron impact in the ion source, thanks to a filament emitting electrons at $\sim 45$~eV, before being accelerated into the ion optics. In the `ion' mode, the filament is not powered and ions are directly admitted into the ion optics. The ion and neutral modes are not operated simultaneously but, for both of them, the total integration time for each individual spectrum is 19.8~s made of 3000 exposures of 6.6~ms. For both modes, DFMS may operate in Low or High mass-per-charge Resolution (hereafter referred to as LR and HR, respectively). HR mode, for which $m/\Delta m>3000$ at the 1\% peak height level for 28~u\textperiodcentered q$^{-1}$ \citep{Balsiger2007}, allows separation of ions that have very close mass-per-charge ratios (e.g. $^{13}$C$^+$ and CH$^+$, H$_2$O$^+$ and NH$_4^+$, CO$^+$ and N$_2^+$), which is not possible in LR mode, for which $m/\Delta m\approx500$. However, the sensitivity at a given gain step is significantly higher in LR than in HR, therefore the LR mode was of particular interest during periods of low outgassing and when fewer ion species can be detected due to limited ion-neutral chemistry \citep[e.g.][]{Fuselier2015}. By comparison, the HR mode was of particular interest for periods at high outgassing activity, such as near perihelion, when ion-neutral chemistry takes place and many new species are present and need to be separated \citep[e.g.][]{Beth2016}.

A typical sequence of acquisition is as follows. Firstly, the first commanded (instructed to the instrument when operating) mass-per-charge ratio is 18~u\textperiodcentered q$^{-1}$ both in LR and in HR. Secondly, the second commanded mass-per-charge ratio is the lowest one which the instrument can perform: 13.65~u\textperiodcentered q$^{-1}$ in LR, 13~u\textperiodcentered q$^{-1}$ in HR. Thirdly, the commanded mass-per-charge ratio is then incremented: exponentially in LR, linearly in HR, with $m_0(i)$ the $i$-th commanded mass-per-charge ratio $m_0$ ($i\geq2$) defined as:
\[m_{0,\text{LR}}(i)\approx11.27798\times1.1^{i}  \text{ u\textperiodcentered q}^{-1}, \]
\[m_{0,\text{HR}}(i)=11+i\text{ u\textperiodcentered q}^{-1}.\]
Fourthly, the penultimate commanded mass-per-charge ratio: 134.4~u\textperiodcentered q$^{-1}$ ($i=26$) in LR, 100~u\textperiodcentered q$^{-1}$ ($i=89$) or 50~u\textperiodcentered q$^{-1}$ ($i=39$) in HR (100 was used as an upper limit during the first half of the mission but nothing was detected above 50~u\textperiodcentered q$^{-1}$, this limit was then lowered in July 2015). Finally, the last commanded mass-per-charge ratio is 18~u\textperiodcentered q$^{-1}$.

The three measurements of 18~u\textperiodcentered q$^{-1}$ during a sequence helped in monitoring the variability of the ambient plasma conditions and/or the effective DFMS geometrical factor in the ion mode which depended on the spacecraft potential. A full sequence lasts between 10 and 20 minutes, depending on the resolution and the number of commanded mass-per-charge ratios. LR and HR modes differ in terms of u\textperiodcentered q$^{-1}$ coverage since the mass-per-charge coverage for a given mass-per-charge ratio $m_0$ is roughly $0.1\ m_{0,\text{LR}}$ in LR and $0.016\ m_{0,\text{HR}}$ in HR (see Eq.~\ref{eq1} below). Therefore, successive LR spectra overlap and cover the full range from 13 to 141~u\textperiodcentered q$^{-1}$. HR successive spectra may overlap only at high masses from 64~u\textperiodcentered q$^{-1}$ onwards. Finally, less spectra are required in LR to cover the same mass-per-charge range because several u\textperiodcentered q$^{-1}$ may be covered in a given spectrum. However, in the latter case, the peaks fall on different locations on the detector, while, in HR, peaks fall close to the centre of the detector.

Thanks to the high resolution of DFMS, the ions species presented in this paper were identified by a detailed and accurate data analysis without the need to rely on photo-chemical models. The models presented in Section~\ref{sec4} only aim at understanding the variability of the cations throughout the escort phase for those confirmed.

\subsection{Data analysis\label{analysis}}

\begin{figure}
		\centering
		\includegraphics[height=\linewidth,trim=25cm 1cm 15cm 1cm,clip,angle=90 ]{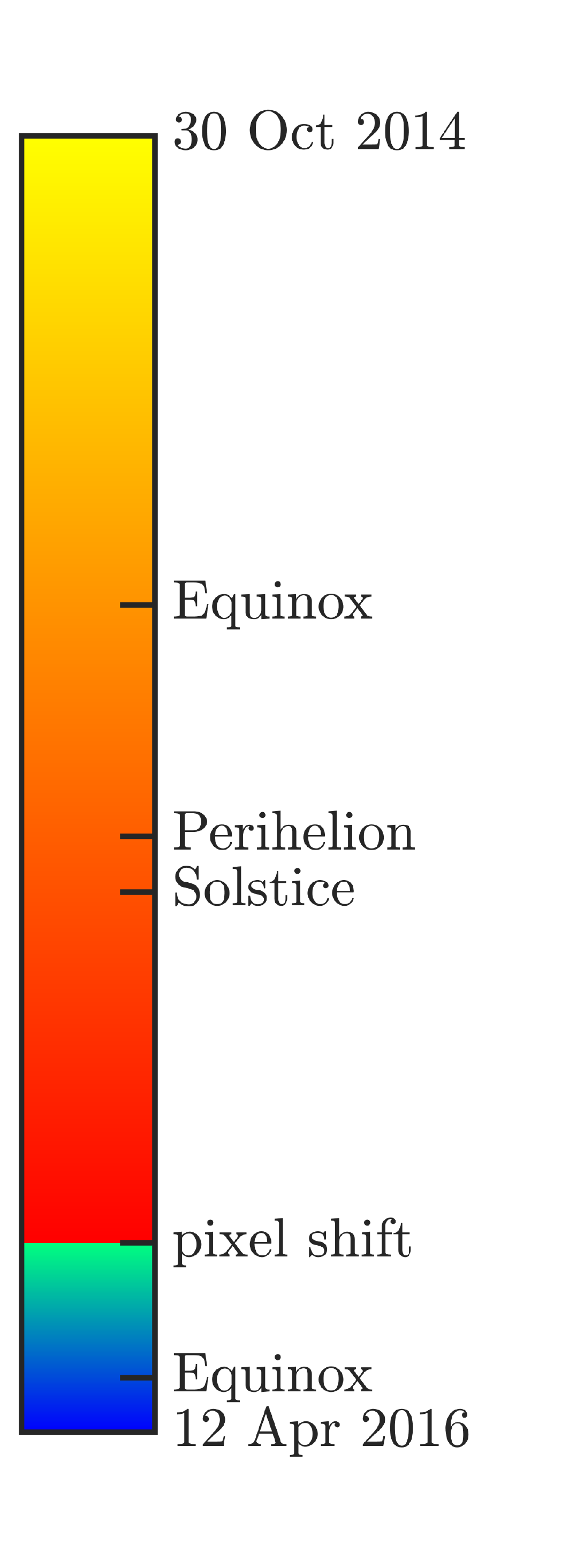}\\
		\includegraphics[height=\linewidth,trim=25cm 1cm 20cm 1cm,clip,angle=-90]{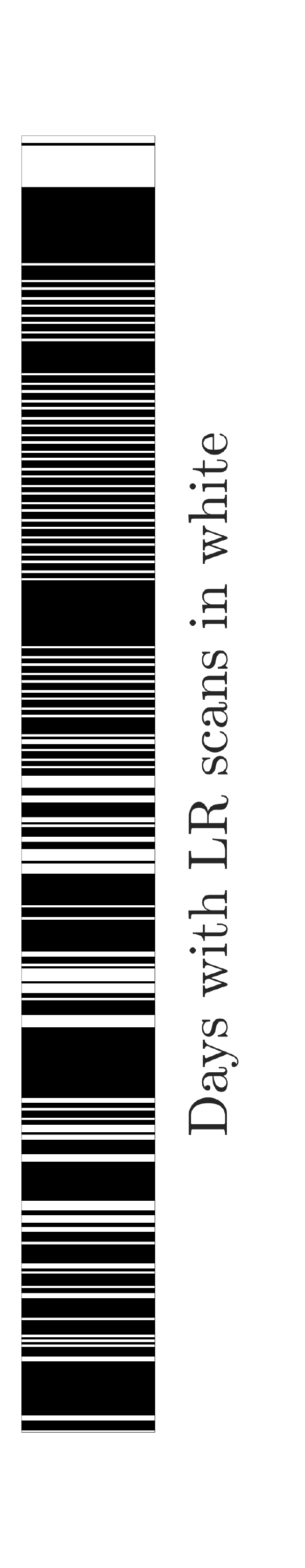}\\
		\includegraphics[height=\linewidth,trim=25cm 1cm 20cm 1cm,clip,angle=-90]{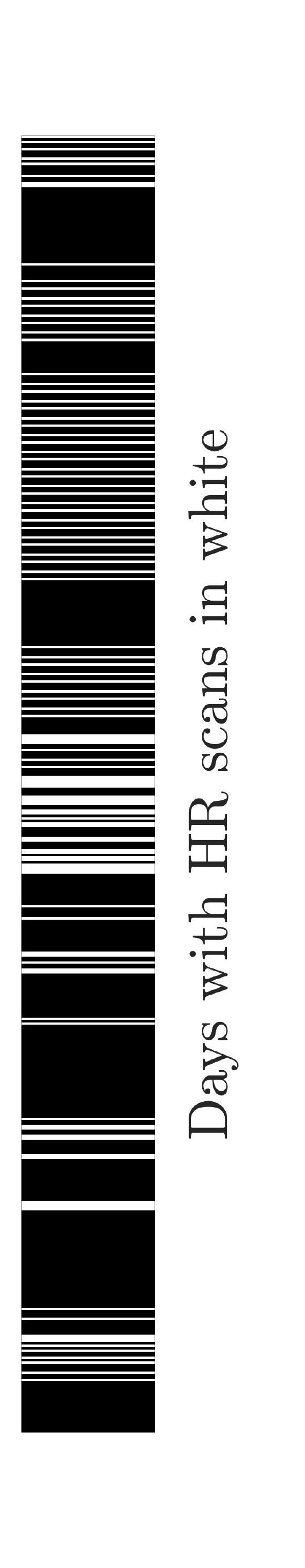}\\
		\includegraphics[width=\linewidth,trim=1.6cm 0cm 1.6cm 0cm,clip]{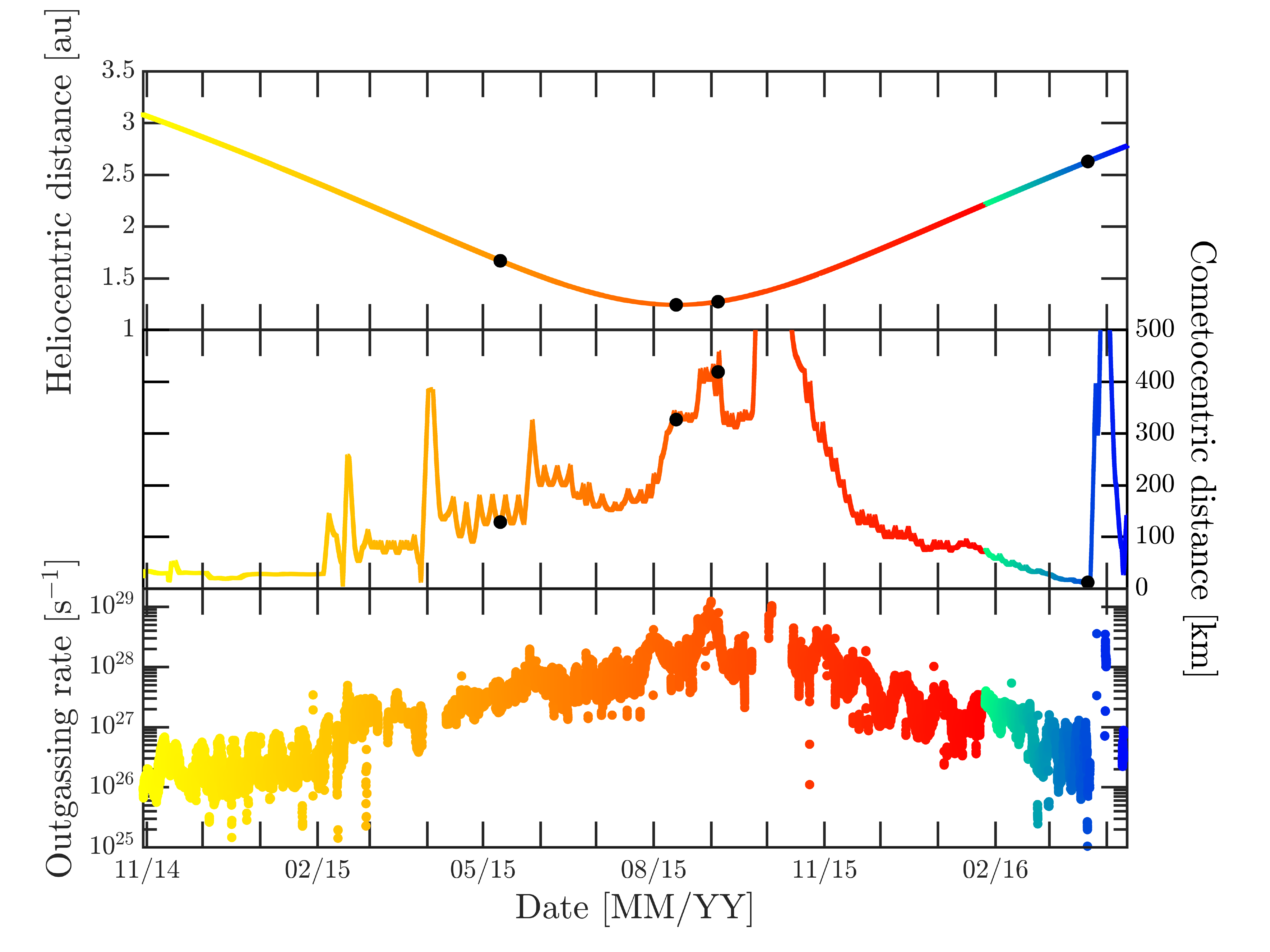}
		\caption{First panel: colour bar used for all spectra in LR and in HR. Spectra have been acquired between the 30$^{th}$ of October 2014 and the 12$^{th}$ of April 2016. A separation has been set on 27 January 2016 corresponding to the time when $p_0$ has been voluntarily shifted in DFMS (see text and Appendix~\ref{AppA}). Colour bars representing the time coverage of DFMS spectra in LR (second panel) and HR (third panel) ion mode are also displayed. White means that sequences of scans are performed on that day and black none. Solstice refers to the Summer Solstice over the Southern Hemisphere (solar latitude = $-52^\circ$). Fourth panel: heliocentric distance, cometocentric distance, and local outgassing rate ($\approx n_nr^2\varv_n$) as a function of time for the period of interest. Black dots correspond to (from left to right) the inbound Equinox, Perihelion, Solstice, and outbound Equinox. An outgassing speed $\varv_n$ of 1~km\textperiodcentered s$^{-1}$ has been assumed for the outgassing. \label{fig1}}
\end{figure}

The HR mode requires the utmost care for its mass calibration, that is determining the exact relation between the location of the pixel $p$ on the detector and the associated mass $m(p)$. This relation is given by \citep{Leroy2015}:
\begin{equation}
m(p)=m_0\exp\left[\dfrac{C}{D}\dfrac{(p-p_{0,m_0})}{z_{m_0}}\right]\approx m_0\left[1+\dfrac{C}{D}\dfrac{(p-p_{0,m_0})}{z_{m_0}}\right],
\label{eq1}
\end{equation}
where $C=25$~$\mu$m is the centre-to-centre distance between adjacent pixels, $D=127000~\mu$m the dispersion factor, $p_{0,m_0}$ the location of the commanded $m_0$\textperiodcentered q$^{-1}$ on the detector, and $z_{m_0}$ the zoom factor (1 for LR). Eq.~\ref{eq1} is linearisable because the argument inside the exponential is $\ll 1$. As indicated by their subscript in Eq.~\ref{eq1}, both $p_{0,m_0}$ and $z_{m_0}$ depend on the commanded mass-per-charge $m_0$ and, as aforementioned, on the magnet temperature since the exit location of a given u\textperiodcentered q$^{-1}$, hence the pixel on the detector, depends on the magnetic field intensity. If the variation of $p_{0,m_0}$ and $z_{m_0}$ between two adjacent u\textperiodcentered q$^{-1}$ (e.g. 18~u\textperiodcentered q$^{-1}$ and 19~u\textperiodcentered q$^{-1}$) are very small, they may be significant for widely different ~u\textperiodcentered q$^{-1}$ such as 13~u\textperiodcentered q$^{-1}$ and 40~u\textperiodcentered q$^{-1}$. To achieve a perfectly accurate spectrum analysis, both parameters should be reassessed for each sequence of acquisition of DFMS which is possible when two species, with mass $m_1$ and $m_2$ located at pixel $p_1$ and $p_2$ respectively, are present in the same spectrum. The zoom factor $z_{m_0}$ can be derived from: 
\begin{equation}
z_{m_0}=\dfrac{p_2-p_1}{\dfrac{C}{D}\log\left(\dfrac{m_2}{m_1}\right)},
\end{equation}
and then $p_{0,m_0}$ is inferred from one of the two species from Eq.~\ref{eq1}. Although this procedure may work well in neutral mode, it is seldom applicable in ion mode since spectra with two well-shaped and separated peaks are only observed for a few u\textperiodcentered q$^{-1}$ and favourable observation periods such as at 18~u\textperiodcentered q$^{-1}$ and at perihelion. Indeed, in ion mode, the count rates on the detector are much smaller than in the neutral mode because of the effective geometrical factor of DFMS for cometary ions, which is lower than that for neutrals due to several combined factors (e.g. large neutral number density, high ion source efficiency, ions accelerated by the spacecraft potential). As a matter of fact, all the ion mode spectra were acquired with the highest gain step to ensure the maximum sensitivity for the instrument. Following the findings of \citet{DeKeyser2015}, we have set the zoom factor $z$ to 5.5 for 13, 14, and 15~u\textperiodcentered q$^{-1}$ and to 6.4 otherwise. Recent analysis of the spectra in neutral mode showed that the zoom factor is slightly lower at 13, 14, and 15~u\textperiodcentered q$^{-1}$ confirming that 5.5 is appropriate. For $p_{0,m_0}$, we used the value determined from the most proximate spectrum either at 18~u\textperiodcentered q$^{-1}$ or 19~u\textperiodcentered q$^{-1}$ during the same sequence of acquisition of DFMS, that is either $p_{0,m_0}\approx p_{0,18}$ or $p_{0,m_0}\approx p_{0,19}$. Indeed, spectra at 18 and 19~u\textperiodcentered q$^{-1}$ show strong peaks throughout the escort phase attributed to H$_2$O$^+$ and H$_3$O$^+$. However, as there is also NH$_4^+$ at 18~u\textperiodcentered q$^{-1}$, we preferred to use 19 ($p_{0,m_0}\approx p_{0,19}$) to remove any ambiguity. This approach for deriving $p_{0,m_0}$ works well, except for 13, 14, and 15~u\textperiodcentered q$^{-1}$, discussed in Appendix~\ref{AppA}. $p_{0,m_0}$ is less constrained than $z$ and varies more significantly in comparison. One may evaluate the uncertainty of the mass $\delta m$ from those of $p_0$, $\delta p_0$, and $z$, $\delta z$:
\begin{equation}
\dfrac{\delta m}{m}\approx\dfrac{C}{Dz}\left(\dfrac{\delta z}{z} (p-p_0)+\delta p_0 \right).
\end{equation}
We found that the main source of uncertainty is $\delta p_0$. In the dataset generated by the ROSINA team, the default value for $|\delta p_0|$ is set to 10. The reader may find additional information in the ROSINA User Guide. For the identification in high resolution, we proceeded as follows. Firstly, we selected a u\textperiodcentered q$^{-1}$-range within which species may be found. As u\textperiodcentered q$^{-1}$ increases, the range does as well. Secondly, we performed an additional visual inspection if needed for low counts to remove any suspicious spectrum (e.g. not-flat spectrum baseline, spurious peak far from any known ion species). Thirdly, we over-plotted spectra (from a few tens to hundreds, depending on the mass-per-charge ratio with colour coding which depends on the time of acquisition through the mission, see Fig.~~\ref{fig1}). Similar studies may be performed with different variables (e.g. latitude).

In addition to ion identification, one of the main goals is also to assess in which conditions these ions have been detected: low and high outgassing activity, close and large heliocentric distance, close and large cometocentric distance. Because of all of these variables, we decided to colour spectra as a function of the time of acquisition during the mission. Fig.~\ref{fig1} shows the colour code used as a function of time for the spectra, the time coverage of DFMS in LR and HR as well as the heliocentric distance, cometocentric distance, and outgassing rate with the corresponding colour. Yellow corresponds to the early phase of the mission, with Rosetta far from the Sun ($>2.2$~au), close to the nucleus ($<50$~km), and 67P with a low outgassing rate ($Q<10^{27}$~s$^{-1}$). Orange corresponds to the period before perihelion with Rosetta close to the Sun ($<2.2$~au), between 100~km and 200~km from the nucleus, and 67P with an intermediate outgassing rate ($10^{27}<Q<10^{29}$~s$^{-1}$). Red corresponds to the period after perihelion with Rosetta close to the Sun ($<2.2$~au), farther than 200~km from the nucleus, and 67P with an intermediate outgassing rate ($10^{27}\lesssim Q<10^{29}$~s$^{-1}$). Green and blue correspond to the period after the pixel shift with Rosetta far from the Sun ($>2.2$~au) and 67P with a low outgassing rate. We strongly advise the reader to refer to Fig.~\ref{fig1} for the interpretation of the figures in Section~\ref{sec3}.

\section{Ion spectra\label{sec3}}

\subsection{Overview}

\begin{figure*}
	\centering
	\stackinset{r}{0.6cm}{b}{3.4cm}{\fbox{LR}}{%
	\includegraphics[width=.4933\linewidth,trim=0 1.9cm 0 5cm,clip]{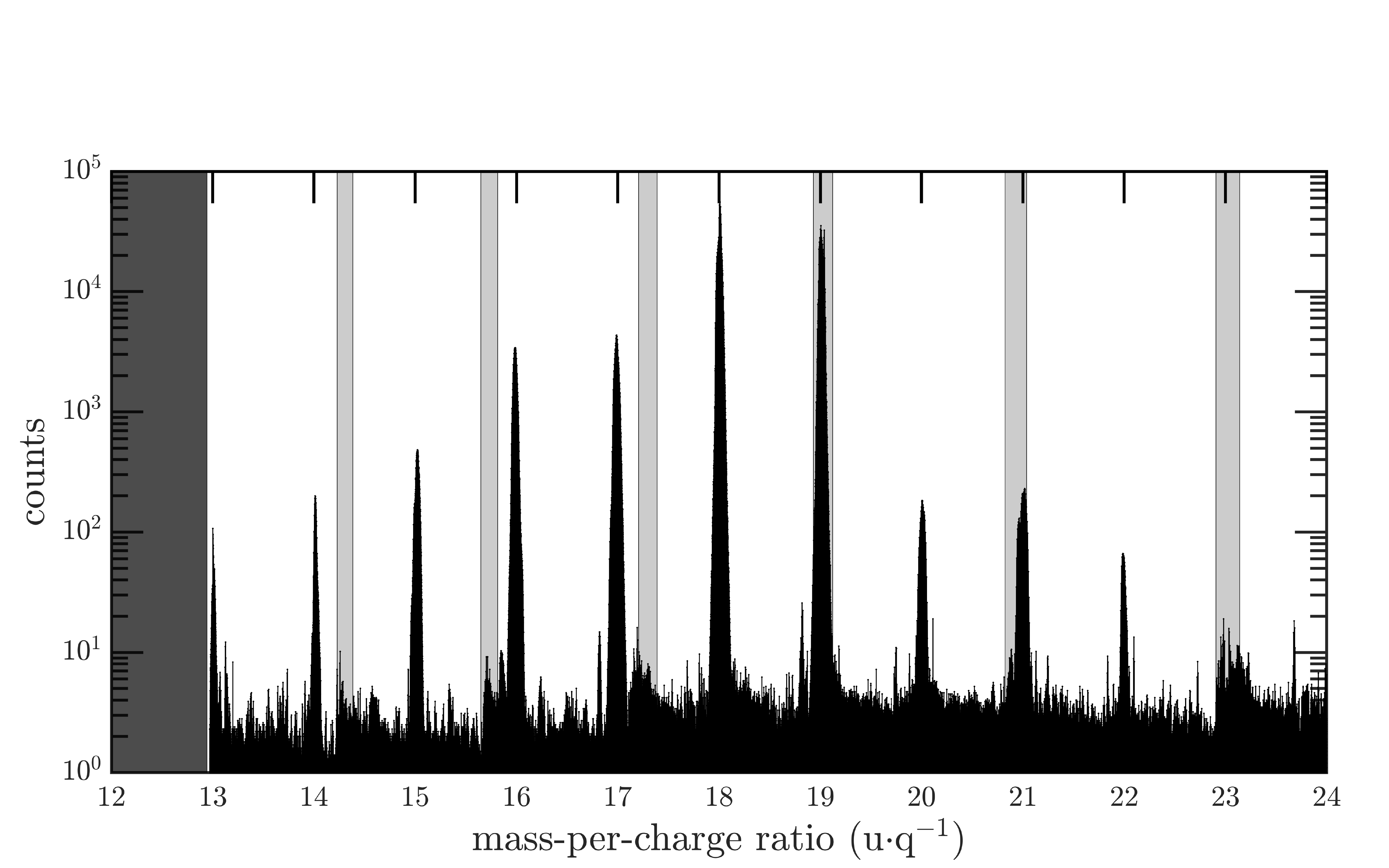}}
	\stackinset{r}{0.6cm}{b}{3.4cm}{\fbox{LR}}{%
	\includegraphics[width=.4933\linewidth,trim=0 1.9cm 0 5cm,clip]{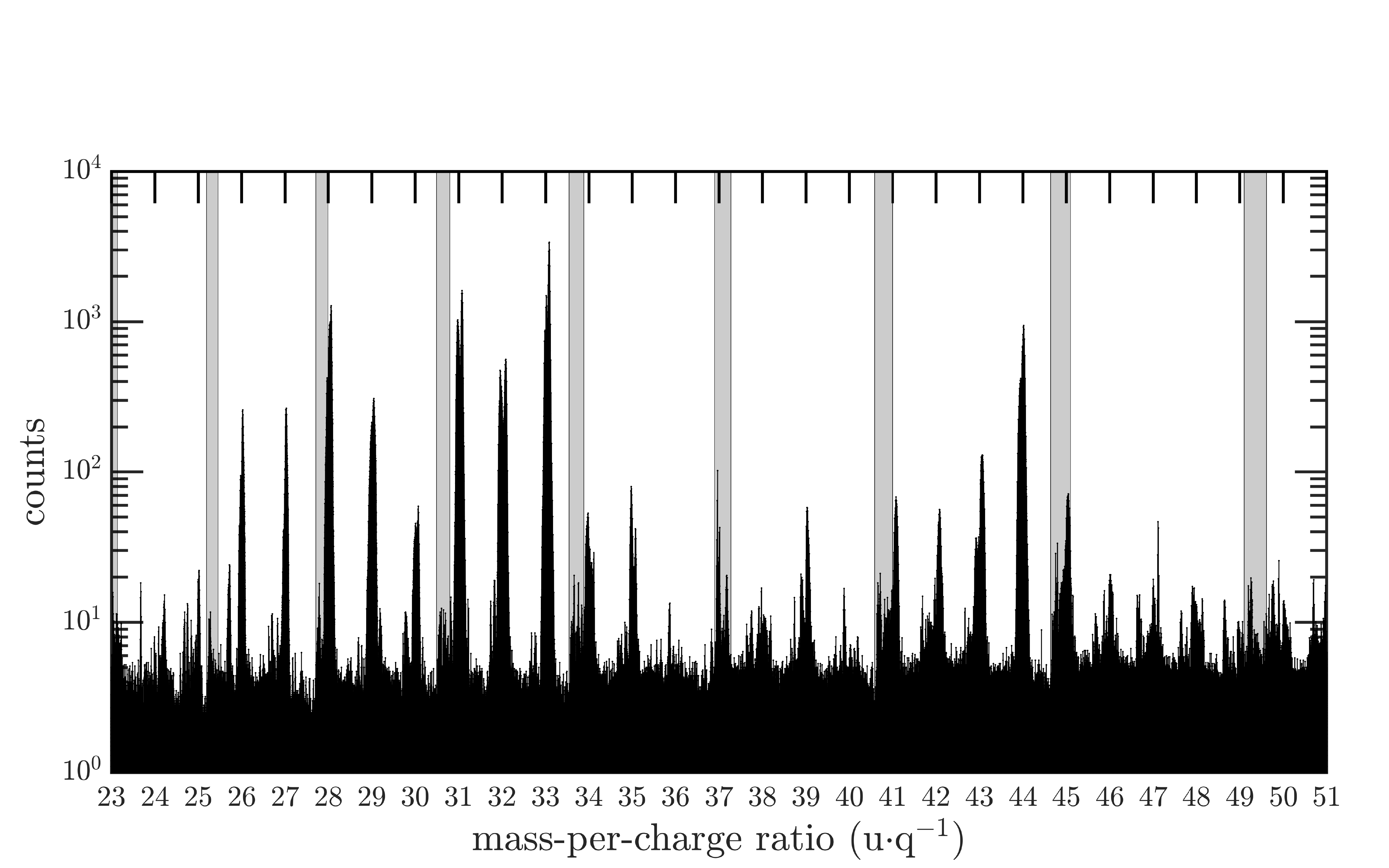}}\\
	\stackinset{r}{0.6cm}{b}{3.4cm}{\fbox{LR}}{%
	\includegraphics[width=.4933\linewidth,trim=0 1.9cm 0 5cm,clip]{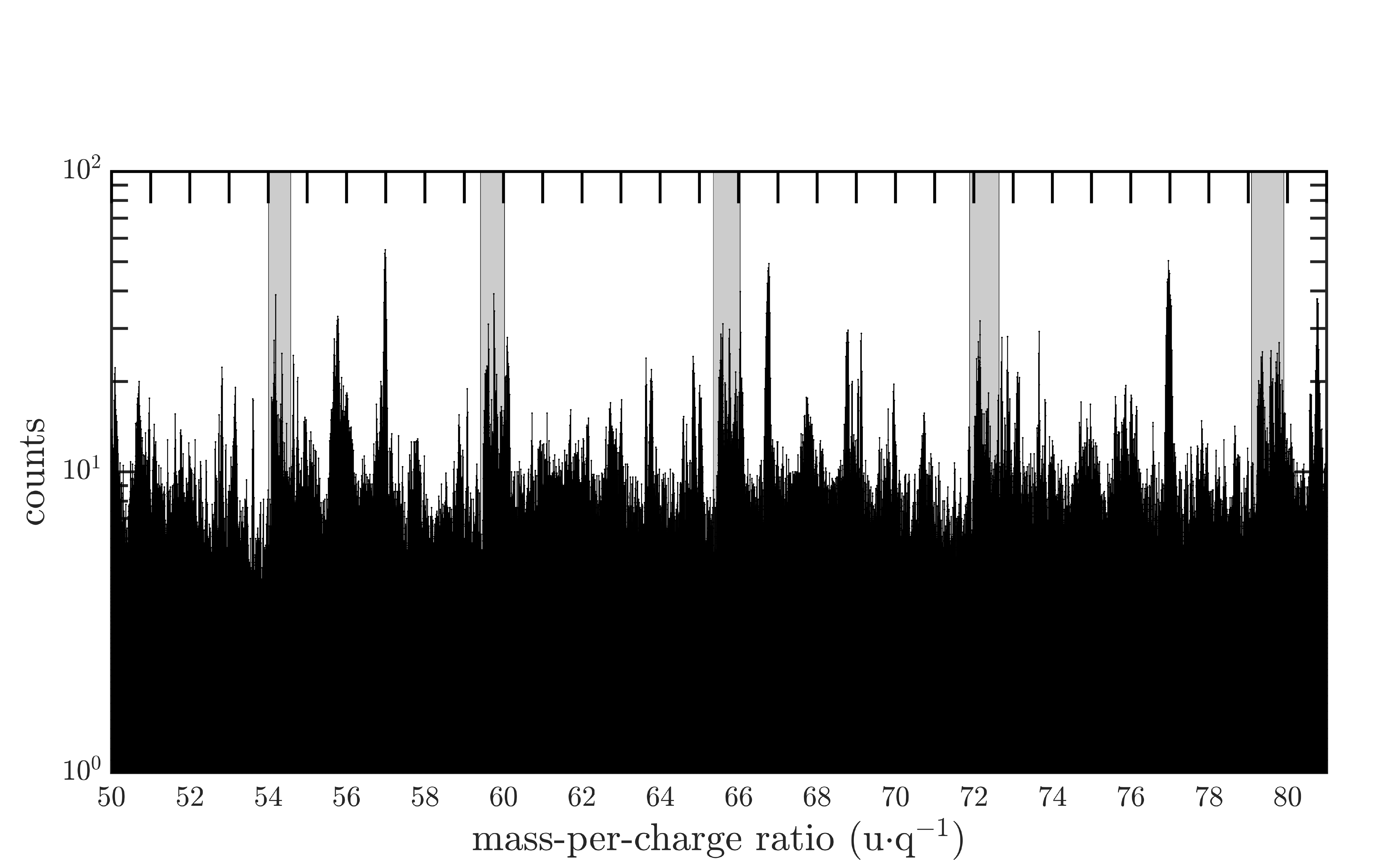}}
	\stackinset{r}{0.6cm}{b}{3.4cm}{\fbox{LR}}{%
	\includegraphics[width=.4933\linewidth,trim=0 1.9cm 0 5cm,clip]{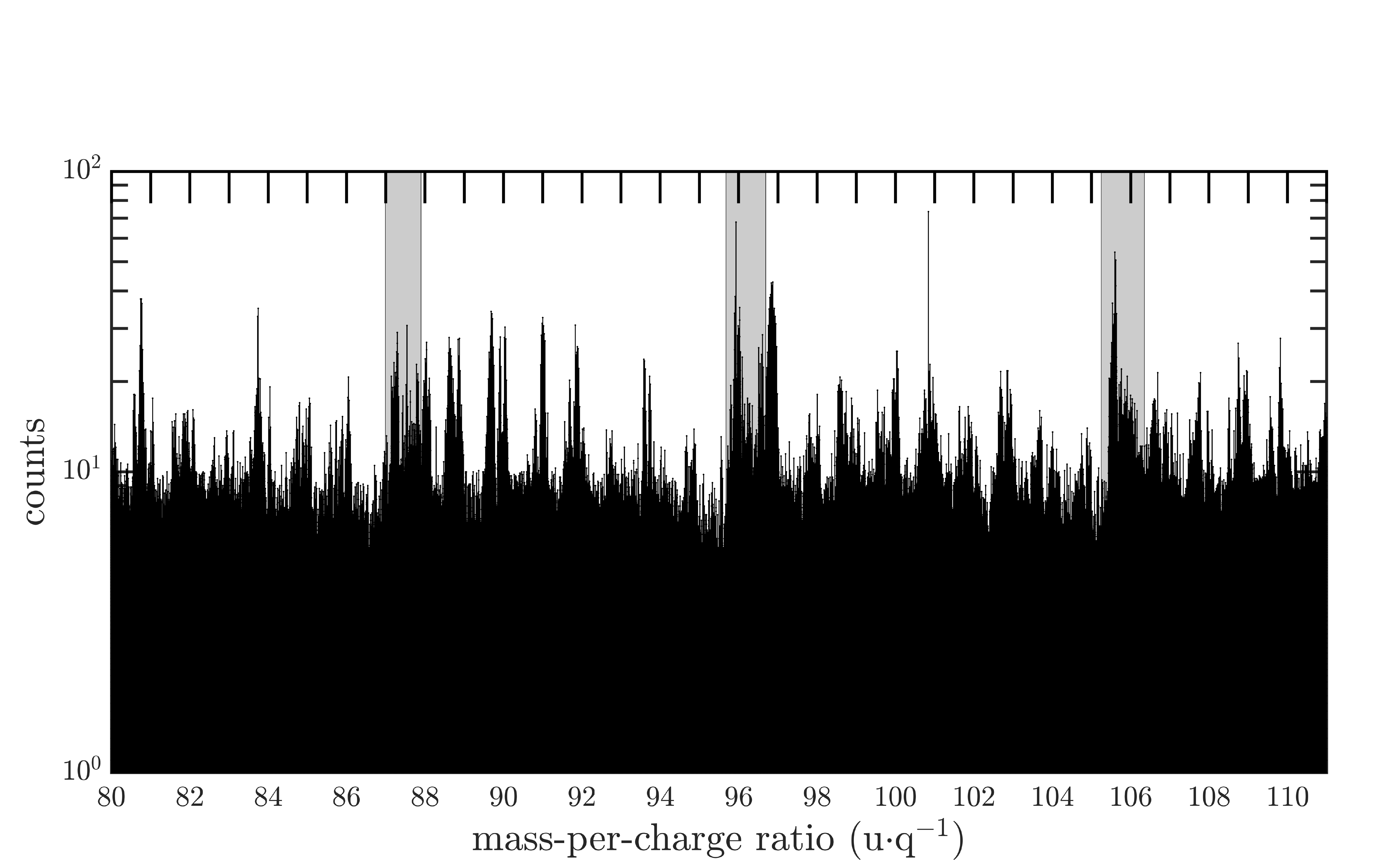}}\\
	\stackinset{r}{0.6cm}{b}{0.33667905cm+3.4cm}{\fbox{LR}}{%
	\includegraphics[width=.4933\linewidth,trim=0 0cm 0 5cm,clip]{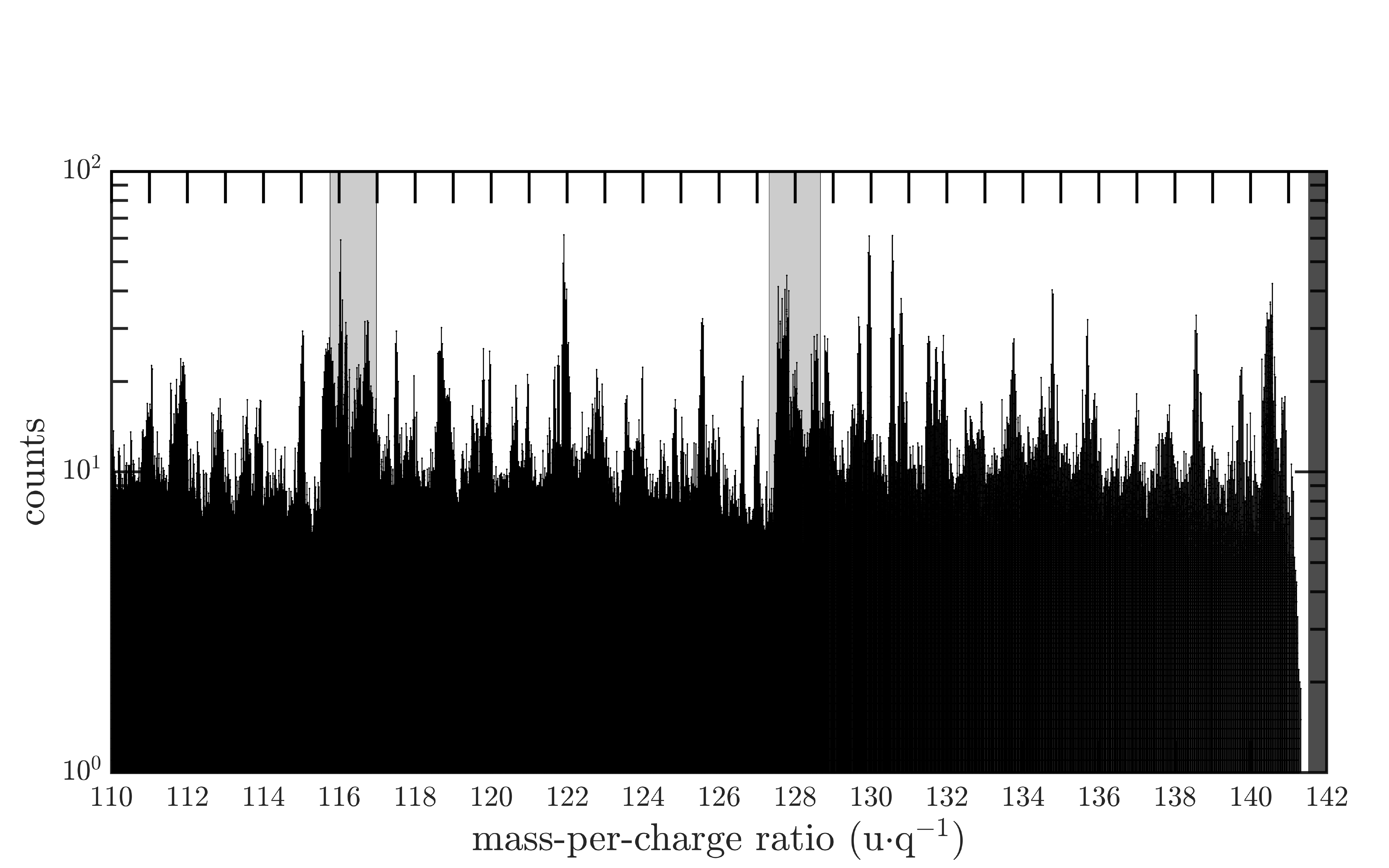}}
	\caption{Concatenation of spectra recorded at each channel in ion low resolution mode. This covers mass-per-charge ranging from 13~u\textperiodcentered q$^{-1}$ to 141~u\textperiodcentered q$^{-1}$. Dark grey regions represent ranges which are not covered by the instrument (below 13~u\textperiodcentered q$^{-1}$ and above 141~u\textperiodcentered q$^{-1}$). Light grey regions represent ranges where two consecutive scans overlap, meaning that these ranges are covered by the edge of the detector.\label{fig2}}
\end{figure*}

Thanks to its great sensitivity, ROSINA-DFMS allowed probing the ion composition in LR up to very high masses for the first time. Fig.~\ref{fig2} shows a series of selected spectra from 13 to 141~u\textperiodcentered q$^{-1}$. Above 72~u\textperiodcentered q$^{-1}$, the mass calibration is not as good as for lower masses {because a different post-acceleration is applied within the instrument}, which explains why peaks are not centred correctly. The highest counts are recorded at $\sim18$~u\textperiodcentered q$^{-1}$ and at $\sim19$~u\textperiodcentered q$^{-1}$, where H$_2$O$^+$ and H$_3$O$^+$ are found. Other high count regions are also observed at $\sim28$~u\textperiodcentered q$^{-1}$ (e.g. CO$^+$) and at $\sim44$~u\textperiodcentered q$^{-1}$ (e.g. CO$_2^+$). We note that we have gaps, low signals, or non-detections for instance at 36~u\textperiodcentered q$^{-1}$ and around 51~u\textperiodcentered q$^{-1}$, similar to those showed by \citet{Mitchell1992}, already described in Section~\ref{sec1}. However, in contrast, we have strong peaks at 21~u\textperiodcentered q$^{-1}$ and 22~u\textperiodcentered q$^{-1}$ where no combination of C, H, O, and N to form a monocation may fit. As the commanded mass-per-charge ratio increases, the signal-to-noise ratio decreases together with the signal (physical) and sensitivity (instrumental). Moreover, at high mass-per-charge ratios, an insidious effect decreases the width of the peak. With a constant $\Delta m / m$, the mass difference between two successive pixels $m(p+1)-m(p)$ increases with $m_0$ such that ions are focused and spread over fewer and fewer pixels, down to a single pixel in extreme cases. This focusing results in sharp peaks, with high counts for one pixel (spikes), instead of broad ones, which may be misinterpreted as `ghost' peaks, that is sharp and spurious peaks at the location of one pixel with high counts compared with surrounding pixels. However, over-plotting several spectra reveals that these spikes are located around each integer mass-per-charge ratio up to 141~u\textperiodcentered q$^{-1}$ and are thus real. Above 40~u\textperiodcentered q$^{-1}$, the exact species identification cannot be achieved due to the lack of peaks in HR ion mode as a consequence of the decreased sensitivity. The following sections are dedicated to the identification of ion species detected in the range of 13 to 39~u\textperiodcentered q$^{-1}$.

\subsection{Ion mass-per-charge range 13 – 21~u\textperiodcentered q$^{-1}$\label{section32}}

\begin{figure}
	\stackinset{r}{0.6cm}{b}{3.4cm}{\fbox{LR}}{%
	\includegraphics[width=\linewidth,trim=0cm 1.9cm 0cm 5cm,clip]{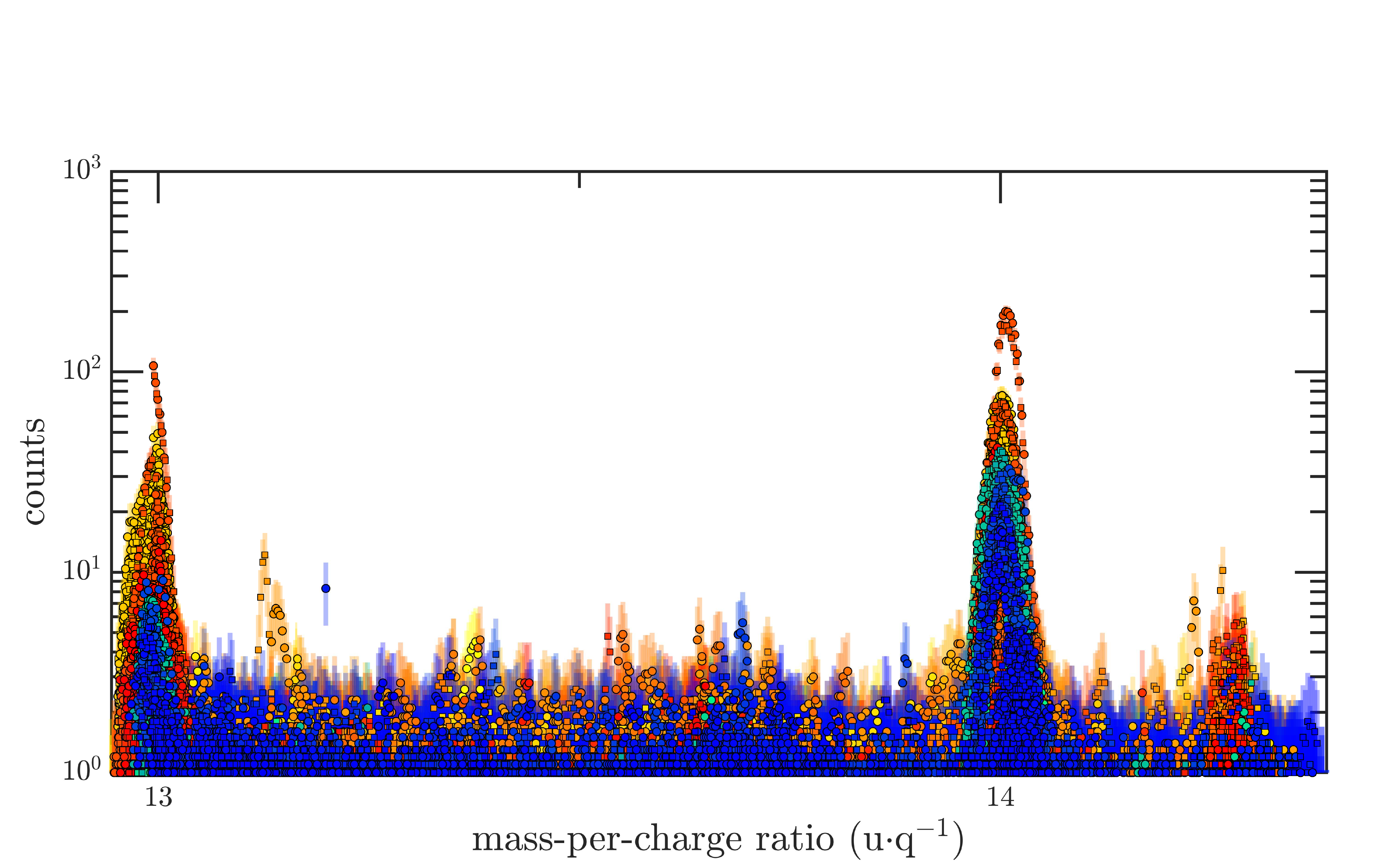}\\
}
\stackinset{r}{0.6cm}{b}{3.4cm}{\fbox{HR}}{%
	\includegraphics[width=\linewidth,trim=0cm 1.9cm 0cm 0cm,clip]{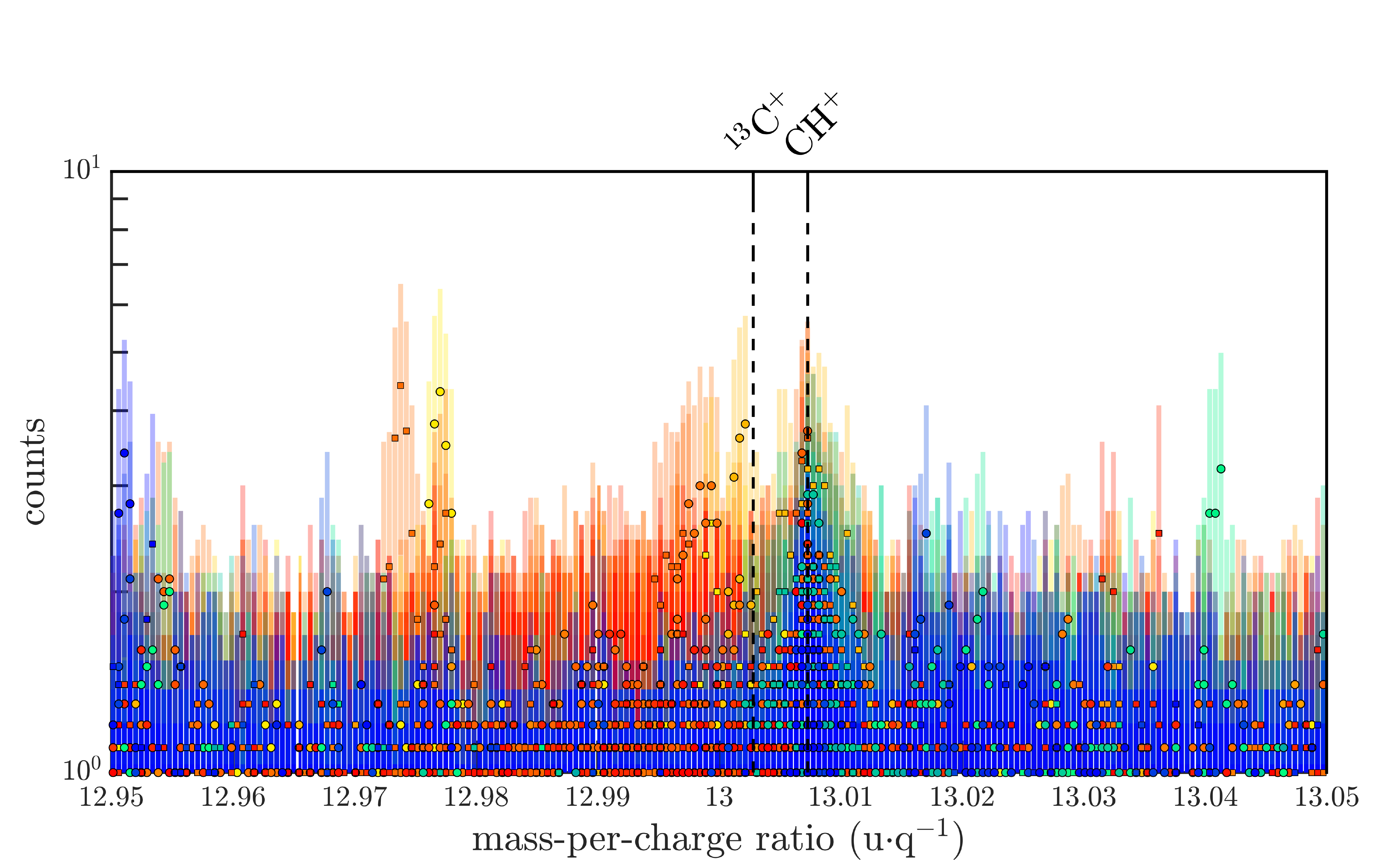}\\
}
\stackinset{r}{0.6cm}{b}{0.33667905cm+3.4cm}{\fbox{HR}}{%
	\includegraphics[width=\linewidth,trim=0cm 0.0cm 0cm 0cm,clip]{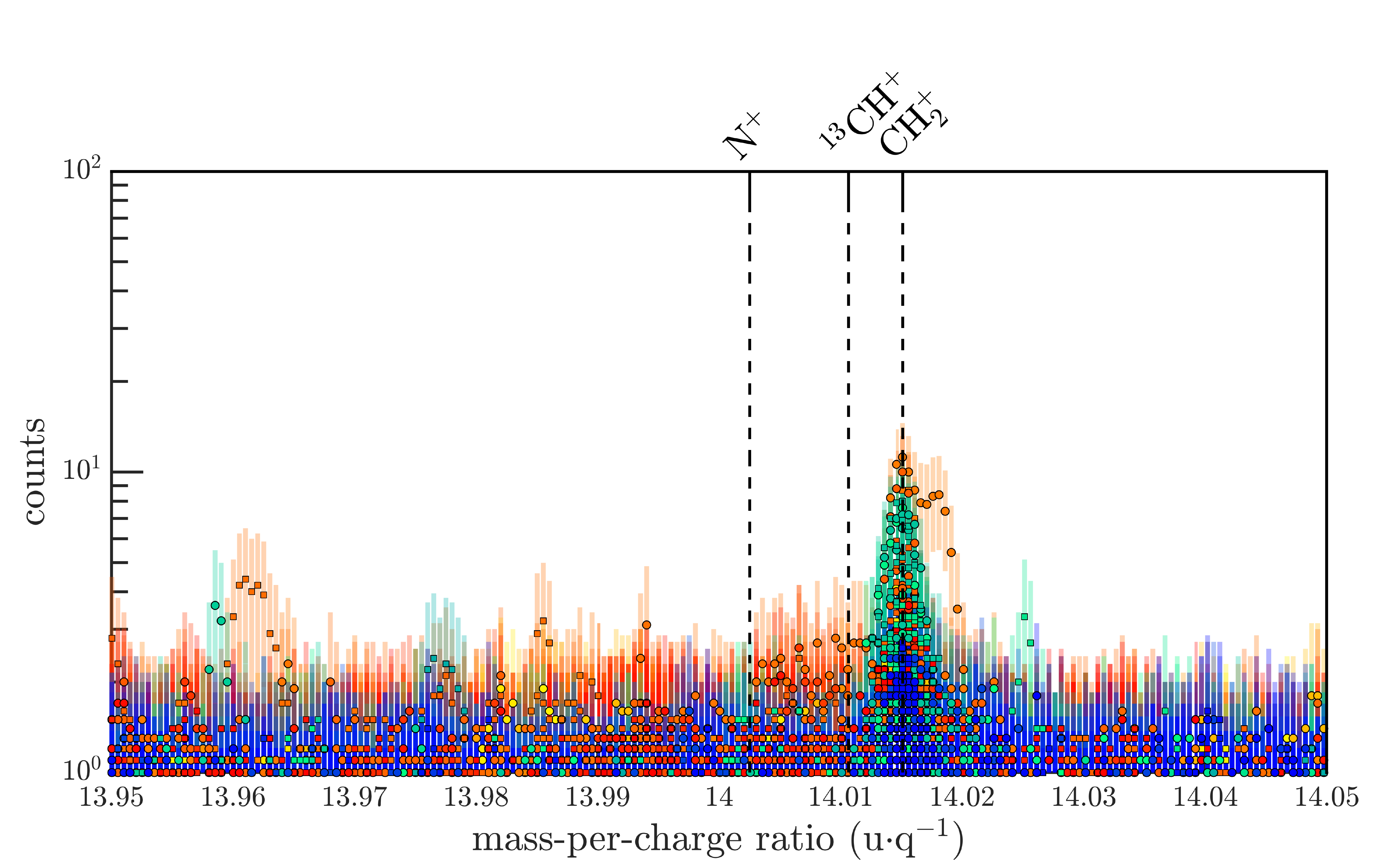}\\
}
	\caption{Stacked individual spectra covering 13 and 14~u\textperiodcentered q$^{-1}$ from the channel A ($\circ$) and channel B ($\square$) of the MCP, each of them represents the accumulated counts during 19.8~s at the maximum gain step. The top panel shows stacked spectra in low resolution covering both ranges (mass-per-charge ratios are indicated by the long ticks on the upper axis, half-integer ones by the short ticks). Panels underneath are HR spectra at specific u\textperiodcentered~q$^{-1}$: 13 (middle) and 14 (bottom), corrected from the pixel shift (see Fig.~\ref{figa1}). A statistical vertical error bar of $\pm\sqrt{N}$, where $N$ stands for the number of counts, is superimposed to the counts for information. The colour coding is given by the colour bar in Fig.~\ref{fig1} and relies on the time of acquisition during the escort phase. The mass-per-charge ratios of expected ions from Table~\ref{table1} are also indicated and given in App.~\ref{AppC}, Table~\ref{tablec1}. \label{fig3}}
\end{figure}

Fig.~\ref{fig3} shows spectra for the range 13-14~u\textperiodcentered q$^{-1}$. In LR, two distinct peaks are present at each integer. In HR at 13~u\textperiodcentered q$^{-1}$, there are two candidates: $^{13}$C$^+$ and CH$^+$. Once the correction described in Appendix~\ref{AppA} has been applied, spectra at mass-per-charge 13~u\textperiodcentered q$^{-1}$ show a very faint signal attributed to CH$^+$ (see Fig.~\ref{fig3}, middle). A weak but stronger peak is also visible at 14~u\textperiodcentered q$^{-1}$ (see Fig.~\ref{fig3}, bottom) and attributed to CH$_2^+$. There is no evidence for N$^+$. From LR spectra, it appears relatively difficult to identify the most favourable periods for the detections of these ions. At 13~u\textperiodcentered q$^{-1}$, even though some peaks appeared around perihelion, the most favourable conditions seem to be met at large heliocentric distances (in yellow, blue, and green). One should not be misguided by the relatively low signal in LR: after the pixel shift on 27 January 2016, the peak 13~u\textperiodcentered q$^{-1}$ shifted to the left edge of the detector such that the left part of the peak is lost. At 14~u\textperiodcentered q$^{-1}$, there is a similar behaviour and overall, the highest counts occurred on average at large heliocentric distances.

\begin{figure}
	\stackinset{r}{0.6cm}{b}{3.4cm}{\fbox{LR}}{%
	\includegraphics[width=\linewidth,trim=0 1.9cm 0 5cm,clip]{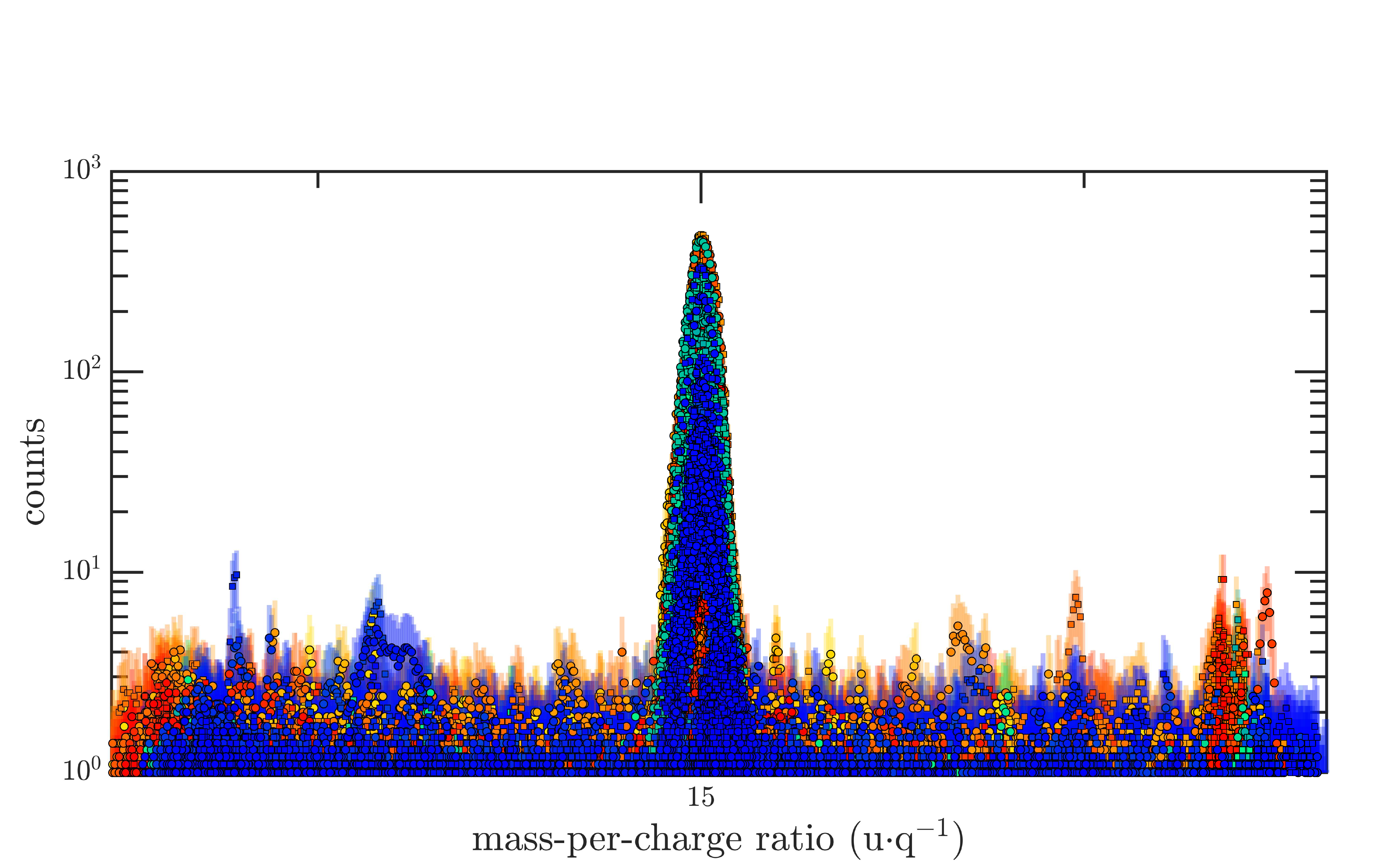}\\
}
	\stackinset{r}{0.6cm}{b}{0.33667905cm+3.4cm}{\fbox{HR}}{%
	\includegraphics[width=\linewidth,trim=0 0.0cm 0 0cm,clip]{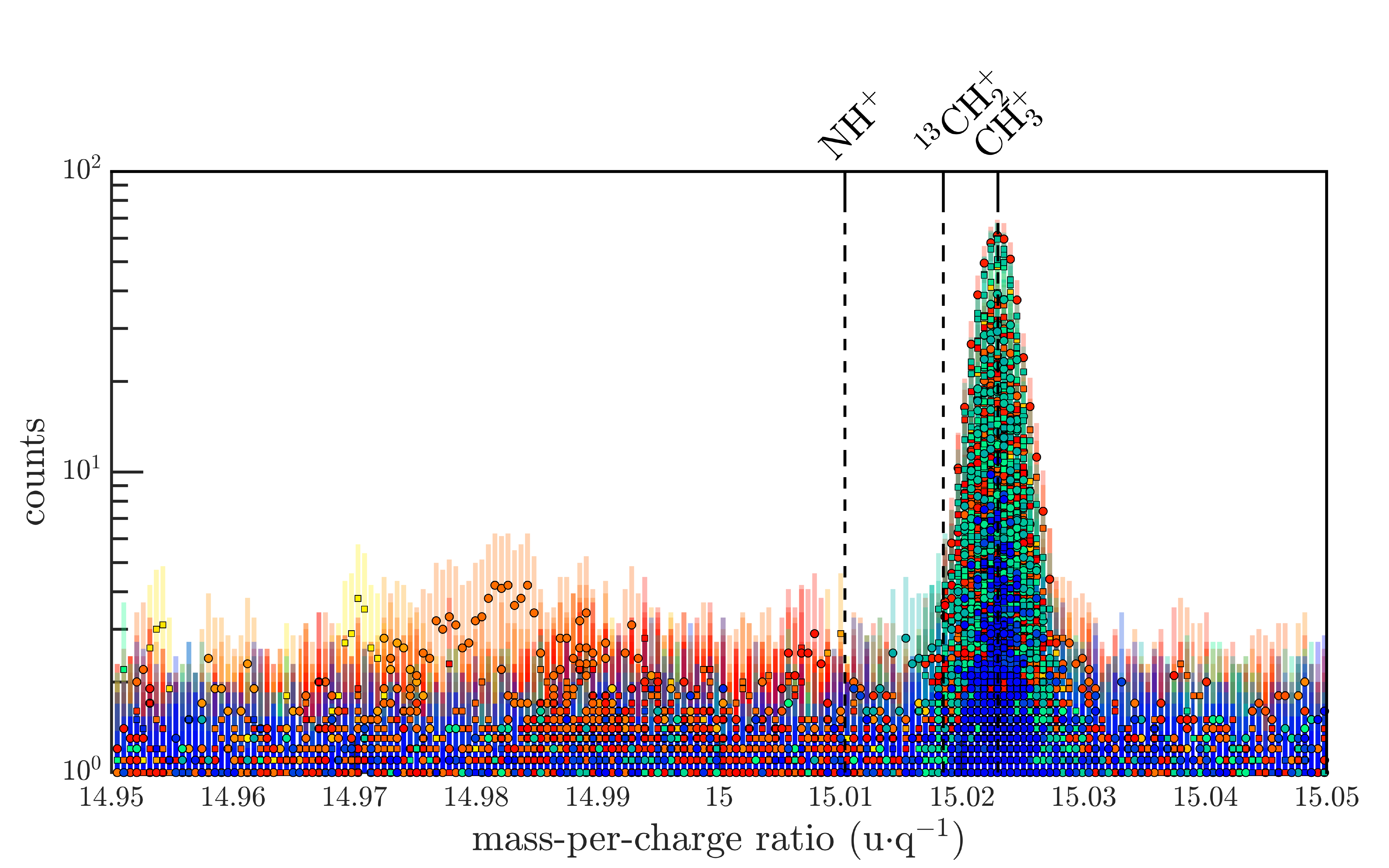}\\
}
	\caption{Same as Fig.~\ref{fig3}, but for 15~u\textperiodcentered q$^{-1}$ only. Stacked spectra in low resolution (upper) and HR (bottom).\label{fig4}}
\end{figure}

\begin{figure}
	\stackinset{l}{0.3115cm+0.6cm}{b}{3.4cm}{\fbox{LR}}{%
	\includegraphics[width=\linewidth,trim=0 1.9cm 0 5cm,clip]{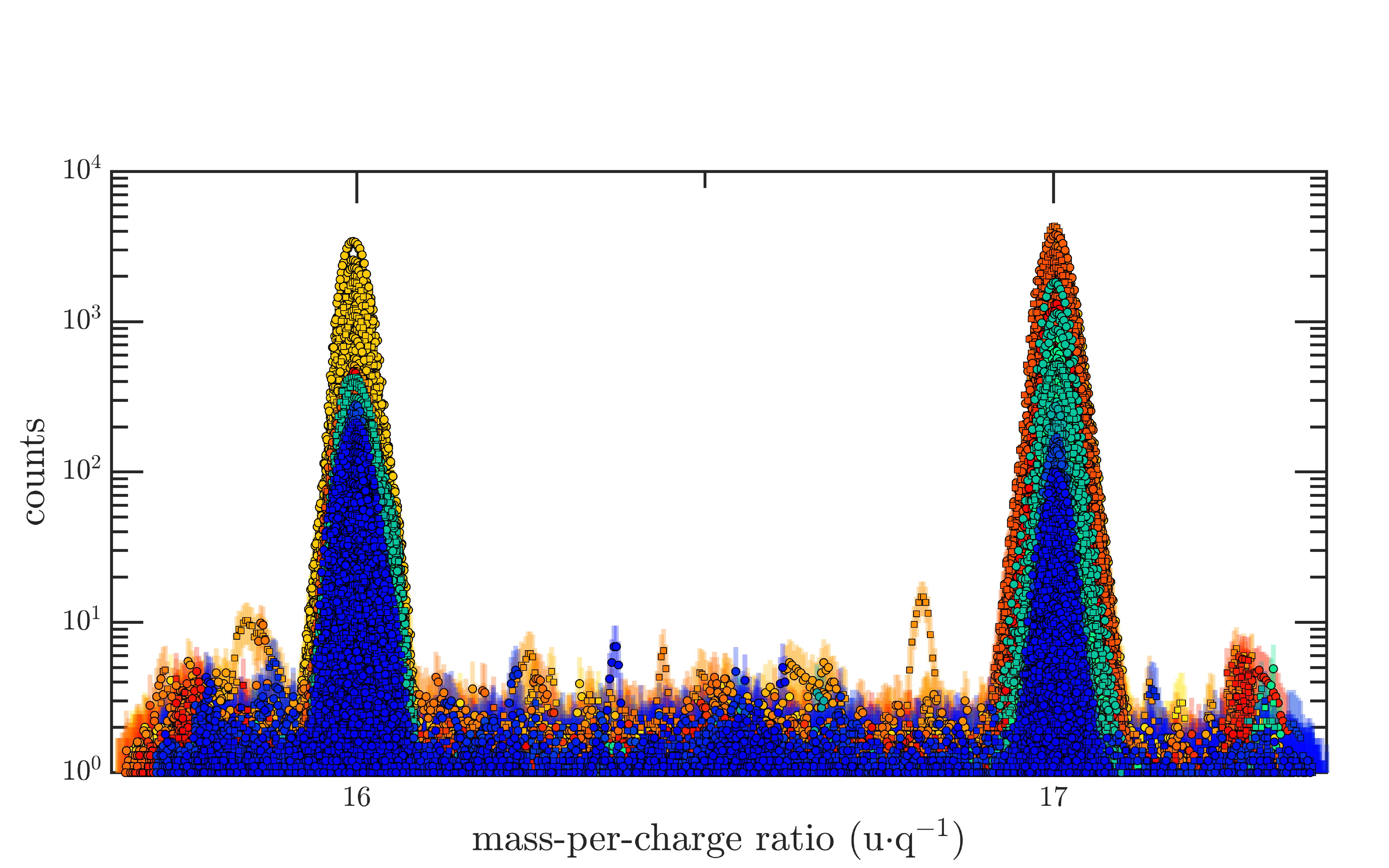}}\\
	\stackinset{l}{0.3115cm+0.6cm}{b}{3.4cm}{\fbox{HR}}{%
	\includegraphics[width=\linewidth,trim=0 1.9cm 0 0cm,clip]{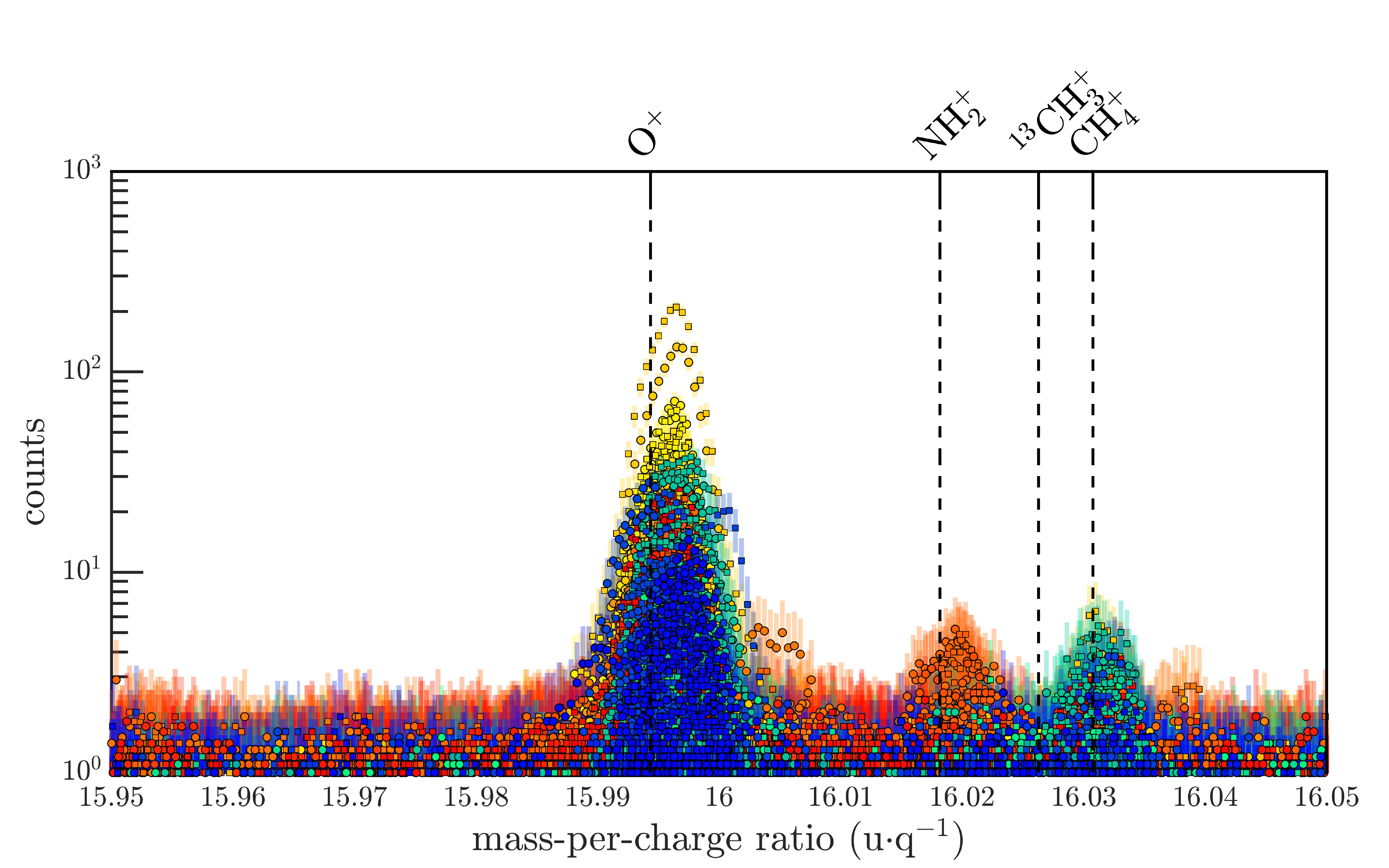}}\\
	\stackinset{l}{0.3115cm+0.6cm}{b}{0.33667905cm+3.4cm}{\fbox{HR}}{%
	\includegraphics[width=\linewidth,trim=0 0.0cm 0 0cm,clip]{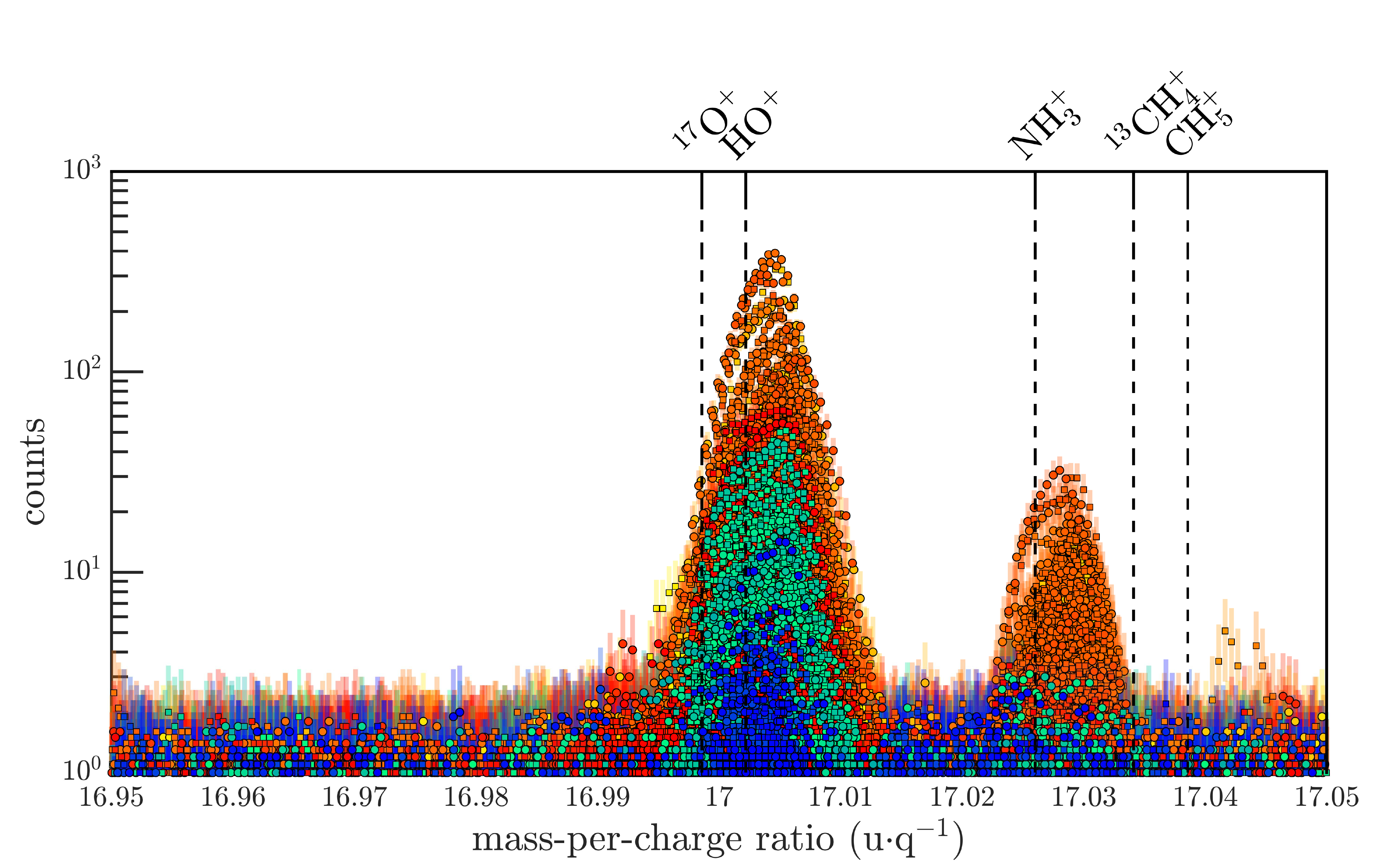}}\\
	\caption{Same as Fig.~\ref{fig3}, but for 16-17~u\textperiodcentered q$^{-1}$. Stacked spectra in low resolution (upper) and HR at 16~u\textperiodcentered q$^{-1}$ (middle) and at 17~u\textperiodcentered q$^{-1}$ (bottom).\label{fig5}}
\end{figure}

Fig.~\ref{fig4} shows mass-per-charge 15~u\textperiodcentered q$^{-1}$ in LR (top) and HR (bottom). It is one of the rare commanded mass-per-charge ratio for which LR spectra (with 18~u\textperiodcentered q$^{-1}$ sometimes) only cover one integer in u\textperiodcentered q$^{-1}$. The peak is associated with CH$_3^+$, as seen in HR and its intensity is quite strong compared with CH$^{+}$, CH$_2^{+}$ (see Fig.~\ref{fig3}), and CH$_4^+$ (see Fig.~\ref{fig5}). LR and HR spectra show that the detection is not controlled by cometary conditions, in particular the outgassing rate. The main reason is that CH$_3^+$ is barely destroyed through ion-neutral reactions with the dominant cometary neutral species, namely H$_2$O, CO$_2$, and CO, as the corresponding kinetic rates are $\leq 10^{-11}$~cm$^3$\textperiodcentered s$^{-1}$ \citep{Bates1983,Herbst1985,Luca2002}.

Fig.~\ref{fig5} shows spectra for the range 16-17~u\textperiodcentered q$^{-1}$. Three ions are identified at 16~u\textperiodcentered q$^{-1}$ in HR: O$^+$, NH$_2^+$, and CH$_4^+$. The O$^+$ signal is stronger prior to spring equinox than near perihelion/winter solstice. At large heliocentric distances, the source of ions is mainly driven by ionisation of the neutral molecules by electron-impact \citep{Heritier2018}. Ion-neutral chemistry is limited or even negligible \citep{Galand2016}. The major sources of O$^+$ are ionisation of CO$_2$, followed by ionisation of H$_2$O, based on their respective ionisation rate and volume mixing ratios. Indeed, although the CO$_2$ abundance is spatially depending on the sub-spacecraft latitude \citep{Hassig2015,Gasc2017}, the photo-ionisation rate yielding O$^{+}$ is an order of magnitude higher than that of H$_2$O \citep{Huebner2015}. Alongside O$^+$, two other ions are present: NH$_2^+$ and CH$_4^+$, while there is no evidence of $^{13}$CH$_3^+$. As $^{13}$CH$_3^+$ should slowly react with H$_2$O like CH$_3^+$, if by any chance $^{13}$CH$_n^+$ ($n=1-4$) would be detectable, $^{13}$CH$_3^+$ would be the best candidate. The non-detection of $^{13}$CH$_3^+$ implies that $^{13}$CH$^+$, $^{13}$CH$_2^+$, and $^{13}$CH$_4^+$ would not be detected either, which is indeed the case. According to the isotopic ratio $^{13}$C/$^{12}$C derived by \citet{Hassig2017}, $^{13}$CH$_3^+$ should be at 1\% height peak level from $^{12}$CH$_3^+$, that is about 0.6 counts in the best case, therefore preventing its detection. The CH$_4^+$ signal (Fig.~\ref{fig5}) is much weaker than that of CH$_3^+$ (Fig.~\ref{fig4}), from fivefold to tenfold, and is only detected at large heliocentric distances. The electron-impact ionisation of CH$_4$ is expected to slightly favour CH$_4^+$ compared to CH$_3^+$, as the associated cross sections are alike \citep{Song2015}, while the ionisation potential is lower for the production of CH$_4^+$ \citep[12.61 eV for CH$_4^+$, 14.25~eV for the dissociative ionisation of CH$_4$ into CH$_3^+$ at 0~K,][]{Samson1989}. Assuming that the ionisation of CH$_4$ is the main source of CH$_n^+$ ($n=1-4$), CH$_3^+$ however dominates over CH$_4^+$ as it almost does not react with H$_2$O, CO, and CO$_2$, as found for 1P/Halley at 0.9~au \citep{Allen1987}. The photo-ionisation rate of CH$_4$ by EUV leading to CH$_4^+$ is roughly twice the corresponding value for CH$_3^+$ \citep{Huebner2015}. The very high count ratio of CH$_3^+$ over CH$_4^+$, especially near perihelion is a clear signature of ion-neutral chemistry occurring in the coma. Possible other sources of CH$_3^+$, and not of CH$_4^+$, are the dissociative ionisation and ionisation following fragmentation of saturated hydrocarbons {(excluding CH$_4$)}, found at 67P \citep{Schuhmann2019}, or the protonation of CH$_2$ like at 1P \citep{Altwegg1994}. 

In contrast to CH$_4^+$, NH$_2^+$ is detected near perihelion, when the photo-ionisation rate and outgassing rates are larger, and not at large heliocentric distances. NH$_2^+$ results from the dissociative ionisation of NH$_3$, but its yield is fourfold less than that of NH$_3^+$ \citep{Huebner2015}. In addition, as NH$_2^+$ is lost through ion-neutral chemistry with H$_2$O, this indicates that its detection at perihelion stems from a higher production rate from NH$_3$. The peaks in LR at 16 ~u\textperiodcentered q$^{-1}$ show a right `shoulder' and, at times, a double peak. While this behaviour may be associated with some instrumental effects \citep{DeKeyser2015}, it more likely results from the contribution of two ion species, for instance O$^+$ and CH$_4^+$ ($\sim11$ pixels apart in LR) and/or O$^+$ and NH$_2^+$ ($\sim7$ pixels apart in LR). Overall, in view of the HR spectra, the main contributor at 16~u\textperiodcentered q$^{-1}$ is O$^+$ whose predominance occurred at large heliocentric distances prior the inbound Equinox, whereas NH$_2^+$ appeared at perihelion when photo-ionisation is much stronger. Interestingly, CH$_4^+$ is more abundant near the outbound Equinox as a possible consequence of the evolution of the neutral composition: \citet{Schuhmann2019} showed that there was a clear enhancement in the CH$_4$/H$_2$O ratio by a factor $\sim20$ between May 2015 and May 2016.

There exists diverse causes of spacecraft pollution specific to Rosetta \citep{Schlappi2010} for nitrogen bearing components which cannot be completely excluded for observations performed after and close to spacecraft manoeuvres since UV photolysis of hydrazine N$_2$H$_4$ can be a source of N$_2$H$_3$, N$_2$H$_2$, and, to a lesser extent, of NH$_3$ and NH$_2$ \citep{Biehl1991,Vaghjiani1993}. As it might in turn affect the detection of NH$_4^+$ (through the ion-neutral reaction NH$_3$+H$_3$O$^+$$\longrightarrow$NH$_4^+$+H$_2$O), in particular after manoeuvres, it might affect NH$_3^+$ and NH$_2^+$ as well \citep{Beth2016}.

At 17~u\textperiodcentered q$^{-1}$ (Fig.~\ref{fig5}), two ions, HO$^{+}$ and NH$_{3}^+$, have been detected. Both are mainly produced by ionisation of their respective parent neutral molecules, H$_2$O and NH$_3$ respectively. HO$^+$ follows the same pattern as the water production with increased intensity as 67P gets closer to the Sun. The NH$_{3}^+$ signal is quite strong, mainly near perihelion. In addition to the ionisation of NH$_3$, NH$_{3}^+$ can be produced through charge transfer between H$_2$O$^+$ and NH$_3$. Although NH$_{3}^+$ may be lost through the reverse charge exchange reaction with H$_2$O, the reaction is slow (rates of about 10$^{-10}$ cm$^3$\textperiodcentered s$^{-1}$) and therefore its contribution to the ion composition remains {negligible} compared with others ions reacting with H$_2$O (see details in Section~\ref{sec41}). To summarise, HO$^+$ is seen throughout the escort phase with a maximum in intensity near perihelion, when outgassing rate and photoionisation are strong. NH$_3^+$ follows the same pattern with high counts near perihelion but cannot be detected at large heliocentric distances because its parent molecule NH$_3$ is much less abundant than H$_2$O. For information, we have indicated the location of $^{17}$O$^+$ (see Fig.~\ref{fig5}, bottom): even if it might be present, its closeness to HO$^+$ ($\sim7$ pixels apart) and the peak deformation would prevent its detection. In addition, according to the isotopic ratio $^{17}$O/$^{16}$O derived by \citet{Schroeder2019} and the counts for $^{16}$O$^+$, $^{17}$O$^+$ is below the background level.

Peaks in HR do not fall at the exact location of a species and may present a distorted shape. This phenomenon is symptomatic of spectra at 16, 17, and, to a lesser extent, of 18~u\textperiodcentered q$^{-1}$ in HR \citep{DeKeyser2015}. Without corrections, the peak is not symmetric and the maximum of the peak is slightly shifted to the right ($\lesssim5$  pixels) due to the DFMS' characteristic double peak structure for this subset of masses \citep{DeKeyser2015}. We did not apply the correction proposed by \citet{DeKeyser2015} as it would not provide further insight on the ion identification.

\begin{figure}
	\stackinset{l}{0.315cm+0.6cm}{b}{3.4cm}{\fbox{LR}}{%
	\includegraphics[width=\linewidth,trim=0 1.9cm 0 5cm,clip]{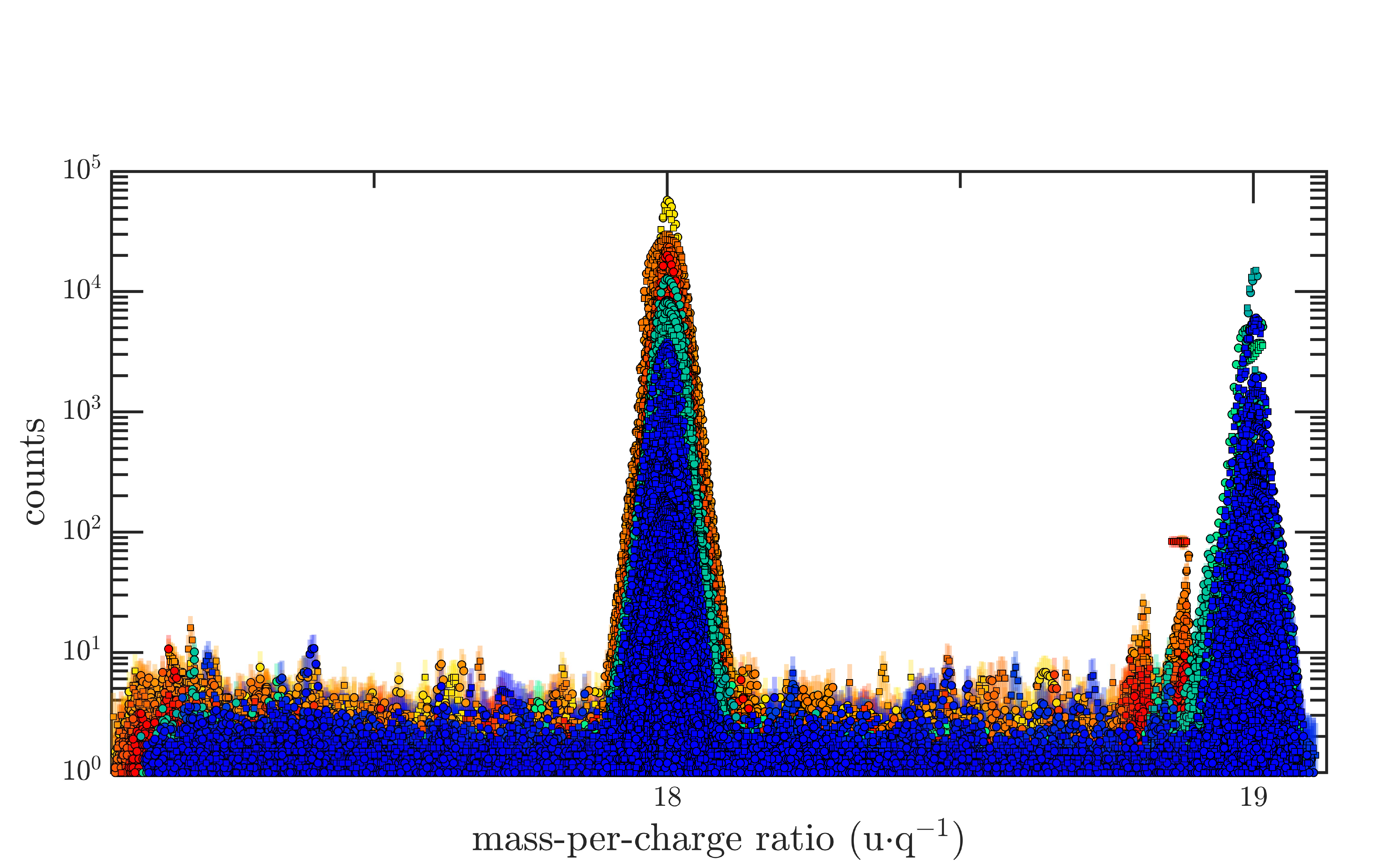}}\\
	\stackinset{l}{0.315cm+0.6cm}{b}{3.4cm}{\fbox{HR}}{%
	\includegraphics[width=\linewidth,trim=0 1.9cm 0 0cm,clip]{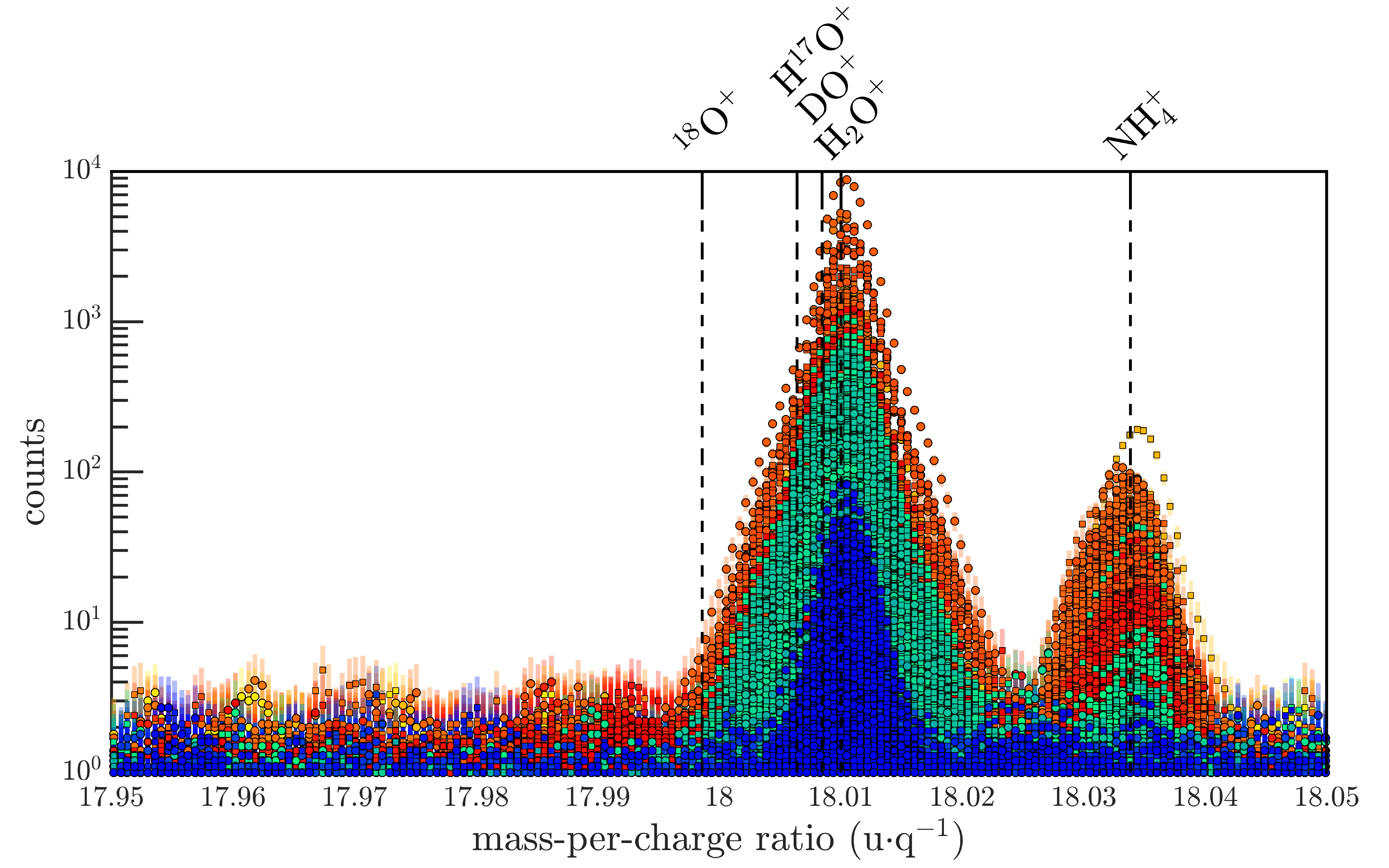}}\\
	\stackinset{l}{0.315cm+0.6cm}{b}{0.33667905cm+3.4cm}{\fbox{HR}}{%
	\includegraphics[width=\linewidth,trim=0 0.0cm 0 0cm,clip]{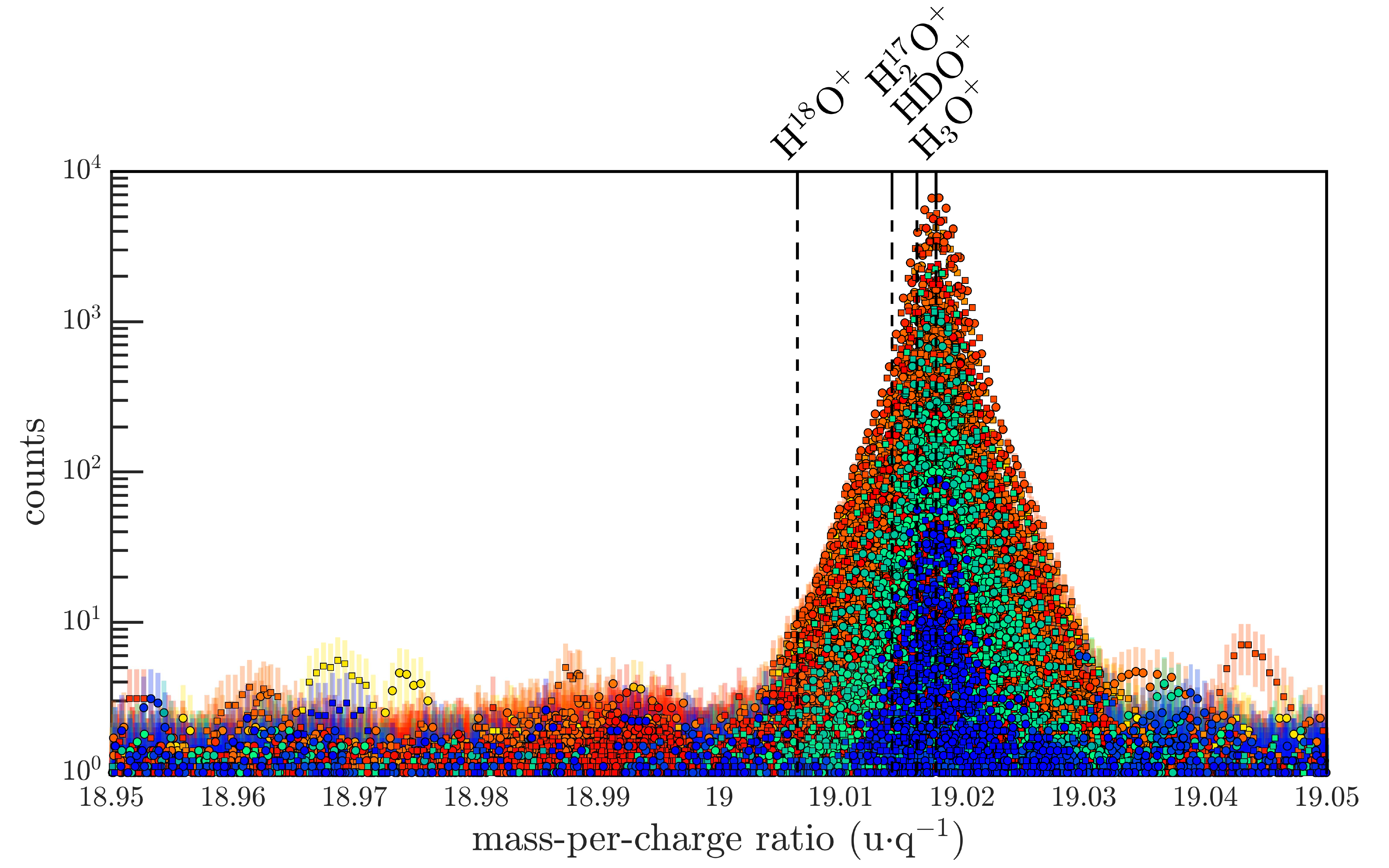}}\\
	\caption{Same as Fig.~\ref{fig3}, but for 18-19~u\textperiodcentered q$^{-1}$. Stacked spectra in low resolution (top panel) and HR at 18~u\textperiodcentered q$^{-1}$ (middle) and 19~u\textperiodcentered q$^{-1}$ (bottom). Isotopic ions have been indicated for information. As one may see, the DFMS mass-per-charge resolution cannot separate: neither H$^{17}$O$^+$ and DO$^+$ from H$_2$O$^+$, nor H$_2$$^{17}$O$^+$ and HDO$^+$ from H$_3$O$^+$, and the individual contributions have to be fitted numerically as done in the neutral mode \citep[cf.][]{Altwegg2015,Schroeder2019}. In addition, the peak in LR at 19~u\textperiodcentered q$^{-1}$ is at the edge of the detector such that it is only partially resolved. \label{fig6}}
\end{figure}

Fig.~\ref{fig6} shows spectra for the range 18-19~u\textperiodcentered q$^{-1}$ and two comments must be made. First, the range around 18~u\textperiodcentered q$^{-1}$ in both LR and HR was scanned threefold more often than any other mass ranges as a result of the organisation of the DFMS measurement sequence: while each u\textperiodcentered q$^{-1}$ range was scanned successively and increasingly, each sequence started and ended by scanning 18~u\textperiodcentered q$^{-1}$. In addition, in LR, both spectra centred on 18 and 18.16~u\textperiodcentered q$^{-1}$ have the 19~u\textperiodcentered q$^{-1}$ peak at the edge of the detector with its right part often lost out of the useful pixel range. HR spectra at 18~u\textperiodcentered q$^{-1}$ show clear signatures of H$_2$O$^{+}$ and NH$_4^{+}$ which have been already reported by \citet{Fuselier2016} and \citet{Beth2016}. The weak signal of NH$_4^{+}$ seen at large heliocentric distances is attributed to hydrazine during manoeuvres \citep{Schlappi2010}. However, near perihelion, it is produced through ion-neutral chemistry within the coma due to the high proton affinity of NH$_3$, higher than that of H$_2$O \citep{Beth2016}. Candidate ion isotopologues are also indicated, such as $^{18}$O$^{+}$, H$^{17}$O$^{+}$, and DO$^{+}$. The H$_2$O$^{+}$ peak is so strong and therefore so widely spread that it prevents their detection as their signal is expected to be weak if present. Indeed, for the strongest recorded signal, the peak covers almost 0.30~u\textperiodcentered q$^{-1}$ in total, over $\sim56$ pixels ($\sim28$ pixels each side). For comparison, $^{18}$O$^{+}$, H$^{17}$O$^{+}$, and DO$^{+}$ are located at 21, 7, and 3 pixels on the left (shortwards) of H$_2$O$^+$ respectively. {Again, assuming the isotopic ratios D/H, $^{17}$O/$^{16}$O, and $^{18}$O/$^{16}$O from \citet{Schroeder2019}, $^{18}$O$^{+}$, H$^{17}$O$^{+}$, and DO$^{+}$ should be less than 1 count, below the noise level in HR.}

H$_2$O$^+$ and NH$_4^+$ increase as the heliocentric distance decreases. H$_2$O$^{+}$ results from direct ionisation of H$_2$O, either by EUV or by electron impact \citep{Galand2016}. However, it reacts with H$_2$O yielding H$_3$O$^+$ (see Fig.~\ref{fig6}, bottom panel). H$_2$O$^{+}$ is a peculiar ion as its production and loss depend on the H$_2$O density in the coma. Photochemical equilibrium, for which production balances chemical losses, is reached at the location of Rosetta or closer to the nucleus depending on the outgassing activity (see Section~\ref{sec41}). Under such a condition, the H$_2$O$^+$ number density is given by: $\nu/k\sim10\sqrt{T[\text{K}]}/r_h[\text{au}]^2$ cm$^{-3}$ (as dissociative recombination is a negligible loss) where $\nu$ is the H$_2$O ionisation rate, $r_h$ is the heliocentric distance of 67P, $k$ is the reaction rate constant of: H$_2$O+H$_2$O$^{+}$$\longrightarrow$ H$_3$O$^{+}$+HO (see Appendix~\ref{AppB}), $k$ being relatively constant, and $T$ is the temperature of the gas. Overall, the H$_2$O$^{+}$ peak intensity exhibits this trend, increasing with decreasing heliocentric distance \citep[and hence increasing $\nu$, at least for the photoionisation,][]{Heritier2018}, though its variability over a day or during a month, in particular near perihelion is still puzzling \citep{Beth2016}. Possible reasons, not investigated in this paper, include the variability of the generally-negative spacecraft potential \citep{Odelstad2018}, the interaction between corotating interaction regions and coronal mass ejections \citep{Hajra2018} with the 67P's ionosphere, and to the proximity of the diamagnetic cavity \citep{Goetz2016a} or plasma boundaries \citep{Mandt2019}. NH$_4^+$ is produced by proton transfer between NH$_3$ and protonated molecules, mainly H$_2$O$^+$ and H$_3$O$^+$ \citep{Geiss1991,Beth2016}. Amongst neutral cometary species, NH$_3$ has the highest proton affinity and, therefore, the ability to steal a proton from other protonated molecular ions \citep{Heritier2017a}. NH$_4^+$ may be only lost through transport and, to a lesser extent, dissociative recombination, making it quite stable together with other ions, such as H$_2$O$^+$. Due to the abundance of its parent molecule H$_2$O, H$_2$O$^+$ is observed during the whole mission with highest counts near perihelion, whereas NH$_4^+$ is mainly detected near perihelion because the necessary conditions of its yielding, a large NH$_3$ number density and a large ion-neutral collision frequency, are only met during this period.

Spectra at 19~u\textperiodcentered q$^{-1}$ (see Fig.~\ref{fig6}, bottom panel) exhibit a large and strong peak associated with H$_3$O$^+$. As H$_2$O has a proton affinity higher than that of HO, once H$_2$O$^+$ is produced, it readily reacts with H$_2$O to yield H$_3$O$^+$, dominating the ion composition at short cometocentric distances \citep{Fuselier2016,Beth2016} in the absence of NH$_3$. Nevertheless, when the NH$_3$ number density is large enough, NH$_4^+$ may become the dominant ion species at distances of a few to tens of kilometres above the nucleus' surface depending on the NH$_3$ volume mixing ratio of a few percent. Close to the nucleus, H$_3$O$^+$ dominates, whereas H$_2$O$^+$ becomes the major ion at larger cometocentric distances. At the location of Rosetta near perihelion ($\sim150-200$~km), H$_3$O$^+$ was expected to dominate over NH$_4^+$ as is indeed observed. In Fig.~\ref{fig6} bottom, we have indicated cations containing minor isotopologues, namely H$^{18}$O$^+$, H$_2$$^{17}$O$^+$, and HDO$^{+}$. Like for H$_2$O$^+$, the H$_3$O$^+$ peak is so strong and wide, covering $\sim0.30$ ~u\textperiodcentered q$^{-1}$ ($\sim52$ pixels slightly lower than at 18~u\textperiodcentered q$^{-1}$) that the small signals corresponding to H$^{18}$O$^{+}$, H$_2$$^{17}$O$^{+}$, and HDO$^{+}$, located at 20, 6, and 3 pixels on the left side of H$_3$O$^+$ cannot be distinguished. Like for NH$_4^+$, the H$_3$O$^+$ peak is stronger near perihelion as expected.

\begin{figure}
	\stackinset{r}{0.6cm}{b}{3.4cm}{\fbox{LR}}{%
	\includegraphics[width=\linewidth,trim=0 1.9cm 0 5cm,clip]{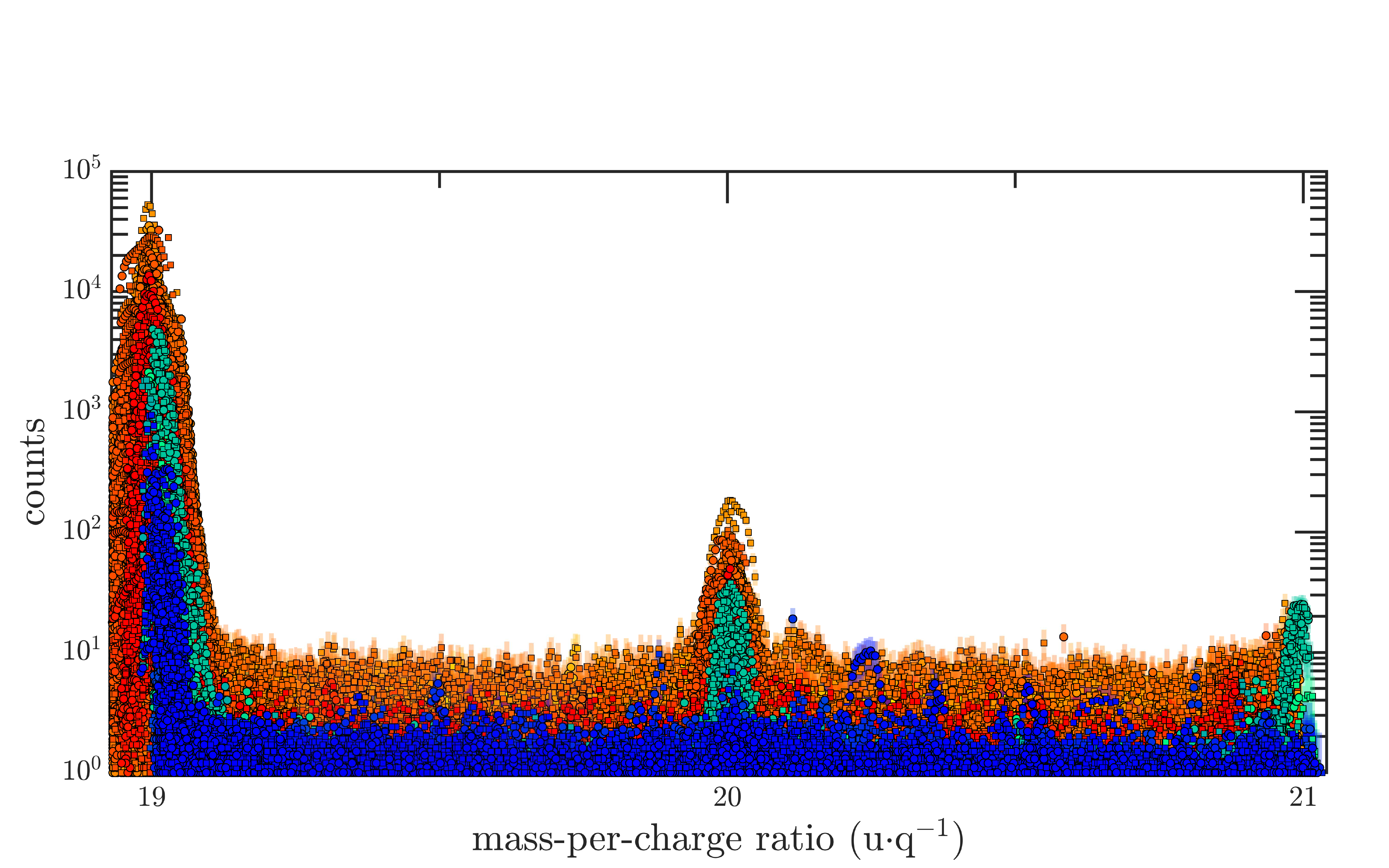}}\\
	\stackinset{r}{0.6cm}{b}{3.4cm+0.33667905cm}{\fbox{HR}}{%
	\includegraphics[width=\linewidth,trim=0 0cm 0 0cm,clip]{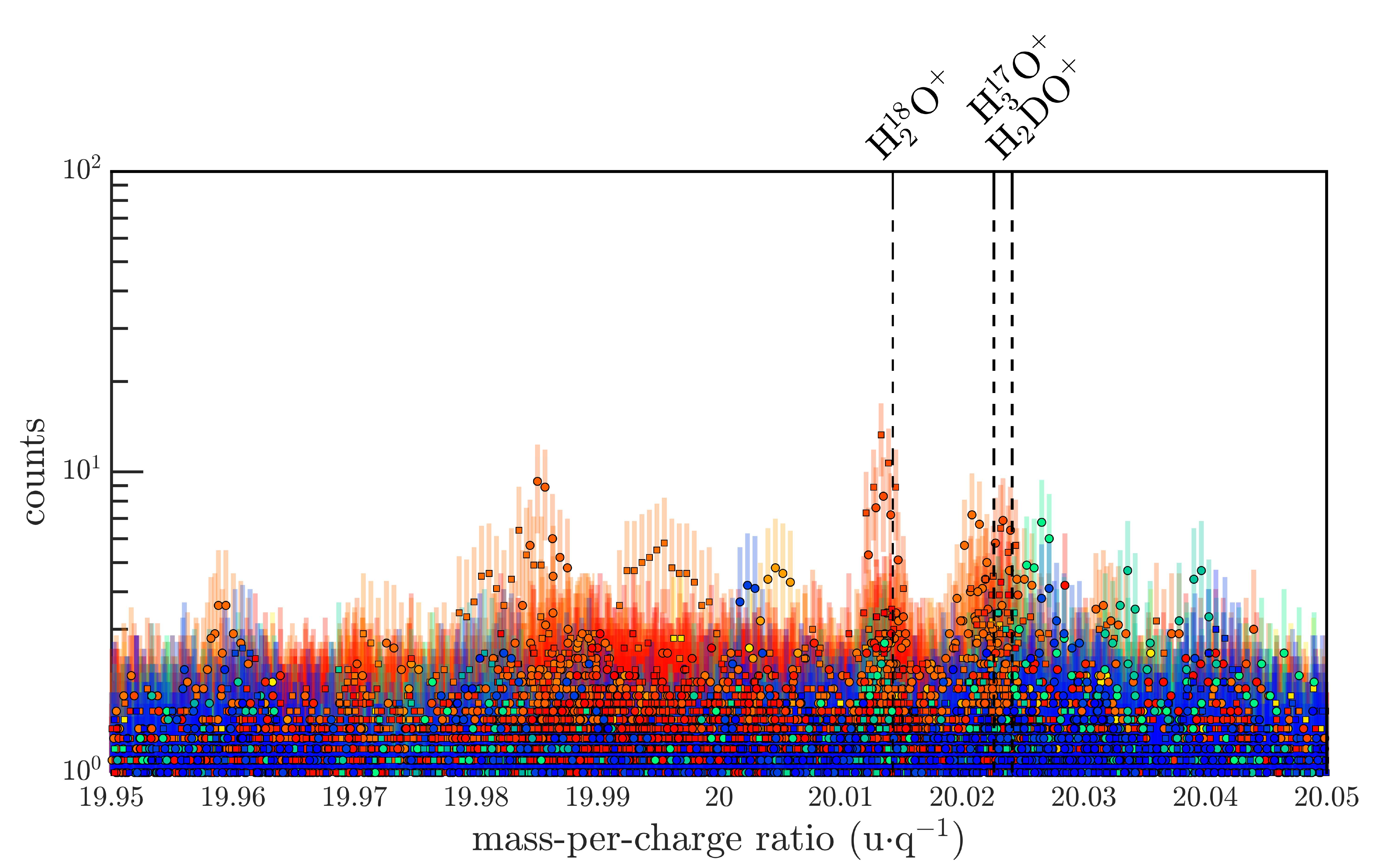}}\\
	\caption{Same as Fig~~\ref{fig3}, but for 20~u\textperiodcentered q$^{-1}$. Stacked spectra in low resolution (top panel) and in HR at 20~u\textperiodcentered q$^{-1}$ (bottom). The HR spectra at 20~u\textperiodcentered q$^{-1}$ are already shown in Fig.~\ref{fig6}. As one may see, the DFMS mass-per-charge resolution cannot separate: H$_ 3$$^{17}$O$^+$ from H$_2$DO$^+$. In addition, the peak in LR at 19~u\textperiodcentered q$^{-1}$ is at the edge of the detector such that it is only partially resolved. \label{fig7}}
\end{figure}

Fig.~\ref{fig7} shows LR spectra at 19-20~u\textperiodcentered q$^{-1}$ and HR spectra at 20~u\textperiodcentered q$^{-1}$. In HR, mass-per-charge ratio 19~u\textperiodcentered q$^{-1}$ is at the edge of the detector, on the left side (shortwards). The peak attributed to H$_3$O$^+$ is sometimes fully resolved before the pixel shift has been applied, depending on the magnet temperature, but not afterwards. Indeed, the reference pixel $p_0$ has been moved towards the left by 70 pixels on the 27$^{th}$ of January 2016 such that a peak located at the pixel $p_1$ before the shift was relocated at the pixel $p_1-70$ afterwards. Consequently, the lower and upper limits for each spectrum in terms of mass-per-charge increase by $\sim1.4\%$ in LR and $\sim0.23\%$ in HR after the shift.

A striking difference, when comparing these spectra with LR spectra at lower u\textperiodcentered q$^{-1}$, is the noise level: it reaches 10~counts, whereas on lower u\textperiodcentered q$^{-1}$, it never exceeded 3-4 counts. The HR spectra at 20~u\textperiodcentered q$^{-1}$ exhibit two peaks: one is unambiguously associated with H$_2$$^{18}$O$^{+}$ and another with either H$_3$$^{17}$O$^{+}$ or H$_2$DO$^{+}$ (or both). The resolving power is not sufficient to separate these species, only $2\times10^{-3}$~u\textperiodcentered q$^{-1}$ apart (3-4~pixels). However, we favour H$_2$DO$^{+}$ against H$_3$$^{17}$O$^{+}$: according to \citet{Schroeder2019}, in neutral mode, the H$_2$$^{17}$O signal is too low and buried in the shoulder of the HDO peak and thus we expect a similar behaviour for the corresponding protonated ions observed in ion mode. In addition, an estimation might be inferred from the DFMS-derived isotopic ratios of neutral species \citep{Altwegg2015,Schroeder2019}: D/H=$(5.3\pm0.7) \times 10^{-4}$, $^{18}$O/$^{16}$O=$(2.25\pm0.18) \times 10^{-3}$, and $^{17}$O/$^{16}$O=$(4.58\pm0.36) \times 10^{-4}$. Assuming that the ion-neutral chemistry of H$_2$O$^+$ and H$_3$O$^+$ and their isotopologues relies on proton transfer and that there is no difference in reaction rate constants, one may calculate ion abundance ratios. For instance, H$_2$D$^{16}$O$^{+}$/H$_3$$^{18}$O$^{+}$ should be $\approx$3(D/H)/($^{18}$O/$^{16}$O)=$(0.88\pm0.12)$ and H$_2$D$^{16}$O$^{+}$/H$_3$$^{17}$O$^{+}\approx$3(D/H)/($^{17}$O/$^{16}$O)=$(4.3\pm1.6)$. As the counts never exceeded $\sim7$, the peak, which is the sum of both ions, is likely dominated by H$_2$D$^{16}$O$^{+}$ and not H$_3$$^{17}$O$^{+}$. Moreover, one should note that the peak 20~u\textperiodcentered q$^{-1}$ in LR looks asymmetric for some spectra, with a right shoulder which might be caused by the simultaneous presence of H$_2$$^{18}$O$^+$ and H$_2$DO$^+$. Overall, a peak, symmetric or not, is present throughout the mission.

\begin{figure}
	\stackinset{r}{0.6cm}{b}{3.4cm}{\fbox{LR}}{%
	\includegraphics[width=\linewidth,trim=0 1.9cm 0 5cm,clip]{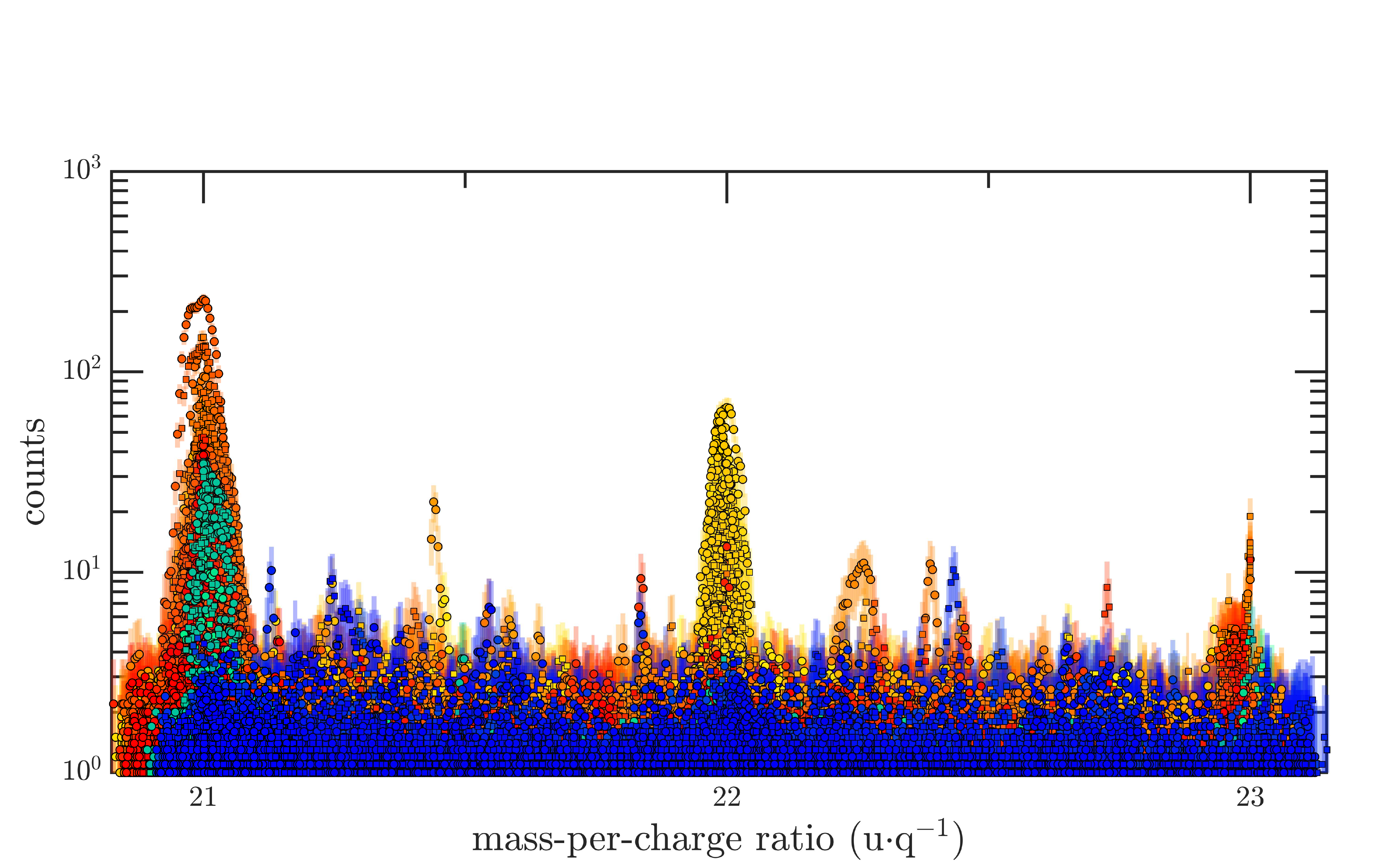}}\\
	\stackinset{r}{0.6cm}{b}{0.33667905cm+3.4cm}{\fbox{HR}}{%
	\includegraphics[width=\linewidth,trim=0 0.0cm 0 0cm,clip]{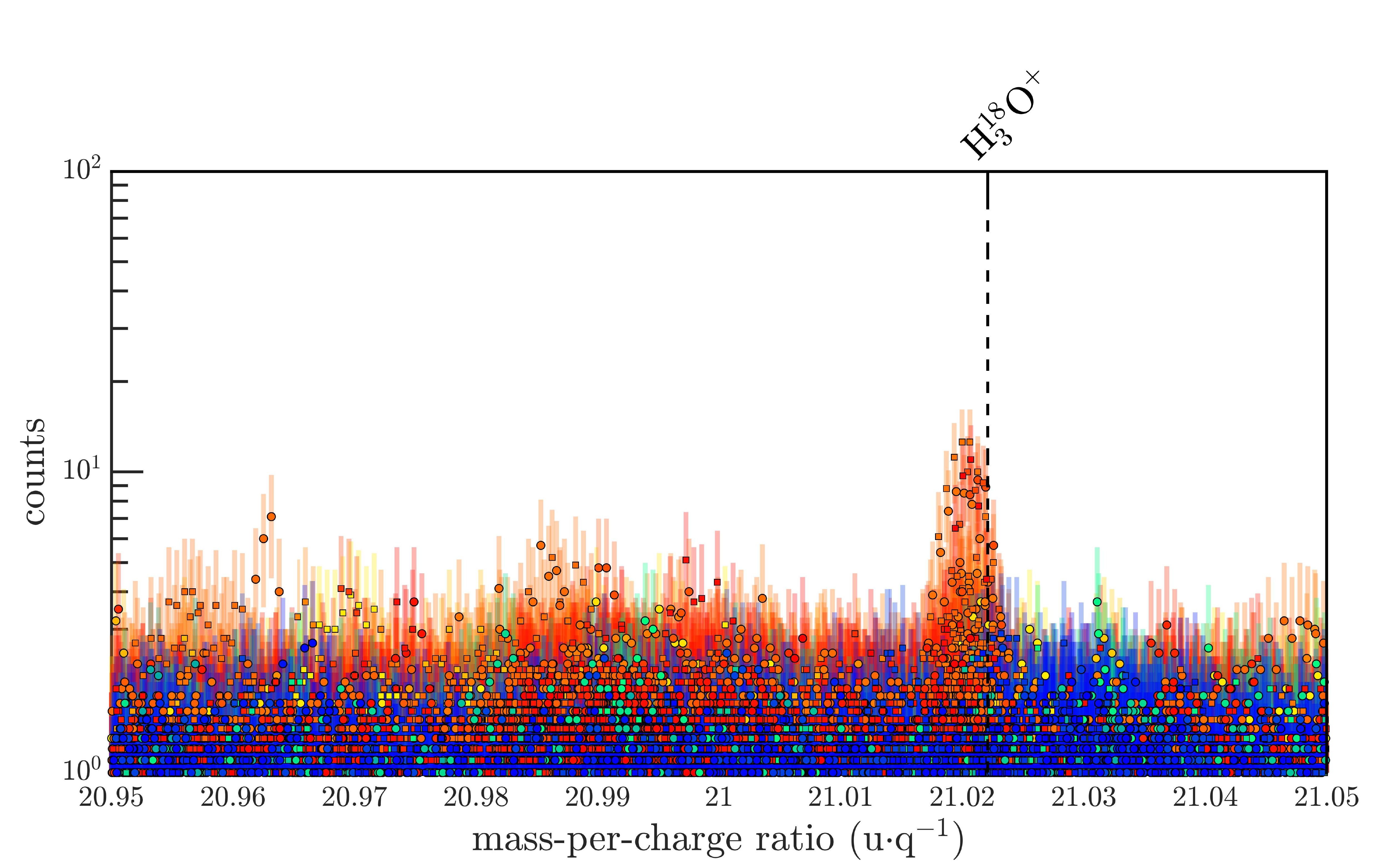}}\\
	\caption{Same as Fig.~\ref{fig3}, but for 21-22~u\textperiodcentered q$^{-1}$. Stacked spectra in low resolution (top panel) and in HR at 20~u\textperiodcentered q$^{-1}$ (bottom). There is no evidence for a species at 22~u\textperiodcentered q$^{-1}$ in high resolution and the associated spectra are not shown. In addition, the peak in LR at 23~u\textperiodcentered q$^{-1}$ is at the edge of the detector such that it is barely caught on its left edge (see also Fig.~\ref{fig2}).\label{fig8}}
\end{figure}

Fig.~\ref{fig8} shows spectra in LR and HR for the range 21-23~u\textperiodcentered q$^{-1}$. There is no ambiguity on the peak at 21~u\textperiodcentered q$^{-1}$, which corresponds to H$_3$$^{18}$O$^+$, detected only near perihelion. The single chemical path leading to H$_3$$^{18}$O$^+$ is the proton transfer between H$_2$$^{18}$O and H$_2$O$^+$. The H$_3$$^{18}$O$^+$/21~u\textperiodcentered q$^{-1}$ should follow in principle the same trend as the H$_3$O$^+$/19~u\textperiodcentered q$^{-1}$ peak. However, it could not be observed around/after the outbound Equinox (in blue) because of its weakness.

\subsection{Ion mass-per-charge 22~u\textperiodcentered q$^{-1}$\label{sec33}}

Fig.~\ref{fig8}, top panel, shows a peak at 22~u\textperiodcentered q$^{-1}$ in LR. The associated mass spectra in HR are not displayed because we do not observe persistent and reliable signals at 22~u\textperiodcentered q$^{-1}$. The presence of a peak in the LR spectra is at first surprising since there is no cometary neutral molecule and no candidate for a protonated one identified at 22~u\textperiodcentered q$^{-1}$. Based on neutral mode observations, we identify the corresponding ion as CO$_2^{++}$. In neutral mode, the ionisation of CO$_2$ in the ion source leads to the production of both CO$_2^+$ and CO$_2^{++}$ which is indeed observed in the data, whereas CO$_2^{++}$ dications detected in the ion mode are naturally produced in the coma. More details on this finding and a discussion on the identification of this dication are presented in Section~\ref{sec44}. It may be noticed that these cations are only detected before the inbound Equinox at large heliocentric distances when the H$_2$O density is low which might indicate that the corresponding ions react with H$_2$O or may only subsist for very low ion-neutral collision rates. The non detection of CO$_2^{++}$ out of this period might be also  linked to the cometocentric distance of Rosetta and the CO$_2^{++}$ lifetime. When CO$_2^{++}$ was detected, Rosetta was around 20-30~km from the nucleus. However, CO$_2^{++}$ is known to have different lifetimes depending on its electronic states, from $4$~s for its ground state \citep{Mathur1995} to $\mu\text{s}$ for excited states \citep{Alagia2009,Slattery2015}. Consequently, in order to have CO$_2^{++}$ be produced in sufficient quantity and detected by DFMS, $(r-r_c)/U$, where $r$ is the cometocentric distance of Rosetta, $r_c$ is the nucleus' radius, and $U$ is the neutral speed, should be of the order of or lower than the lifetime of CO$_2^{++}$. If $(r-r_c)/U\gtrsim 4$~s, only a small fraction of CO$_2^{++}$ ions would have the time to reach the spacecraft after they have been created. During the periods when DFMS was operating in ion mode, this condition was the most favourable in December 2014 - January 2015; for this period, $(r-r_c)/U\approx 20-30$~s (see Fig.~\ref{fig1} assuming $U\sim1$~km\textperiodcentered s$^{-1}$). This also implies that the CO$_2^{++}$ ions which are detected are most likely in the ground state \citep[though the latter is less likely to be produced than the short-lived excited state,][]{Masuoka1994,Alagia2009}.

\subsection{Ion mass-per-charge range 23 – 40~u\textperiodcentered q$^{-1}$}

\begin{figure}
	\stackinset{r}{0.6cm}{b}{3.4cm}{\fbox{LR}}{%
	\includegraphics[width=\linewidth,trim=0 1.9cm 0 5cm,clip]{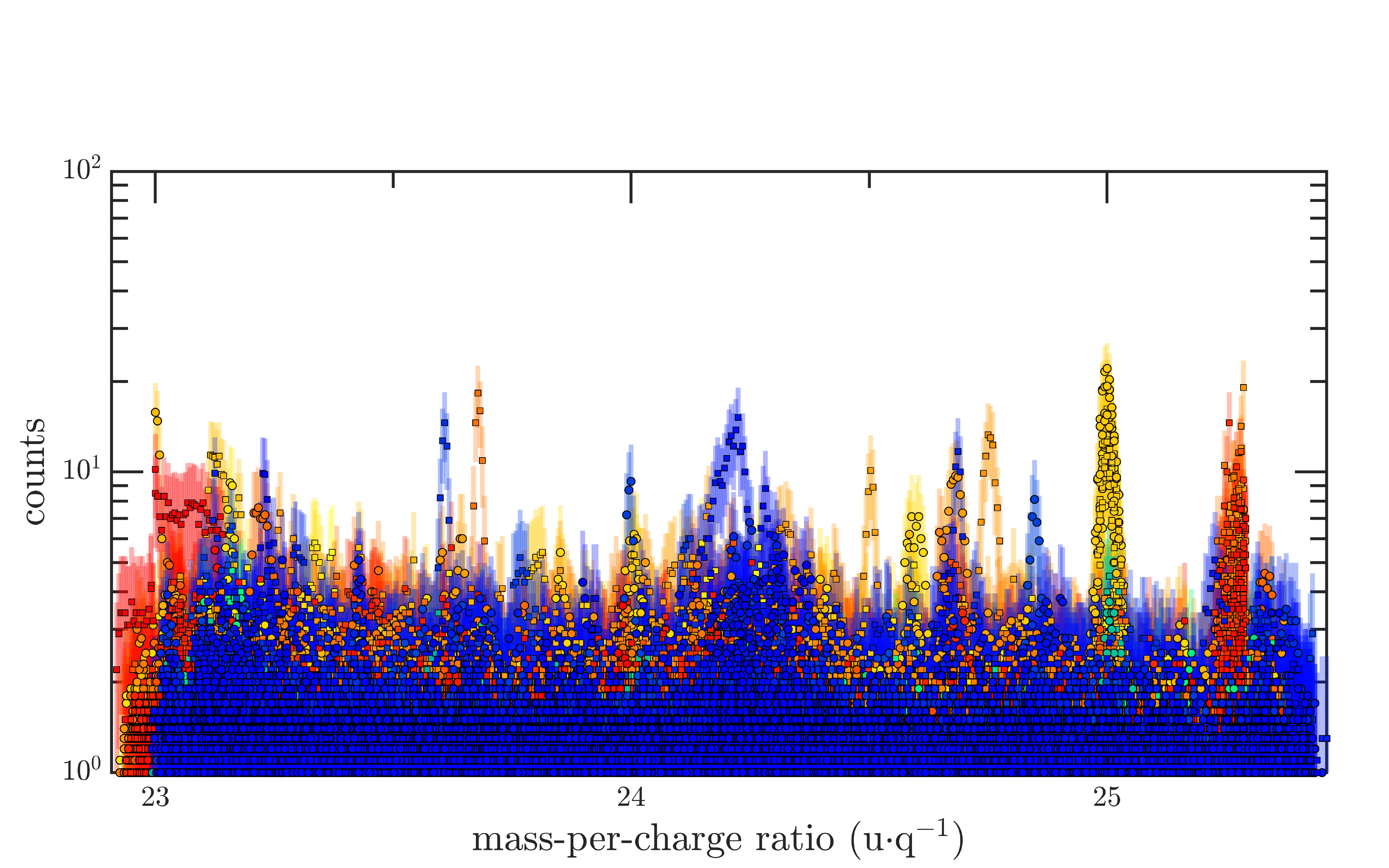}\\
}
\stackinset{r}{0.6cm}{b}{0.33667905cm+3.4cm}{\fbox{HR}}{%
	\includegraphics[width=\linewidth,trim=0 0.0cm 0 0cm,clip]{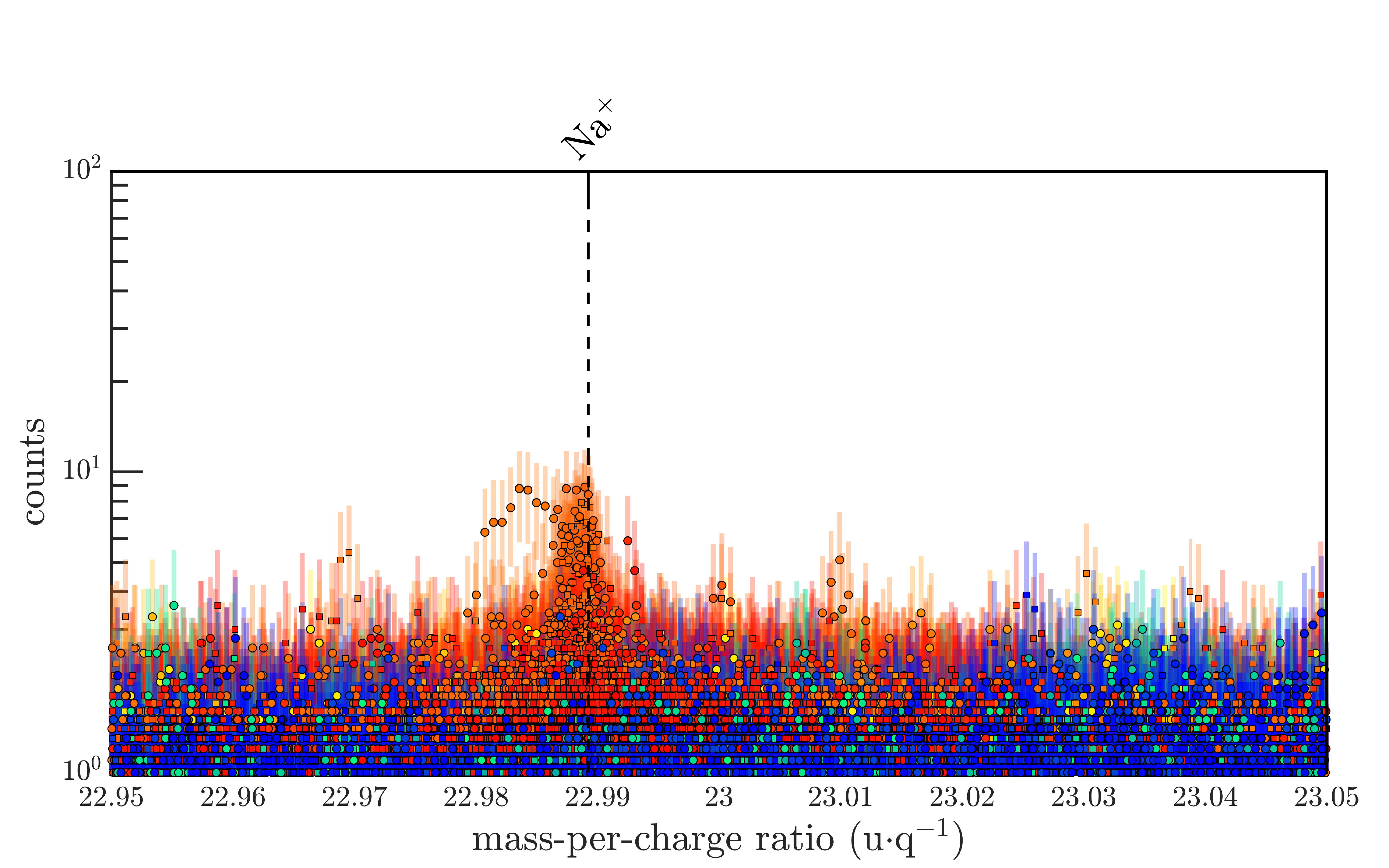}\\
}
	\caption{Same as Fig.~\ref{fig3}, but for 23-25~u\textperiodcentered q$^{-1}$. Stacked spectra in low resolution (top panel) and in HR at 23~u\textperiodcentered q$^{-1}$ (bottom). There is no evidence for a cation neither at 24~u\textperiodcentered q$^{-1}$ nor at 25~u\textperiodcentered q$^{-1}$ and the associated spectra are not shown. In addition, the peak in LR at 23~u\textperiodcentered q$^{-1}$ is at the edge of the detector such that it is barely caught on its right edge (see also Fig.~\ref{fig2}).\label{fig9}}
\end{figure}

In this subsection, we present ions which have been detected in this range; they are not part of the water ion group and are usually minor. Fig.~\ref{fig9} shows spectra for the range 23-25~u\textperiodcentered q$^{-1}$ in LR and at 23~u\textperiodcentered q$^{-1}$ in HR only. HR spectra at 24 and 25~u\textperiodcentered q$^{-1}$ do not exhibit any physical and reliable signal and are thus not displayed. The peak at 23~u\textperiodcentered q$^{-1}$ cannot be fully resolved in LR since, neither in Fig.~\ref{fig8} nor in Fig.~\ref{fig9}, the full extent of the peak is captured by the detector: in LR for scans centred on 21.98~u\textperiodcentered q$^{-1}$ (Fig.~\ref{fig8}), only the left ({lower masses}) tail of the peak is captured before the pixel shift, whereas in LR for scans centred on 24.18~u\textperiodcentered q$^{-1}$ (Fig.~\ref{fig9}) only the right ({higher masses}) tail of the peak is captured. In both cases, the peak is partially, or not at all, covered by the detector (see also Fig.~\ref{fig2}). However, both indicate a strong signal near perihelion. This is consistent with the HR spectra and the detection of Na$^+$ close to perihelion. The associated neutral, Na (sodium), is a refractory element and has been observed by DFMS early during the escort phase. Indeed, \citet{Wurz2015} reported its detection in October 2014, along those of Si (silicium), Ca (calcium), and K (potassium). Their presence results from solar wind sputtering of dust grains on 67P's surface. However, for heliocentric distances below 2~au, the coma of 67P is dense enough to create the so-called solar wind ion cavity within which the nucleus surface and dust grains close to Rosetta are shielded from solar wind ions \citep{Behar2017}. However, the solar wind can still sputter dust grains outside the cavity and an as-yet-unknown mechanism could therefore transport Na/Na$^+$ inwards, towards Rosetta. Another possibility might be sputtering by ENAs (Energetic Neutral Atoms), that is, neutralised solar wind ions, which are expected to be produced in large amount near perihelion, still having access to the nucleus of 67P unlike the solar wind \citep{Simon2019c}. Near perihelion, the source of Na and, therefore, that of Na$^+$ may be different. Na$^+$ was also observed at 21P/Giacobini-Zinner \citep{Geiss1986,Ogilvie1998} and 1P/Halley \citep{Krankowsky1986}. For 21P, \citet{Geiss1986} ruled out sputtering on cometary grains and favoured the idea that Na was either trapped in the ice or sorbed in or on carbonaceous grains. \citet{Ogilvie1998} suggested that Na might come from the evaporation of ice-like grains, containing Na in ionic form. Another scenario was proposed by \citet{Combi1997} for 1P/Halley, Benett C/1969 Y1, and Kohoutek C/1973 D for a near nucleus Na source: the photo-dissociation of the parent molecule, NaOH, characterised by a high photo-dissociation rate \citep[10$^{-3}$~s$^{-1}$ at 1~au,][]{Plane1991}. We note that Na, as an alkali metal, has a low ionisation energy of 5.139~eV, such that it can be ionised by the intense Lyman-$\alpha$, with a typical ionisation rate of $7\times10^{-6}$~s$^{-1}$ at 1~au \citep{Huebner2015}. In addition, it is interesting to point out the peculiar ion-neutral chemistry of Na and Na$^+$. As an alkali metal, Na is a great electron donor and reacts with several cations (e.g. H$_2$O$^+$ and NH$_3^+$) in the coma through charge exchange. It also reacts with protonated molecules. Given a high proton-affinity molecule X (e.g. X = H$_2$O, H$_2$CO, HCN, NH$_3$), and their protonated version XH$^+$, Na$^+$ results from Na + XH$^+$ $\longrightarrow$ Na$^+$ + X + H. This reaction with protonated molecules, which dominate the ion composition near perihelion \citep{Heritier2017a}, is likely a dominant source of Na$^+$. Unfortunately, it is impossible to track the evolution of Na$^+$ throughout the escort phase. In HR, the sensitivity is too low, only allowing the detection near perihelion. In LR, in spite of the higher sensitivity and the unambiguity on the ion species at 23~u\textperiodcentered q$^{-1}$, the bad quality of measurements at the edges of the MCP prevents from obtaining accurate data. However, according to Fig.~\ref{fig8}, top panel, one can conclude that the Na$^+$ number density maximises near perihelion and may reach even higher levels.

Even if peaks are detected in LR at 25~u\textperiodcentered q$^{-1}$, no species have been identified in HR. As shown in LR, the peaks are lower than 20-30~counts and many spurious ones contaminate the spectra. From spectra at lower u\textperiodcentered q$^{-1}$, we may define a rough scaling factor for counts between LR and HR modes. Due to the differences between slits, the sensitivity in HR is about 10\% to 1\% of its level in LR explaining why often no reliable peaks are observed in HR for C$_2^+$ or C$_2$H$^+$. However, the peak at 25~u\textperiodcentered q$^{-1}$ observed in LR at large heliocentric distances before the inbound Equinox is likely due to C$_2$H$^+$ for two reasons: there are no other species close to 25~u\textperiodcentered q$^{-1}$ and the peak disappeared at perihelion, indicating that the corresponding species should react with H$_2$O, which is the case of C$_2$H$^+$ \citep{Prasad1980}.

\begin{figure}
	\stackinset{r}{0.6cm}{b}{3.4cm}{\fbox{LR}}{%
	\includegraphics[width=\linewidth,trim=0 1.9cm 0 5cm,clip]{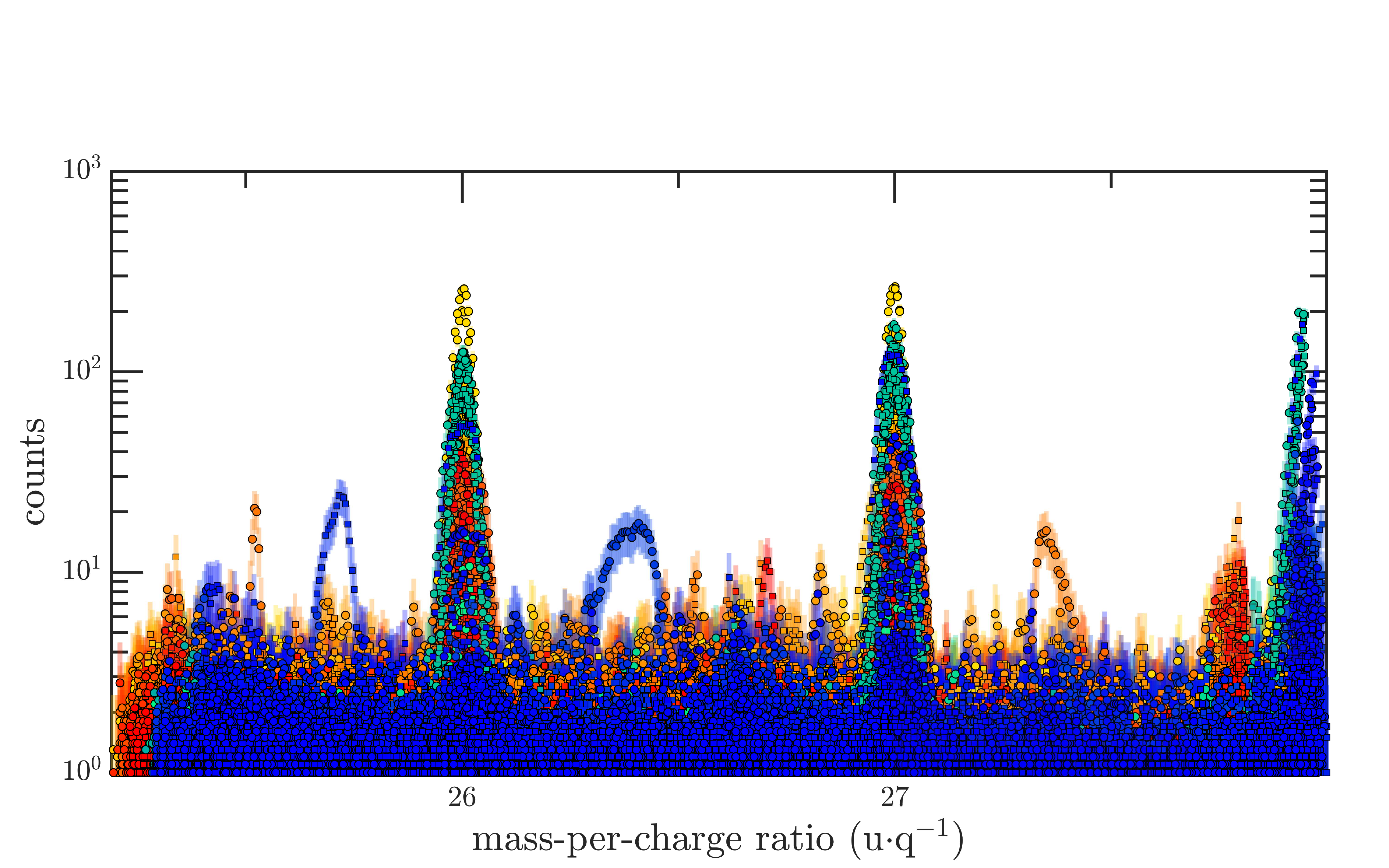}}\\
	\stackinset{r}{0.6cm}{b}{3.4cm}{\fbox{HR}}{%
	\includegraphics[width=\linewidth,trim=0 1.9cm 0 0cm,clip]{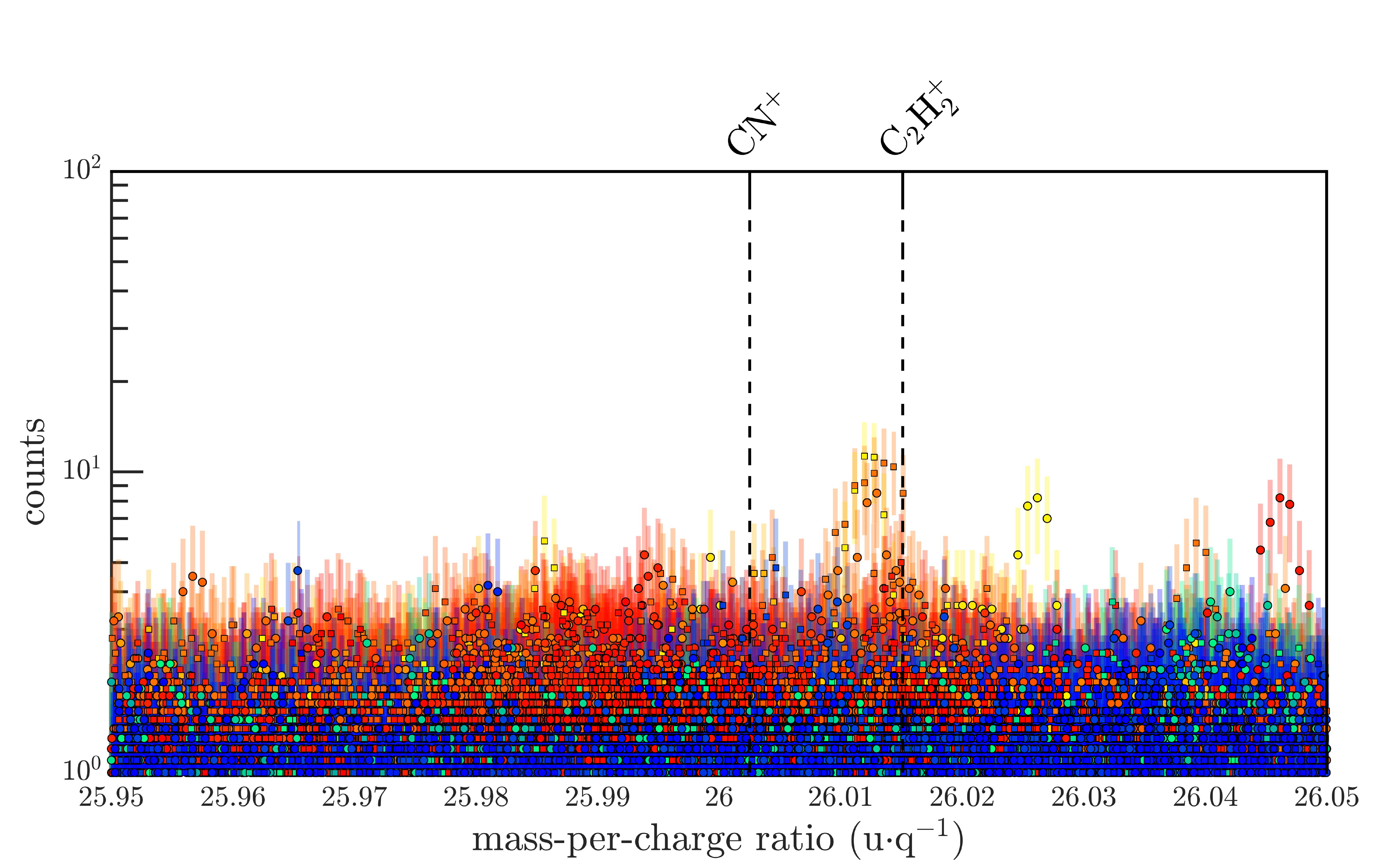}}\\
	\stackinset{r}{0.6cm}{b}{0.33667905cm+3.4cm}{\fbox{HR}}{%
	\includegraphics[width=\linewidth,trim=0 0.0cm 0 0cm,clip]{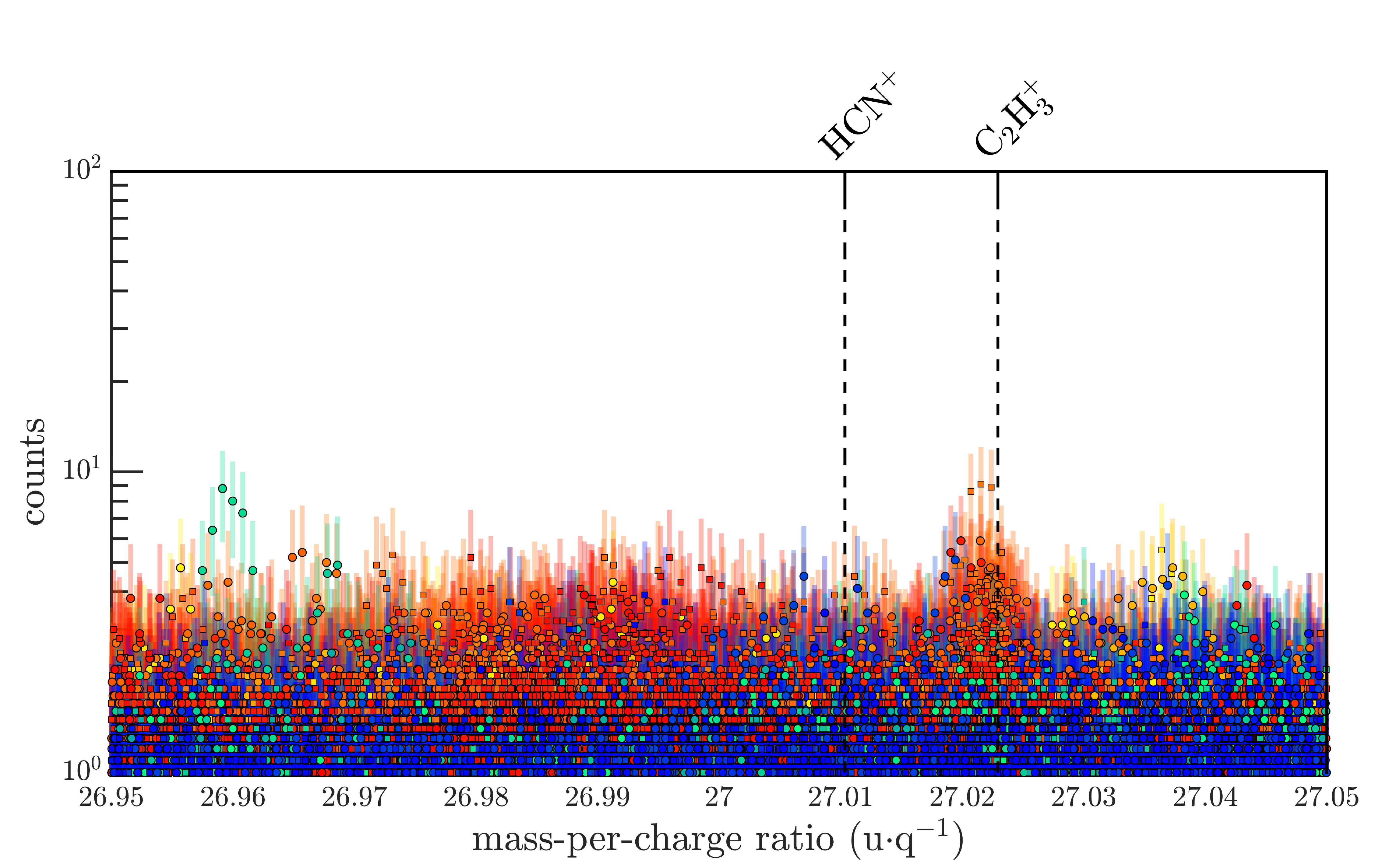}}\\
	\caption{Same as Fig.~\ref{fig3}, but for 26-27~u\textperiodcentered q$^{-1}$. Stacked spectra in low resolution (top panel), in HR at 26~u\textperiodcentered q$^{-1}$ (middle) and at 27~u\textperiodcentered q$^{-1}$ (bottom).\label{fig10}}
\end{figure}

Fig.~\ref{fig10} shows the range 26-27~u\textperiodcentered q$^{-1}$ in LR (top) and HR (middle and bottom) also allowing to distinguish the low u\textperiodcentered q$^{-1}$ part of a peak at 28~u\textperiodcentered q$^{-1}$. In HR, we barely see a peak for C$_2$H$_2^+$ at 26~u\textperiodcentered q$^{-1}$, without evidence of CN$^+$. The analysis for minor species above 23~u\textperiodcentered q$^{-1}$ in HR is made difficult due to the increased background level requiring unfortunately a visual inspection of each spectrum with two main criteria to select reliable peaks: (i) a well-shaped peak (i.e. the peak can be reasonably well-fitted by a single-Gaussian, two-Gaussian, Lorentzian, or a Voigt profile and be not too spiky/sharp) and (ii) the existence of a candidate species with an u\textperiodcentered q$^{-1}$ close enough to the peak location. Not all spectra at 26 ~u\textperiodcentered q$^{-1}$ exhibit a peak at C$_2$H$_2^+$. Only three spectra, out of more than hundreds through the escort phase, reach almost 10 counts at the location of C$_2$H$_2^+$, which is however sufficient to ascertain its presence. There is no evidence for CN$^+$ which can be explained by a reaction rate of CN$^+$ with H$_2$O about tenfold higher than that of C$_2$H$_2^+$ \citep{Anicich2003}. Near perihelion, the high cometary activity leads to the loss of CN$^+$ through chemistry impeding its detection. Stronger signals are observed on LR spectra at large heliocentric distances but the counts barely exceed 100 which explain the poor or lack of detection in HR. This is consistent with the detection of a faint peak at the location of C$_2$H$_3^+$ on the spectra at 27~u\textperiodcentered q$^{-1}$ suggesting that the latter has a common origin with C$_2$H$_2^+$. This might be investigated by correlating both signals which is beyond the scope of this paper. There is no evidence of HCN$^+$ in HR. However, at large heliocentric distances, LR spectra at 27~u\textperiodcentered q$^{-1}$ exhibit two overlapping peaks of similar amplitude separated by a few pixels only. Although the low resolution does not allow to unambiguously separate HCN$^+$ from C$_2$H$_3^+$, the double peak at 27~u\textperiodcentered q$^{-1}$ in LR suggests the presence of both HCN$^+$ and C$_2$H$_3^+$ ions with comparable contributions at large heliocentric distances.

\begin{figure}
	\stackinset{l}{0.315cm+0.6cm}{b}{3.4cm}{\fbox{LR}}{%
	\includegraphics[width=\linewidth,trim=0 1.9cm 0 5cm,clip]{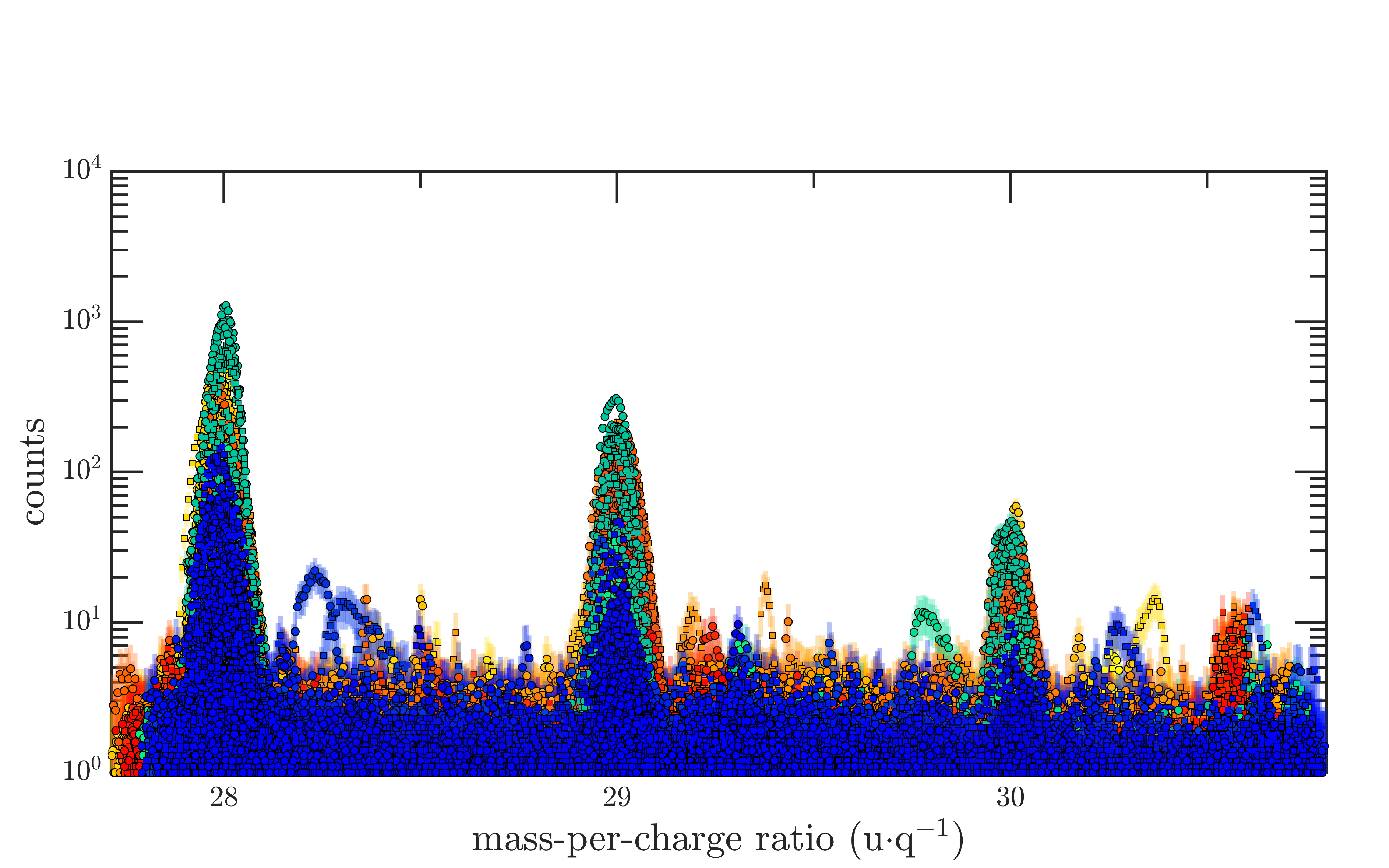}}\\
\stackinset{l}{0.315cm+0.6cm}{b}{3.4cm}{\fbox{HR}}{%
	\includegraphics[width=\linewidth,trim=0 1.9cm 0 0cm,clip]{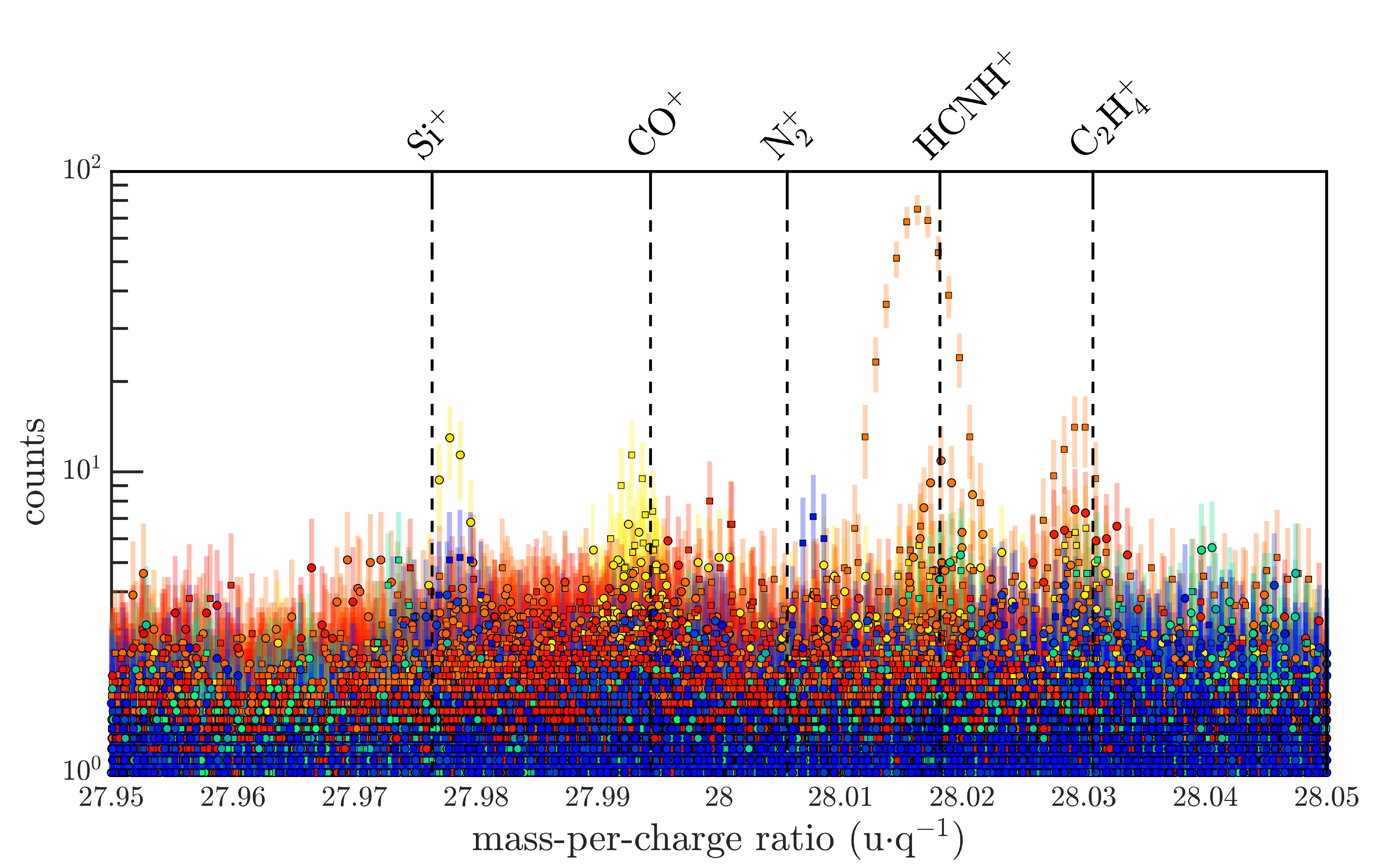}}\\
\stackinset{l}{0.315cm+0.6cm}{b}{0.33667905cm+3.4cm}{\fbox{HR}}{%
	\includegraphics[width=\linewidth,trim=0 0.0cm 0 0cm,clip]{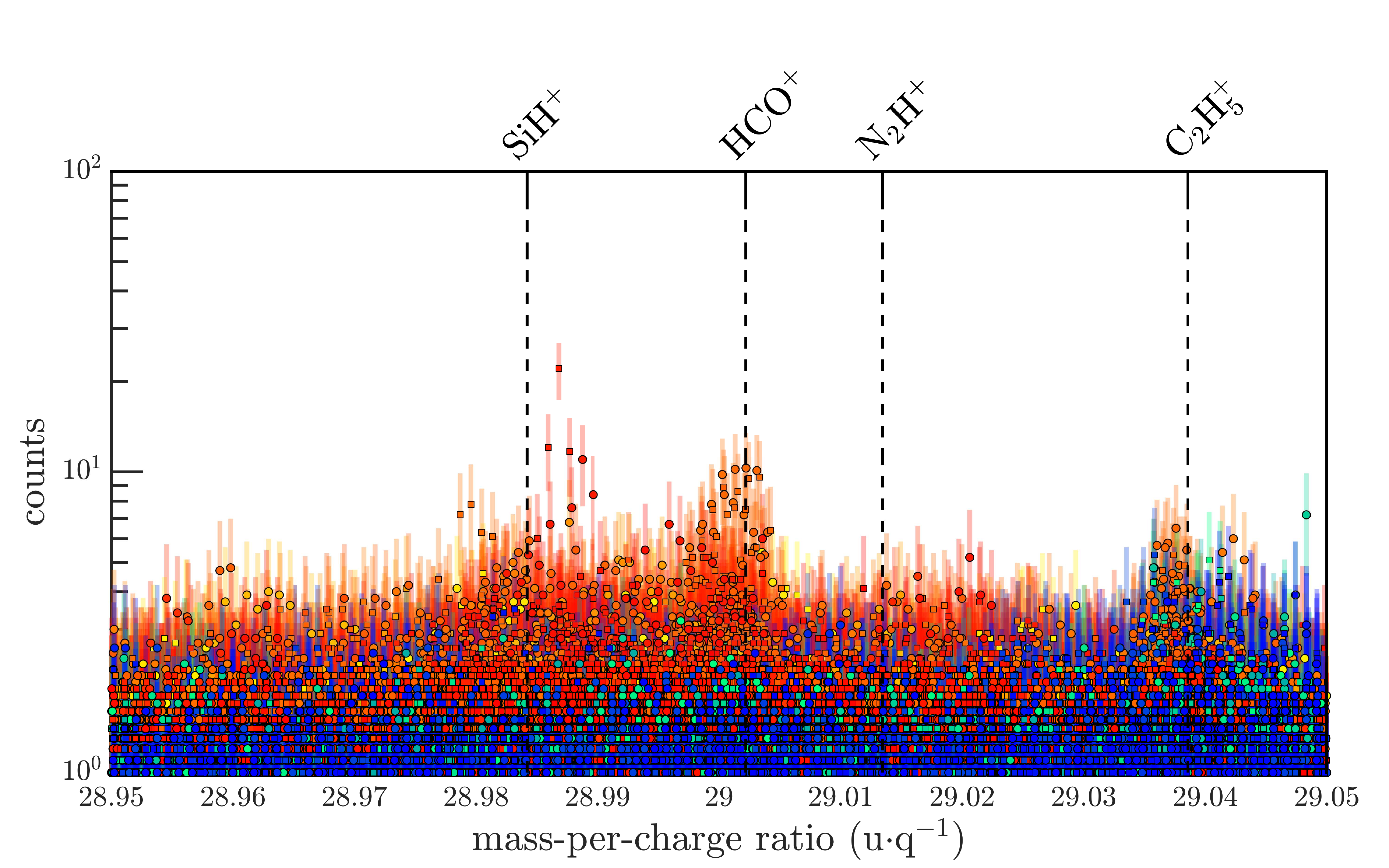}}
	\caption{Same as Fig.~\ref{fig3}, but for 28-30~u\textperiodcentered q$^{-1}$. Stacked spectra in low resolution (top panel), in HR at 28~u\textperiodcentered q$^{-1}$ (middle) and at 29~u\textperiodcentered q$^{-1}$ (bottom). There is no evidence for a cation at 30~u\textperiodcentered q$^{-1}$ and the associated spectra are not shown.\label{fig11}}
\end{figure}

Fig.~\ref{fig11} shows the range 28-30~u\textperiodcentered q$^{-1}$ in LR and HR. Species at 28~u\textperiodcentered q$^{-1}$ are detected at large heliocentric distances as well as near perihelion. However, the contributing species are not the same for both periods. As seen in HR, the detected species at large heliocentric distances are CO$^+$ and, at times, C$_2$H$_4^+$. CO is one of three major components reported by \citet{Hassig2015} along with CO$_2$ and H$_2$O. However, CO$^+$ reacts with H$_2$O and CO$_2$ and disappears or is, at least, strongly attenuated close to perihelion \citep{Heritier2017a}. C$_2$H$_4^+$ is seen during both periods. However, its parent molecule may be different during these two periods. At large heliocentric distances, C$_2$H$_4^+$ can be produced by ionisation of C$_2$H$_4$ or dissociative ionisation of C$_2$H$_6$ since the electron-impact ionisation, dominant at large heliocentric distances, of C$_2$H$_6$ primarily leads to C$_2$H$_4^+$ \citep{Avakyan1998,Tian1998a}. On the contrary, at perihelion, it may be produced either by protonation of C$_2$H$_3$ (C$_2$H$_3$+H$_3$O$^+$) or by charge exchange (C$_2$H$_4$+H$_2$O$^+$). HCNH$^+$ is detected near perihelion as it is produced through proton transfer, between HCN (or HNC) and H$_2$O$^+$/H$_3$O$^+$ \citep{Heritier2017a}. One spectrum early in the mission exhibits a peak at Si$^+$ but it may be a spurious and ghost peak since it is observed on a single spectrum and on one channel only. In fact, both channels are not equally sensitive (due to ageing, temperature, tuning, etc.) and hence for low signals this may happen regularly. No conclusion can be drawn, though Si has been detected by another mass spectrometer of ROSETTA during the same time interval \citep{Wurz2015}. Si was assumed to be produced by sputtering of the nucleus' surface by solar wind ions, still able to access the surface at low outgassing activity \citep{Behar2017}. There is no evidence of N$_2^+$, which is consistent with Earth-based observations of other comets \citep{Rubin2015}. The N$_2^+$/CO$^+$ ratio is of interest in radio-astronomy and has been investigated at several comets \citep{Cochran2000,Cochran2002}. Due to the telluric contamination of N$_2^+$ or a lack of detection, an upper limit of this ratio is usually given of the same order as the ratio of their parent molecules \citep{Rubin2015}, that is less than $\sim1\%$, which explains why a possible N$_2^+$ peak would be buried in the background. Overall, although the peak at 28~u\textperiodcentered q$^{-1}$ is present at any time of the escort phase, the main contributors may have changed during Rosetta's escort phase: CO$^+$ (and maybe C$_2$H$_4^+$) at large heliocentric distances, HCNH$^+$+C$_2$H$_4^+$ near perihelion.

At 29~u\textperiodcentered q$^{-1}$, two ion species, HCO$^+$ and C$_2$H$_5^{+}$, are clearly visible on HR spectra near perihelion. Spectra in LR at 29~u\textperiodcentered q$^{-1}$ show a peak in January 2016 (green) with the same amplitude as near perihelion (orange) while, in HR mode, both species are detected mainly near perihelion. The presence of HCO$^+$ is puzzling as it should be lost through chemistry with H$_2$O. Included in the photochemical model of \citet{Heritier2017a} for 67P at perihelion, HCO$^+$ is produced through ion-neutral chemistry and its contribution is of the order of CO$^+$ which is however not observed at that time by DFMS. At 1P/Halley, \citet{Haider2005} had used the same ion-neutral chemical reactions plus another one: C$^+$+H$_2$O. The latter was calculated to contribute up to 10\% to the total amount of HCO$^+$ in 1P's coma. At 67P, DFMS did not perform detections below 13~u\textperiodcentered q$^{-1}$ such that C$^+$ at 12~u\textperiodcentered q$^{-1}$ cannot be qualitatively assessed. However, we do not expect C$^+$ to be significantly dense enough near perihelion to yield HCO$^+$. Rosetta was close to the nucleus between 150 and 200~km and the potential sources of C$^+$ are limited to carbon-bearing molecules, namely CO$_2$, CO, and, to a much lesser extent, H$_2$CO. That said, the potential direct sources of HCO$^+$ are the dissociative ionisation of H$_2$CO \citep[present in the coma,][]{Heritier2017a}, the photo-dissociation of H$_2$CO into HCO followed by its ionisation, and/or ion-neutral chemistry of H$_2$CO with ions. HCO$^+$ barely reacts with H$_2$O, like CH$_3^+$ \citep{Herbst1985}: perihelion is hence a favourable period for its production with a stronger EUV flux and the limited loss through ion-neutral chemistry. However, it is difficult to assess the contribution of the different chemical pathways, as the most likely parent molecule, H$_2$CO, may be a distributed source like at 1P/Halley \citep{Meier1993}. Due to the poor spatial coverage of Rosetta near perihelion, it is difficult to determine the neutral number density profile of H$_2$CO and determine whether or not it departs from a $\sim1/r^2$ dependency. However, studying distributed sources for neutrals is beyond the scope of this paper dedicated to ions. No HR spectra at 30~u\textperiodcentered q$^{-1}$ are shown as no peak was detected, which is consistent with the relatively weak signal in LR ($\lesssim60$ counts). The LR spectra do not exhibit significant differences between observations at perihelion and at large heliocentric distances, meaning that the main contributor may change throughout the mission and/or may not react with H$_2$O.

\begin{figure}[H]
	\stackinset{l}{0.315cm+0.6cm}{b}{3.4cm}{\fbox{LR}}{%
	\includegraphics[width=\linewidth,trim=0 1.9cm 0 5cm,clip]{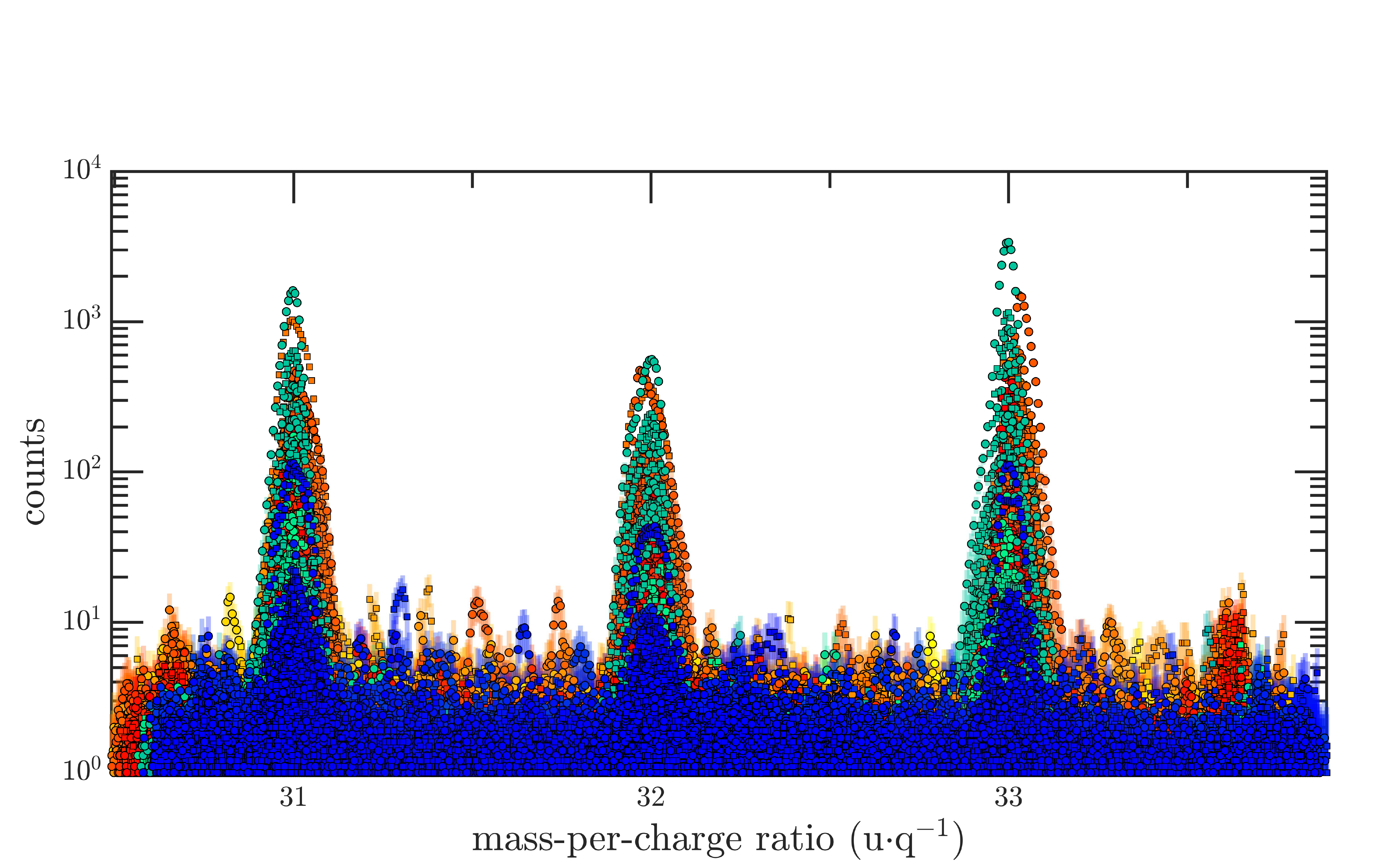}}\\
	\stackinset{l}{0.315cm+0.6cm}{b}{3.4cm}{\fbox{HR}}{%
	\includegraphics[width=\linewidth,trim=0 1.9cm 0 0cm,clip]{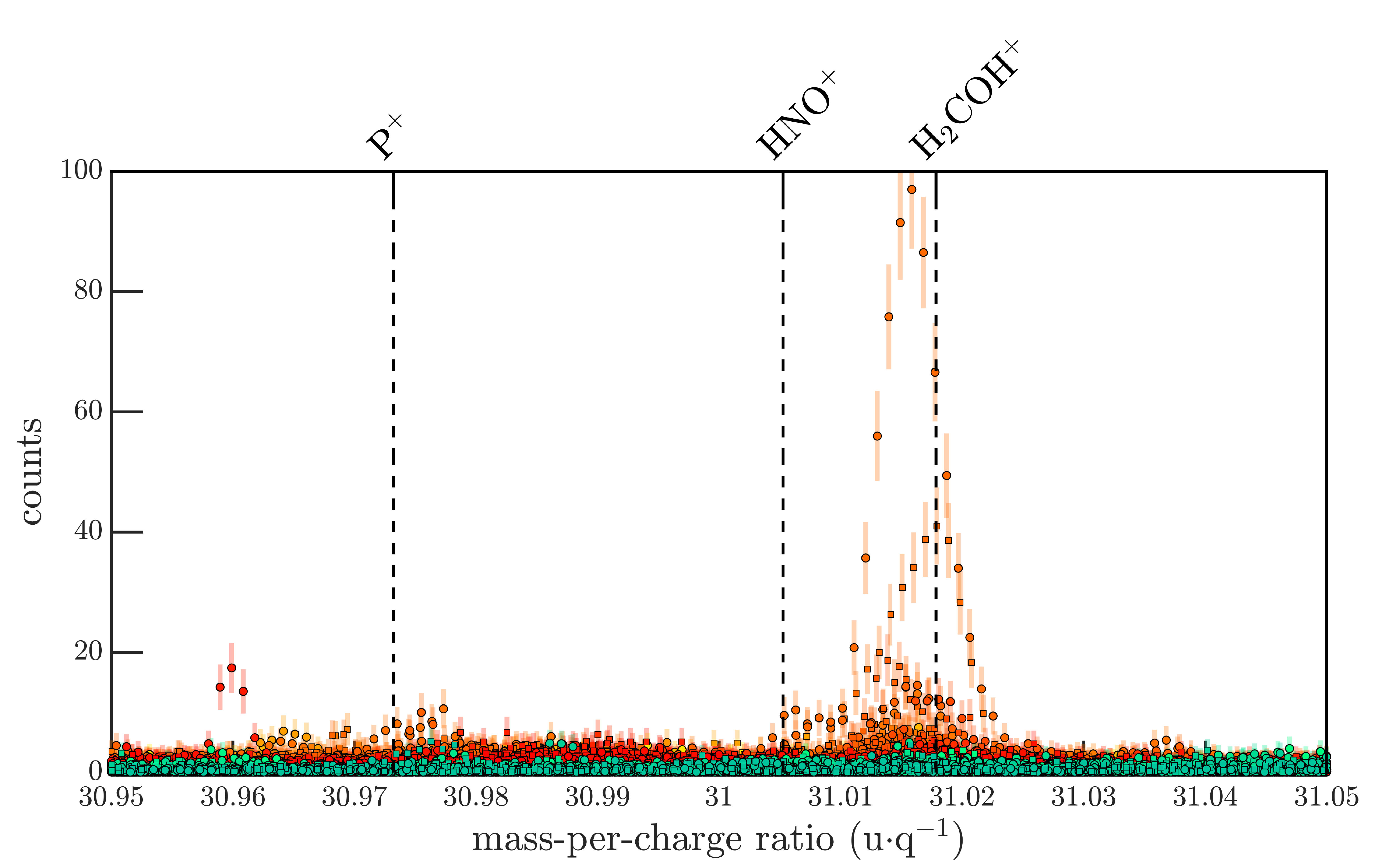}}\\
	\stackinset{l}{0.315cm+0.6cm}{b}{3.4cm}{\fbox{HR}}{%
	\includegraphics[width=\linewidth,trim=0 1.9cm 0 0cm,clip]{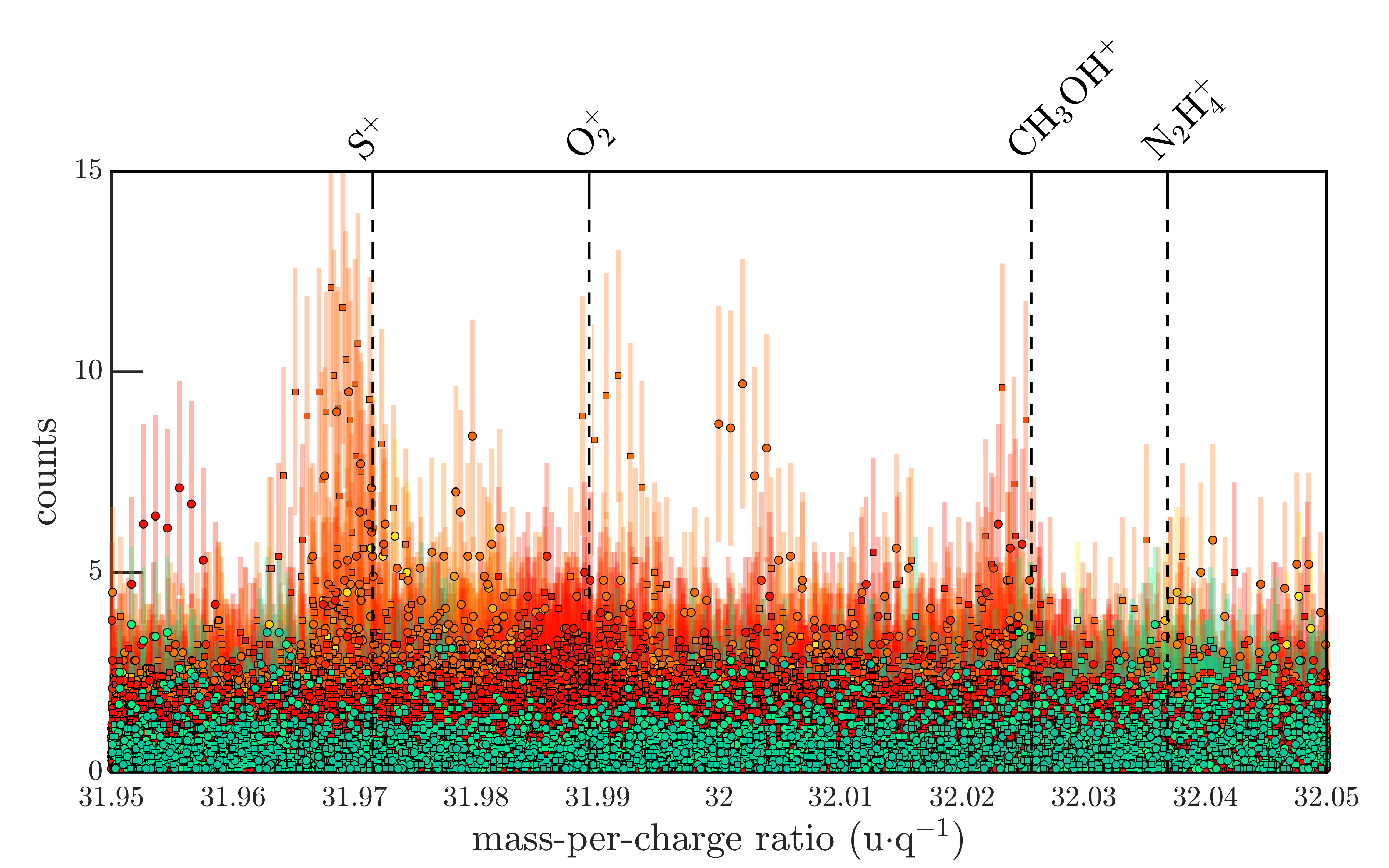}}\\
	\stackinset{l}{0.315cm+0.6cm}{b}{0.33667905cm+3.4cm}{\fbox{HR}}{%
	\includegraphics[width=\linewidth,trim=0 0.0cm 0 0cm,clip]{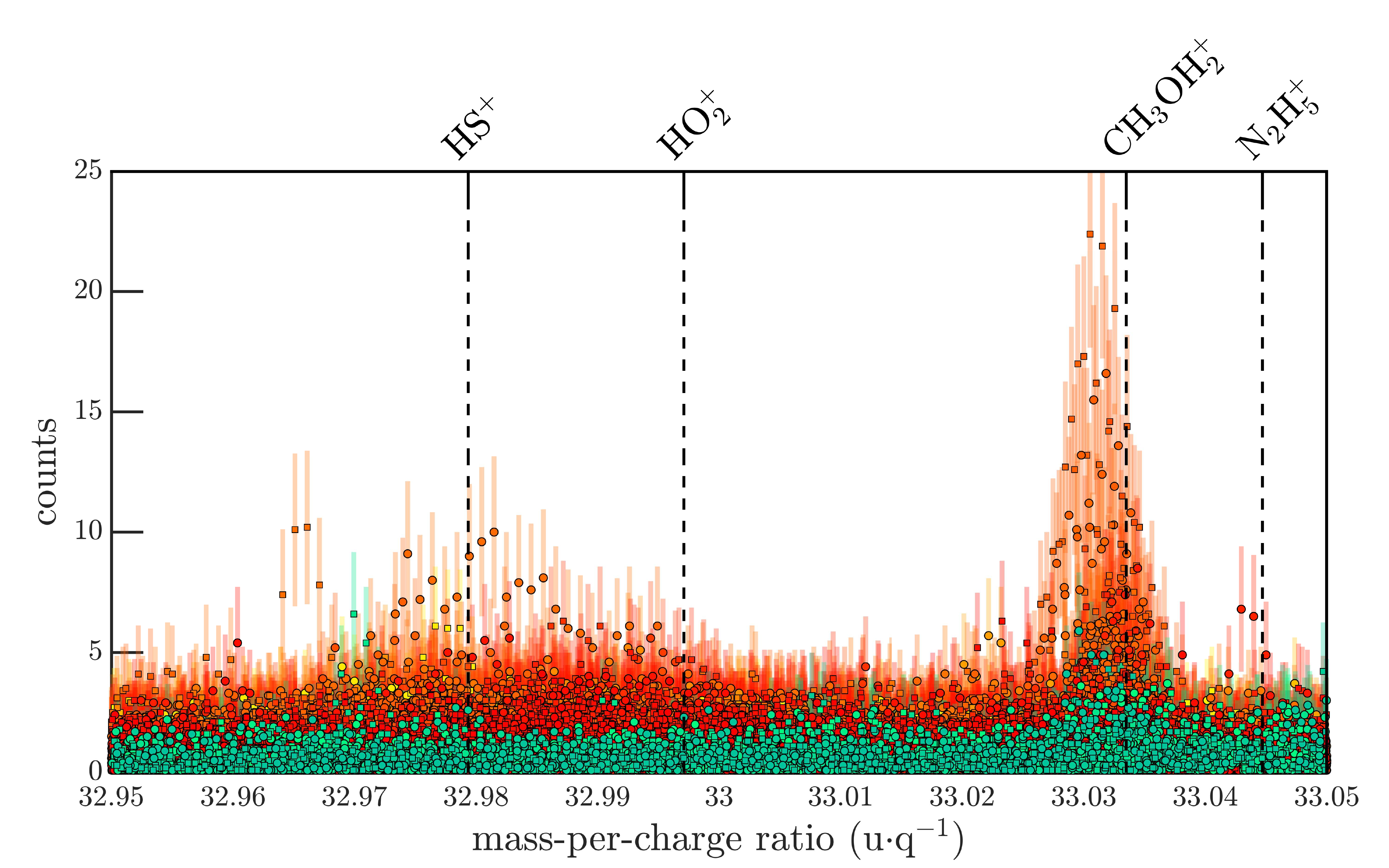}}\\
	\caption{Same as Fig.~\ref{fig3}, but for 31-33~u\textperiodcentered q$^{-1}$. Stacked spectra in low resolution (top panel) and in HR at 31~u\textperiodcentered q$^{-1}$ (second panel), at 32~u\textperiodcentered q$^{-1}$ (third panel), and at 33~u\textperiodcentered q$^{-1}$ (fourth panel).\label{fig12}}
\end{figure}

Fig.~\ref{fig12} shows spectra for the range 31-33~u\textperiodcentered q$^{-1}$. Peaks at 31, 32, and 33~u\textperiodcentered q$^{-1}$ are similar, showing larger intensities at heliocentric distances from 2 to 2.5~au after perihelion (green), in particular at 33~u\textperiodcentered q$^{-1}$, slightly above the levels observed near perihelion (orange). At 31~u\textperiodcentered q$^{-1}$, H$_2$COH$^+$ (protonated formaldehyde) is clearly identified in HR as the major ion species. H$_2$CO has a proton affinity higher than that of H$_2$O, such that H$_2$COH$^+$ is mainly produced through H$_2$CO+H$_2$O$^{+}$ or H$_2$CO+H$_3$O$^{+}$. There are some weak (<5-7~counts) peaks at the location of phosphorus cation P$^+$ near perihelion as seemingly shown by the observations of phosphorus atoms and amino-acids in neutral mode reported by \citet{Altwegg2016}. The faint level of the P$^+$ signal in HR can be explained by the loss of P$^+$ which reacts, even slowly, with the most abundant neutral species such as H$_2$O, CO$_2$, and NH$_3$. For HR spectra at 32~u\textperiodcentered q$^{-1}$, there are several peaks with low intensity (<20 counts per 19.8~s) which can be identified by superimposing a number of spectra. S$^+$ and CH$_3$OH$^+$ are clearly detected. Amongst the sulphur-bearing molecules detected in the coma of 67P by DFMS, the most abundant species near perihelion are H$_2$S, of which dissociative ionisation leads to S$^+$ in part, and neutral S \citep{Calmonte2015}. There is no significant loss of S$^+$ through chemistry as it does not react with the dominant neutral species H$_2$O, CO$_2$, and CO. Regarding CH$_3$OH$^+$, the methanol cation, CH$_3$OH, the parent molecule of the methanol cation CH$_3$OH$^+$, is quite abundant around perihelion, between 0.5\% and 3\% with respect to H$_2$O, based on MIRO \citep[Microwave Instrument for the Rosetta Orbiter,][]{Gulkis2007} sub-millimetre radio-telescope observations \citep{Biver2019}. Nevertheless, these values, derived from measurements of the column density between the surface of the nucleus and Rosetta, differ from those (0.1-0.3\%) measured in situ by DFMS for the same period \citep[see Fig.~10 from][]{Heritier2017a}. This difference may be attributed to the adiabatic expansion of the gas along the line of sight \citep{Heritier2017a} and the differences between the bulk speed of light and heavy species. In addition, there are also disagreements between H$_2$O local measurements by DFMS and integrated column measurements by MIRO and VIRTIS with associated impacts on the relative abundances of cometary species \citep{Hansen2016,Marshall2017,Combi2020}. There is one occurrence for a peak in one channel at the location O$_2^+$, of which the main sources are charge exchange (O$_2$+H$_2$O$^+$) and ionisation of O$_2$. The latter was not considered in \citet{Heritier2017a} such that O$_2^+$ is underestimated in their model. While the conditions were favourable for its production, its detection cannot be confirmed. Indeed, another peak of similar intensity is also observed between O$_2^+$ and CH$_3$OH$^+$ locations but cannot be assigned to a given species, it is most likely a ghost peak, that is an unphysical peak. 

Concerned by spacecraft contamination, in Fig.~\ref{fig12} (third panel) we have added N$_2$H$_4^+$, of which the ionisation potential is 8.1~eV \citep{Meot1984}. No strong signal is detected at this mass, ruling out its contribution to 32~u\textperiodcentered q$^{-1}$. Same precaution has been undertaken for 33~u\textperiodcentered q$^{-1}$ with N$_2$H$_5^+$ because N$_2$H$_4$ has a proton affinity just above that of NH$_3$. As only one peak is observed (see Fig.~\ref{fig12}, bottom panel), this is not conclusive. HR spectra at 33~u\textperiodcentered q$^{-1}$ reveal the presence of a protonated molecule, namely CH$_3$OH$_2^+$ (protonated methanol), produced from CH$_3$OH+H$_2$O$^+$ and CH$_3$OH+H$_3$O$^+$. CH$_3$OH$_2^+$ is more abundant than the ionised methanol and is the main contributor to mass 33~u\textperiodcentered q$^{-1}$. A faint accumulation of signals is observed around HS$^+$. Considering the main sulphur-bearing molecules, H$_2$S and S (see discussion for mass 32~u\textperiodcentered q$^{-1}$), the possible processes to generate HS$^+$ are rather limited. The most likely source is the photo-dissociative ionisation of H$_2$S, though this process is not very efficient \citep{Huebner2015}. Regarding its loss, S has a low proton affinity with respect to water and HS$^+$ easily loses its proton for the benefit of H$_2$O, for instance.

\begin{figure}
	\stackinset{r}{0.6cm}{b}{0.33667905cm+3.4cm}{\fbox{LR}}{%
	\includegraphics[width=\linewidth,trim=0 0cm 0 5cm,clip]{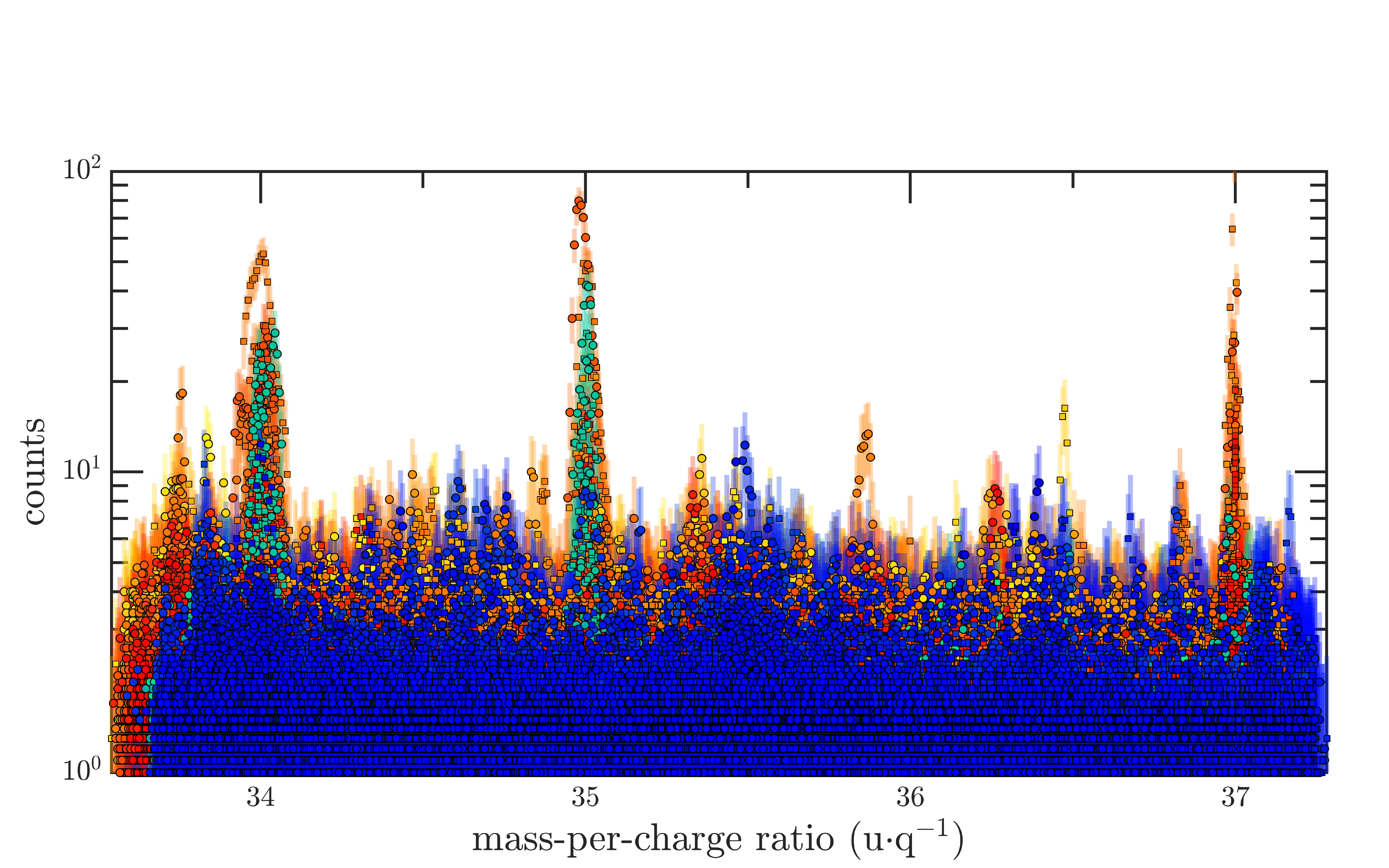}}\\
	\caption{Same as Fig.~\ref{fig3}, but for 34-37~u\textperiodcentered q$^{-1}$. There is no evidence of cations in high resolution and the associated spectra are not shown.\label{fig13}}
\end{figure}

Fig.~\ref{fig13} shows stacked LR spectra for the range 34-37~u\textperiodcentered q$^{-1}$. None of the associated HR spectra present clear peaks over this range. The most curious and surprising absence is that of H$_3$S$^+$, even in LR. H$_2$S is present in the coma of 67P \citep{Calmonte2015} and its proton affinity is higher than that of H$_2$O, though lower than those of H$_2$CO, HCN, CH$_3$OH, and NH$_3$ \citep{Heritier2017a}. It is thus predicted to be detectable by Rosetta near perihelion, a period favourable for proton transfer \citep{Heritier2017a}. However, the sensitivity of DFMS in the ion mode is affected by (i) the instrument energy acceptance window and, to a lesser extent, (ii) the decrease of the detector efficiency at higher energies. As the mass-per-charge ratio u\textperiodcentered q$^{-1}$ increases, the DFMS energy acceptance window decreases. Indeed, to select ions with respect to their mass-per-charge, a specific post-acceleration $V_\text{acc}\propto (\text{u\textperiodcentered q}^{-1}$)$^{-1}$ is applied within DFMS. The energy acceptance window for the electrostatic analyzer is $\sim20 \text{ V}\pm 0.1\%|V_\text{acc}|$ such that ions are filtered through a narrower energy range when the mass-per-charge ratio increases. Further details are given in \citet{Schlappi2011}. In views of previous works at 1P/Halley \citep[e.g.][]{Eberhardt1994}, the main contributors are $^{34}$S$^+$, followed by H$_2$S$^+$ and $^{13}$CH$_3$OH$_2^+$ at 34~u\textperiodcentered q$^{-1}$, and H$_3$S$^+$ at 35~u\textperiodcentered q$^{-1}$. However, for the latter, it is clear that the signal is not stronger in LR near perihelion, which is unexpected for a protonated molecule.

\begin{figure}
	\stackinset{r}{0.6cm}{b}{3.4cm}{\fbox{LR}}{%
	\includegraphics[width=\linewidth,trim=0 1.9cm 0 5cm,clip]{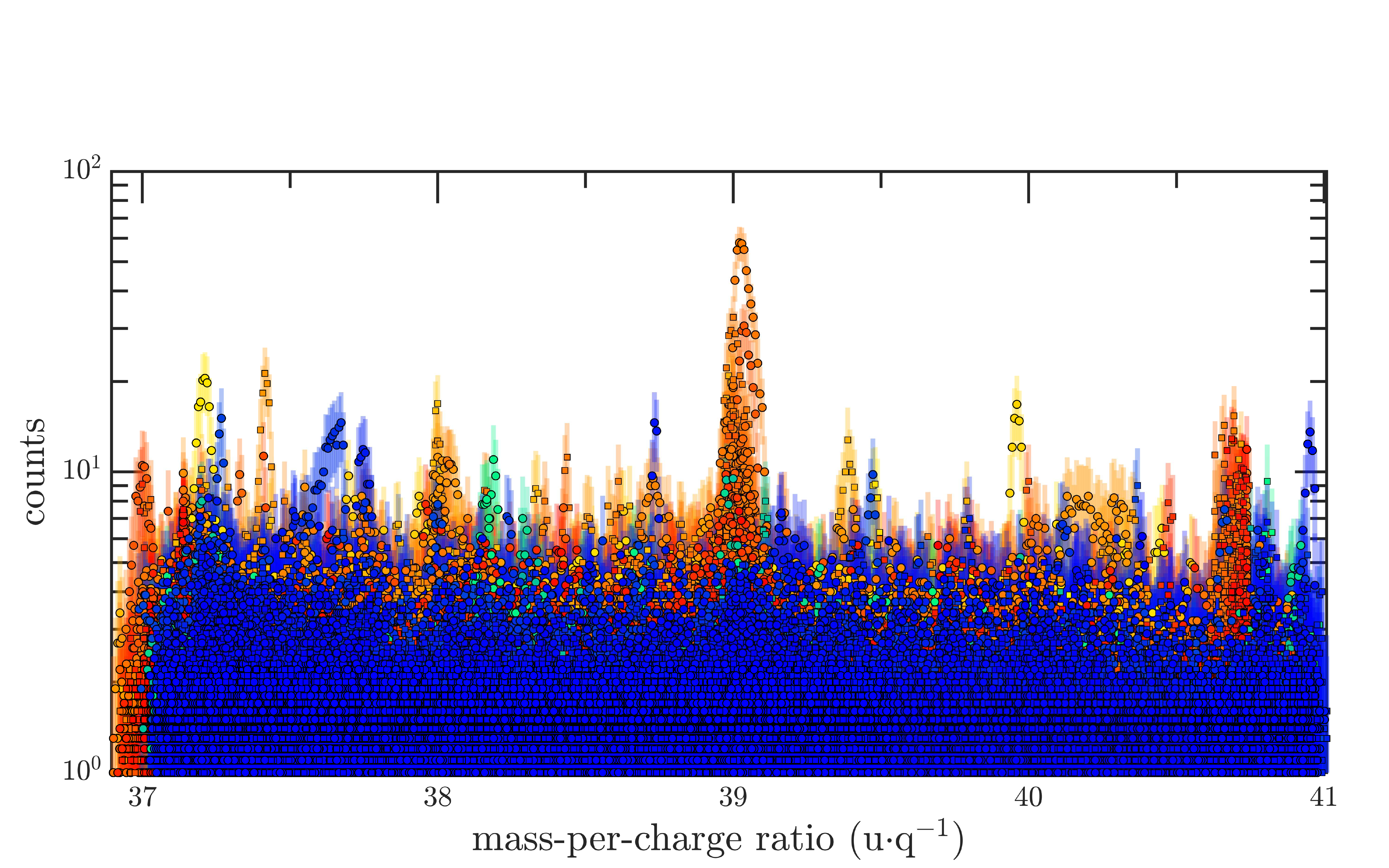}}\\
	\stackinset{r}{0.6cm}{b}{0.33667905cm+3.4cm}{\fbox{HR}}{%
	\includegraphics[width=\linewidth,trim=0 0.0cm 0 0cm,clip]{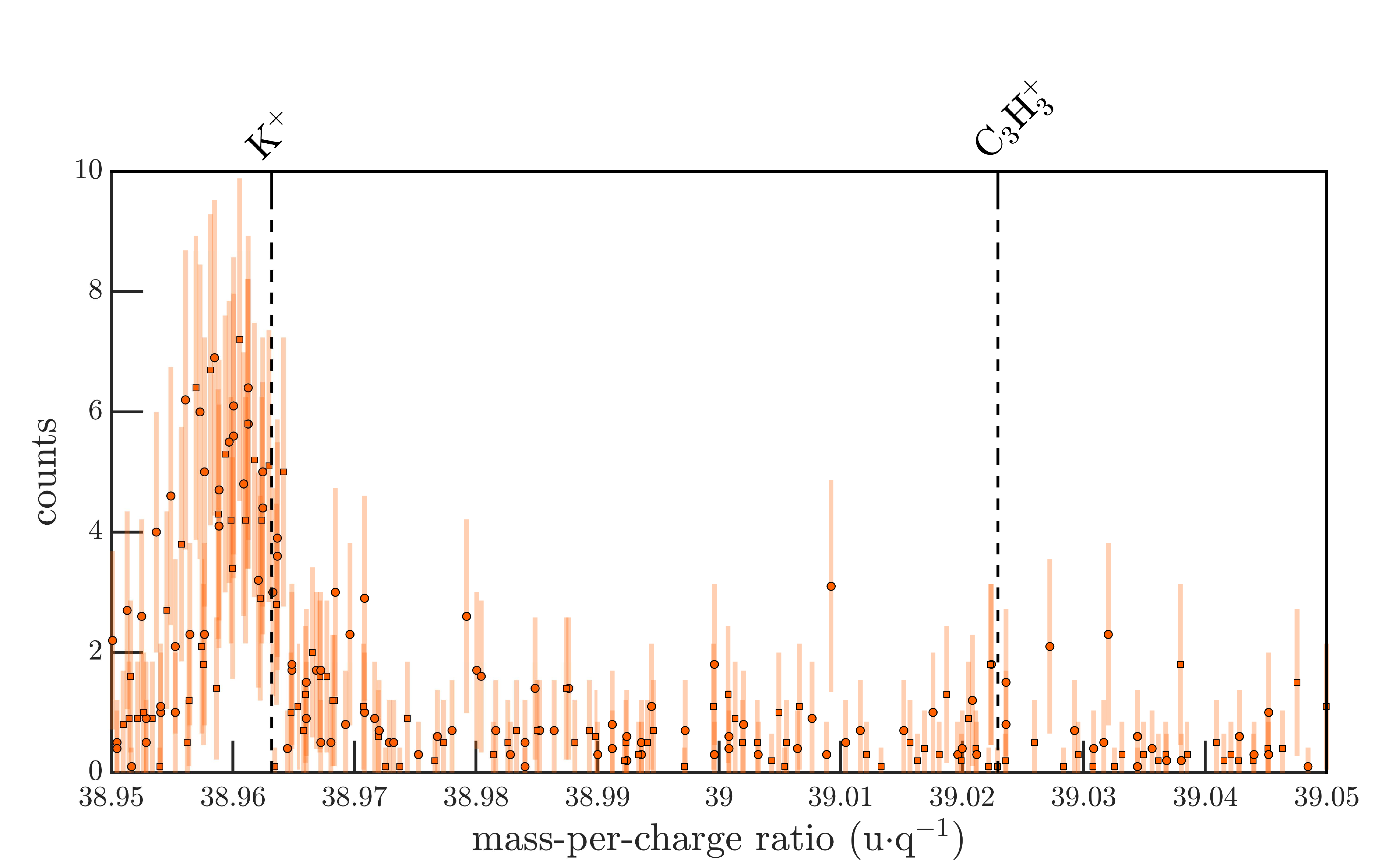}}\\
	\caption{Same as Fig.~\ref{fig3}, but for 37-40~u\textperiodcentered q$^{-1}$. Stacked spectra in low resolution (top panel) and in low resolution at 39~u\textperiodcentered q$^{-1}$ (bottom). There is no evidence of cations in HR at 37~u\textperiodcentered q$^{-1}$ and at 38~u\textperiodcentered q$^{-1}$ and the associated spectra are not shown.\label{fig14}}
\end{figure}

Fig.~\ref{fig14} shows LR spectra (top panel) for the range 37-40~u\textperiodcentered q$^{-1}$ and HR spectra (bottom panel) for 39~u\textperiodcentered q$^{-1}$. No cations have been detected except at 39~u\textperiodcentered q$^{-1}$. Because the signal-to-noise ratio in HR spectra increases with higher masses, we have only kept spectra with signals above 4 counts and present on both channels in order to get rid of contamination by spurious and unreliable signals. Only two species are expected: C$_3$H$_3^+$ and K$^+$. \citet{Korth1989} argued that the peak at 39~u\textperiodcentered q$^{-1}$ at 1P/Halley, detected by PICCA \citep{Korth1987} on board {\it Giotto}, was C$_3$H$_3^+$ and ruled out K$^+$. However, thrice, on the 8$^{th}$, the 9$^{th}$, and 11$^{th}$ of August 2015, K$^+$, and not C$_3$H$_3^+$, was detected. K (potassium) is an alkali metal like Na yet with a lower electronegativity. Ion-neutral reaction rates of K$^+$ and K of interest for astrochemistry are practically inexistent; however, as K belongs to the same group as Na, we can likely assume that K$^+$ will undergo the same interaction with neutrals as Na$^+$. K was detected early in the mission at large heliocentric distances by \citet{Wurz2015}, along with Na. In spite of the rather faint K$^+$ and Na$^+$ peaks in the HR mode, we believe that the presence of these ions in the ionised coma is undoubted, supported, in particular, by the detection of K and Na neutral atoms by ROSINA instruments. The photo-ionisation of K and Na appears as a likely production mechanism. However, the ultimate origin of these neutral species is still debated as already mentioned for Na.

\subsection{Ion mass-per-charge range $>40$~u\textperiodcentered q$^{-1}$}

\begin{figure}
	\stackinset{r}{0.6cm}{b}{0.33667905cm+3.4cm}{\fbox{LR}}{%
	\includegraphics[width=\linewidth,trim=0 0cm 0 5cm,clip]{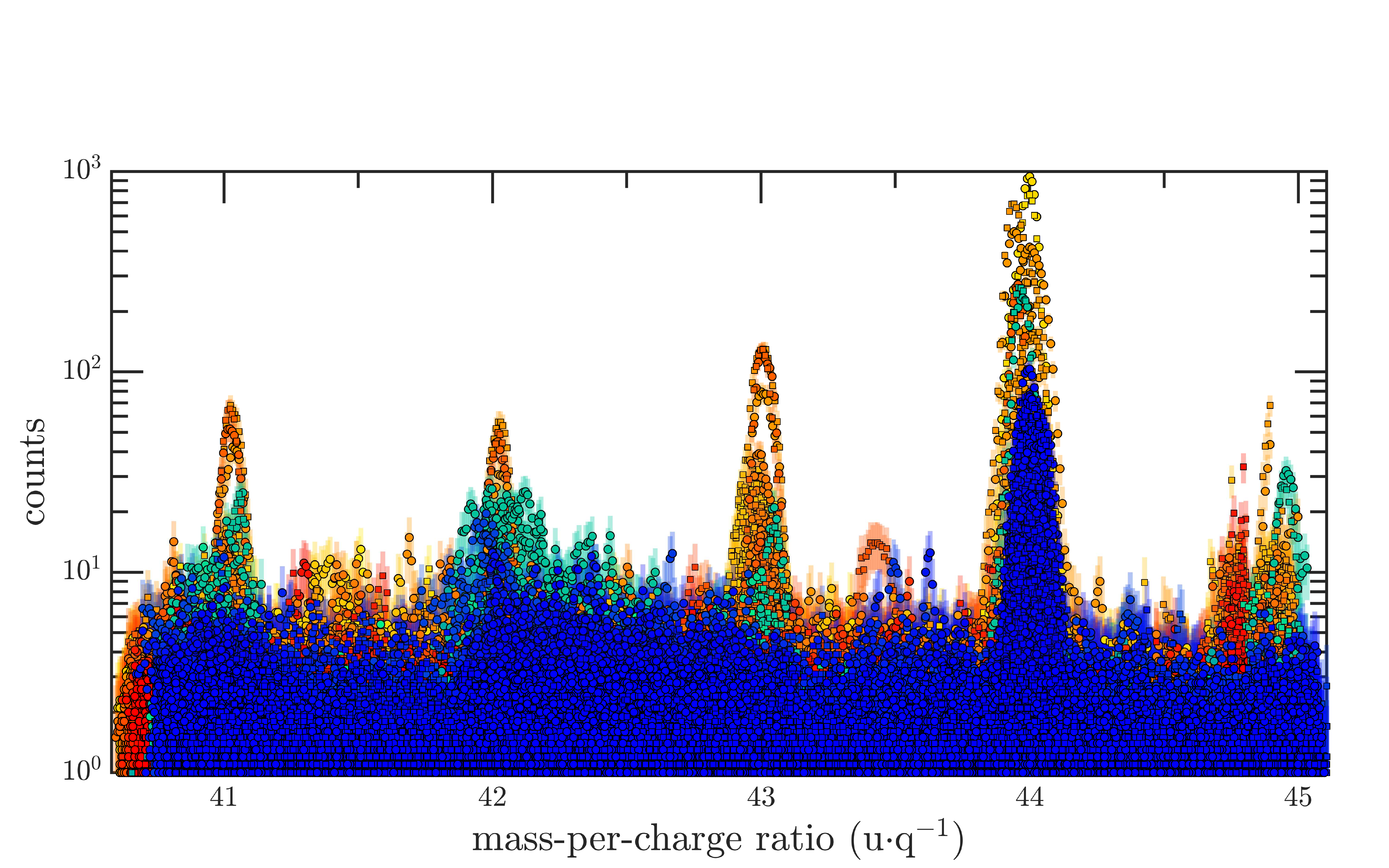}}
	\caption{Same as Fig.~\ref{fig3}, but for 41-44~u\textperiodcentered q$^{-1}$. Stacked spectra in low resolution only.\label{fig15}}
\end{figure}

We have decided not to look at u\textperiodcentered q$^{-1}$ above 40, as the sensitivity of DFMS in HR becomes too low to unambiguously allow a clear identification of the detected species . This is illustrated by a comparison of Fig.~\ref{fig15} with Fig.~\ref{fig8}. The most contributing species at 44~u\textperiodcentered q$^{-1}$ is undoubtedly CO$_2^+$ and consequently CO$_2^{++}$ at 22~u\textperiodcentered q$^{-1}$ at least at large heliocentric distances, especially over the Southern Hemisphere pre-perihelion and over both hemispheres post-perihelion (although no ion DFMS spectra are available, see Fig.~\ref{fig1}) where the neutral coma is dominated by CO$_2$ \citep{Hassig2015,Gasc2017}. As explained in Section~\ref{sec44}, the production rate of the CO$_2^{++}$ dication is a hundredfold lower, at least, than that of the monocation CO$_2^{+}$ but the relative intensity between the peaks at 22~u\textperiodcentered q$^{-1}$ and 44~u\textperiodcentered q$^{-1}$ is only about 10\% which means that the DFMS sensitivity at 44~u\textperiodcentered q$^{-1}$ is about tenfold lower than that at 22~u\textperiodcentered q$^{-1}$, a number consistent with mass-dependency reported by \citet{Schlappi2011}. In addition, CO$_2^+$ is mainly lost through charge exchange with H$_2$O such that near perihelion, it is close to photochemical equilibrium \citep[i.e. its loss is through chemistry, not transport, and its number density barely varies with respect to the cometocentric distance above tens of kilometres, see Fig.~7 and 8 in][]{Heritier2017a}.

\section{Discussion\label{sec4}}

\subsection{Highlights\label{sec41}}

HR ion mode dataset of ROSINA-DFMS has allowed us to directly identify for the first time different ion species and obtain an improved knowledge of the composition of a cometary plasma (columns 4 and 5 in Table~\ref{table1}). Our analysis has revealed the complexity of the cometary ionosphere made of ionised neutral atoms and molecules and also of ionised radicals and protonated molecules (see Section~\ref{sec42}). We also confirm the presence of isotopologues (see Section~\ref{sec43}). Observations from previous cometary missions, such as {\it Giotto}, had to rely on photochemical models to determine the exact nature of numerous ions detected at a given u\textperiodcentered q$^{-1}$. From ground-based observations, only a few ions were directly identified. Table~\ref{table1} shows in the second and third columns a compilation of the previous knowledge on cometary plasma composition from the literature \citep{Delsemme1985,Delsemme1991,Balsiger1995,Lis1997,Huebner1991,Haider2005} from 13~u\textperiodcentered q$^{-1}$ (lower bound for DFMS) to 40~u\textperiodcentered q$^{-1}$. 

We have detected a number of new ion species that have not been predicted before the Rosetta mission such as the alkali metal ions Na$^+$ and K$^+$ and the dication CO$_2^{++}$ (see Section~\ref{sec44}). Some ion species predicted by photo-chemical models have not been detected by DFMS. Their presence cannot be ruled out since favourable conditions for their detection may have not been met, such as a lower outgassing rate of the nucleus compared to 1P during the {\it Giotto} fly-by, lack of DFMS ion mode closer to the nucleus, and, probably of great importance, the detrimental influence of the spacecraft potential on the effective acceptance of DFMS in ion mode.

\begin{table*}
	\begin{tabularx}{\linewidth}{lllclX}
		&\multicolumn{2}{c}{Previous works (see caption)}&&\multicolumn{2}{c}{This work}\\
		\cline{2-3}\cline{5-6}
		u\textperiodcentered q$^{-1}$&Identified species&Predicted species&&Detected species in HR&{Peaks in LR but not in HR}\\
		&&(from modelling)&&&{(new candidate)}\\
		13&CH$^{+}$&$^{13}$C$^+$, CH$^+$&&CH$^+$&\\
		14&&N$^+$, $^{13}$CH$^+$, CH$_2^+$&&CH$_2^+$&\\
		15&&NH$^+$, $^{13}$CH$_2^+$, CH$_3^+$&&CH$_3^+$&\\
		16&&{O$^+$}, NH$_2^+$, $^{13}$CH$_3^+$, {CH$_4^+$}&&{O$^+$}, NH$_2^+$, {CH$_4^+$}&\\
		17&HO$^+$&{HO$^+$}, {NH$_3^+$}, $^{13}$CH$_4^+$, CH$_5^+$&&{HO$^+$}, {NH$_3^+$}&\\
		18&{H$_2$O$^+$}&{H$_2$O$^+$}, {NH$_4^+$}&&{H$_2$O$^+$}, {NH$_4^+$}&\\
		19&{H$_3$O$^+$}&{H$_3$O$^+$}&&{H$_3$O$^+$}&\\
		20&&{H$_2$$^{18}$O$^+$}, {H$_3$$^{17}$O$^+$}, H$_2$DO$^+$ &&{H$_2$$^{18}$O$^+$}, H$_2$DO$^+$&\\
		21&&{H$_3$$^{18}$O$^+$}&&{H$_3$$^{18}$O$^+$}&\\
		22&&&&& \ding{51}(CO$_2^{++}$)\\
		23&&&&Na$^+$&\\
		24&&C$_2^+$&&&\ding{51}\\
		25&&C$_2$H$^+$&&&\ding{51}\\
		26&CN$^{+}$&CN$^+$, {C$_2$H$_2^+$}&& {C$_2$H$_2^+$}&\\
		27&&HCN$^+$, {C$_2$H$_3^+$}&&{C$_2$H$_3^+$}&\\
		28&CO$^{+}$, N$_2^{+}$&{CO$^+$}, N$_2^+$, {HCNH$^+$}, {C$_2$H$_4^+$}&&Si$^{+}$*, {CO$^+$}, {HCNH$^+$}, {C$_2$H$_4^+$}&\\
		29&{HCO$^+$}&{HCO$^+$}, N$_2$H$^+$, {C$_2$H$_5^+$}&&{HCO$^+$}, {C$_2$H$_5^+$}&\\
		30&&NO$^+$, H$_2$CO$^+$, CH$_2$NH$_2^+$, C$_2$H$_6^+$&&&\ding{51}\\
		31&&HNO$^+$, {H$_2$COH$^+$}&&P$^{+}$*, {H$_2$COH$^+$}&\\
		32&&{S$^+$}, O$_2^+$, {CH$_3$OH$^+$}&&{S$^+$}, O$_2^{+}$*, {CH$_3$OH$^+$}&\\
		33&&HS$^+$, HO$_2^+$, {CH$_3$OH$_2^+$}&&{CH$_3$OH$_2^+$}&\\
		34&H$_2$S$^+$&$^{34}$S$^+$, H$_2$S$^+$, $^{13}$CH$_3$OH$_2^+$&&&\ding{51}\\
		35&&H$_3$S$^+$&&&\ding{51}\\
		36&&H$_2$$^{34}$S$^+$, H$_3$$^{33}$S$^+$, C$_3^+$&&&\ding{55}\\
		37&&C$_3$H$^+$, H$_3$O\textperiodcentered H$_2$O$^+$&&&\ding{51}\\
		38&&C$_2$N$^+$, C$_3$H$_2^+$&&&\ding{51}\\
		39&&C$_3$H$_3^+$&&K$^+$&\\
		40&Ca$^+$&CH$_2^+$CN$^+$, C$_3$H$_4^+$&&&\ding{51}\\
	\end{tabularx}
	\caption{Compilation of the ions predicted and detected at comets as a function of the mass. `Identified species' are those detected by UV, IR, visible or radio spectroscopic ground-based observations of comets \citep[][see Section~\ref{sec1}]{Delsemme1985,Delsemme1991,Lis1997}. `Predicted species' are those included in photochemical models for 1P/Halley \citep{Huebner1991,Haider2005} or considered as dominant for the corresponding mass \citep{Balsiger1995}. `Detected species in HR' refers to those detected with ROSINA-DFMS for this study. `Peaks in LR but not in HR' refers to peaks detected in LR (\ding{51}), (\ding{55}) otherwise, while no peaks were present in HR with a strong candidate in parenthesis (see Section~\ref{sec44}). * means for Si$^+$, P$^+$, and O$_2^+$ that, although a peak is located at the correct mass, it is seen once or twice in one of the two channels only, which may cast some doubt on their presence.\label{table1}}
\end{table*}

Most, yet not all, cometary ion species may be sorted into three main families (hereinafter F):
\begin{itemize}
	\item[(F1)] those produced by ionisation of a parent molecule $p$ and lost through transport, like CH$_3^+$ (i.e. reacting slowly or not at all with H$_2$O),
	\item[(F2)] those produced by ionisation of a parent molecule $p$ and lost through chemistry with H$_2$O, the dominant neutral species, like HO$^+$, CH$_4^+$ (i.e. the ion species $X^+$ is reacting with H$_2$O such that $X^++\text{H}_2\text{O}\longrightarrow$ products)
	\item[(F3)] those produced through ion-neutral chemistry only and lost either by transport (e.g. NH$_4^+$), by chemistry, or both (e.g. molecules with a proton affinity between those of H$_2$O and NH$_3$, such as H$_3$O$^+$, CH$_3$OH$_2^+$, and H$_3$S$^+$). Ions produced from high proton affinity neutrals are further discussed in Section~\ref{sec42}.
\end{itemize}
However, not all ion species belong to one of these families. 

For ions produced by ionisation, there is a possibility to quantify which loss process, that is transport (F1) or chemical reactions with water (F2), dominates. \citet{Beth2019} showed that a dimensionless parameter of interest is:
\begin{equation}
\alpha_{X^+}=k_{X^++\text{H}_2\text{O}}\dfrac{n_{\text{H}_2\text{O}}(r)r^2}{U r_c}=k_{X^++\text{H}_2\text{O}}\dfrac{Q_{\text{H}_2\text{O}}}{4\pi U^2 r_c},
\end{equation}
where $k_{X^++\text{H}_2\text{O}}$ stands for the reaction rate constant of $X^++\text{H}_2\text{O}\longrightarrow$ products (see Appendix~\ref{AppB}), $n_{\text{H}_2\text{O}}(r)$ is the local water number density close to the comet (assumed $\propto1/r^2$), $Q_{\text{H}_2\text{O}}$ is the outgassing rate of H$_2$O, $U$ is the ion outward radial speed, assumed to be that of neutrals and constant with cometocentric distance, and $r_c$ is the radius of the nucleus. $\alpha_{X^+}$ gauges which loss process dominates. For $\alpha_{X^+}\gg1$, $X^+$ is mainly lost through chemistry with water close to the surface, while for $\alpha_{X^+}\ll1$, it is through transport. At large cometocentric distances, the loss through transport always dominates such that the ion number decreases in $1/r$ asymptotically. Due to the range of kinetic rates depending on the species as showed in Table~\ref{tableb2} and the evolution of the water number density, $\alpha_{X^+}$ is depending as well as the heliocentric distances. In order to assess the evolution and the variability of some detected ions, one should refer to Appendix~\ref{AppB} for detailed information on the photo-ionisation and kinetic rates.

Electron-ion dissociative recombination is negligible at the location of Rosetta \citep{Heritier2018,Beth2019}; as a result, the ion number density profile of these ions is given by \citep[adapted from Eq. B.3 in][]{Beth2019}:
\begin{align*}
n_{X^+}(r,\alpha_{X^+},\tau_c)=&\dfrac{\nu_pQ_p}{4\pi U^2}\left(\dfrac{r_c}{r}\right)^2\exp\left[\alpha_{X^+}\dfrac{r_c}{r}\right]\\\
&\times\left\{\dfrac{r}{r_c}E_2\left[(\alpha_{X^+}+\tau_c)\dfrac{r_c}{r}\right]-E_2\left[\alpha_{X^+}+\tau_c\right]\right\}\\
\phantom{n_{X^+}(r,\alpha_{X^+},\tau_c)}=&n_{X^+}(r,\alpha_{X^+}+\tau_c,0)\exp\left[-\tau_c\dfrac{r_c}{r}\right],
\end{align*}
where $E_2$ stands for the exponential integral function, $\tau_c$ the mean optical depth at the surface for a water-dominated coma, significant near perihelion \citep{Heritier2017a,Beth2019}, $\nu_p$ the ionisation frequency of the neutral parent molecule $p$ yielding $X^+$, and $Q_p$ the outgassing rate of the parent molecule. For low outgassing activity ($\alpha_{X^+}\ll1$, i.e. $Q\lesssim 10^{25}$~s$^{-1}$, and $\tau_c\ll1$), the main loss for the ion is through transport regardless the cometocentric distance and its number density profile converges towards \citep{Galand2016}:
\begin{equation}
n_{X^+}(r)\approx\dfrac{\nu_pQ_p}{4\pi U^2}\dfrac{r-r_c}{r^2},
\label{transport}
\end{equation}
For high outgassing activity ($\alpha_{X^+}\gg1$), the loss is dominated by chemistry with H$_2$O for distances close to the nucleus; as a result, the ion is in photochemical equilibrium and its density is given by:
\begin{equation}
n_{X^+}(r)\approx\dfrac{\nu_p}{k_{X^++\text{H}_2\text{O}}}\dfrac{Q_p}{Q_{\text{H}_2\text{O}}}\left(\dfrac{\alpha_{X^+}}{\alpha_{X^+}+\tau_c}\right)\exp\left[-\tau_c\dfrac{r_c}{r}\right],
\label{photoeq}
\end{equation}
valid up to tens or hundreds of kilometres above the surface, depending how high $\alpha_{X^+}$ is. Interestingly, $\alpha_{X^+}/4$ corresponds to the ratio between the maximum $X^+$ number density reached with transport-dominated loss (i.e. Eq.~\ref{transport} at $r=2r_c$) and that reached with chemistry-dominated loss, when photo-absorption is ignored (i.e. Eq.~\ref{photoeq} with $\tau_c=0$).

\begin{figure*}
	\centering
	\includegraphics[width=.49\linewidth,trim=0cm 0cm 0cm 3cm,clip]{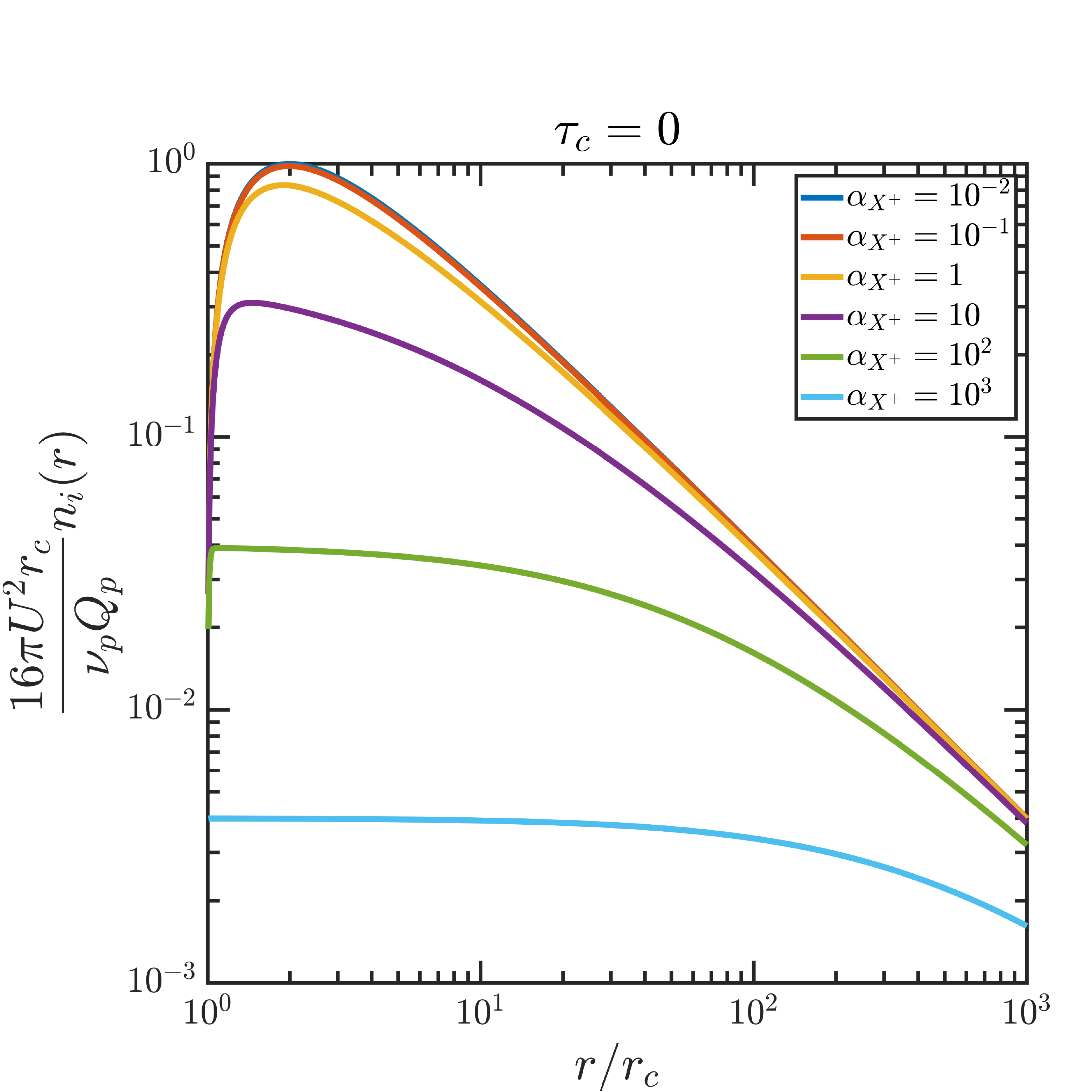}
	\includegraphics[width=.49\linewidth,trim=0cm 0cm 0cm 3cm,clip]{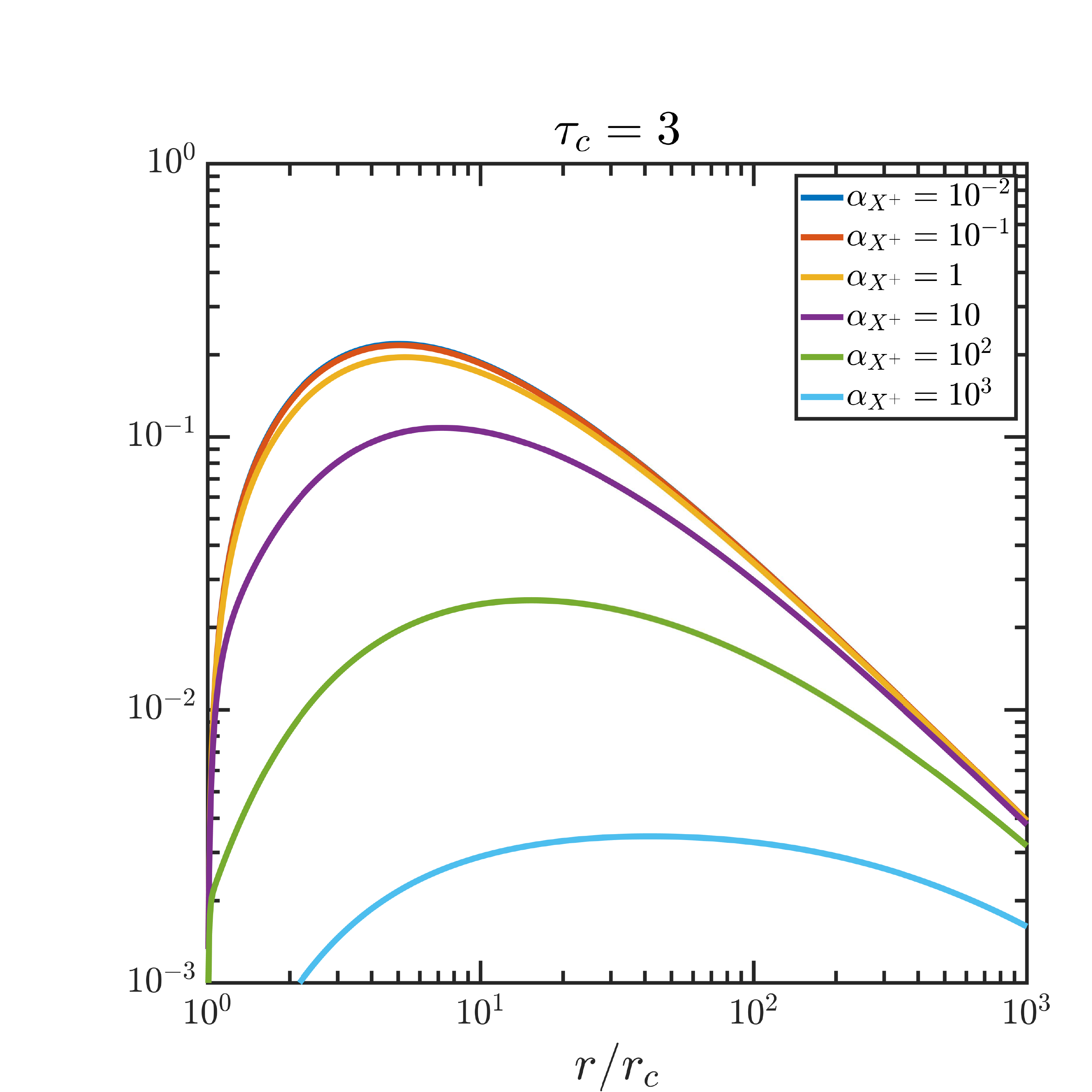}
	\caption{Dimensionless ion number density profile versus scaled cometocentric distance as a function of $\alpha$ with $\tau_c=0$ (left panel) and $\tau_c=3$ (right panel). On the $y$-axis, the values of the ion number density $n_i$ are scaled with respect to the maximum ion number density reached when only transport is considered for a given ion, i.e. $\alpha_{X^+}=0$. As $\alpha_{X^+}$ (coloured curves) increases, more and more ion-neutral reactions take place especially close to the nucleus. This results in damping the number density of ions reacting with H$_2$O. \label{fig16}}
\end{figure*}

Fig.~\ref{fig16} illustrates the effect of increasing the relative importance of reactions with H$_2$O on the ion number density. As $Q_{\text{H}_2\text{O}}$ increases, $\alpha_{X^+}$ increases and the ion density profile is damped and flattened to the photo-chemical value, or lower in presence of photo-absorption (based on Eq.~\ref{photoeq}). At large cometocentric distances, the ion number density profile follows Eq.~\ref{transport}. As 67P got closer to the Sun, $\alpha_{X^+}$ increased from $\ll1$ to $\gg1$, similarly to $\tau_c$. \citet{Beth2019} assessed the importance of the photo-absorption near perihelion. According to the average photo-absorption cross-section of H$_2$O that they derived, the optical depth is $\sim2--3$ near perihelion at the nucleus' surface. This entails a decrease in the effective H$_2$O photo-ionisation rate by 7--20 at the comet's surface. 

\begin{figure}
	\centering
	\includegraphics[width=.85\linewidth,clip,trim=0.5cm 1.9cm 0cm  4.1cm]{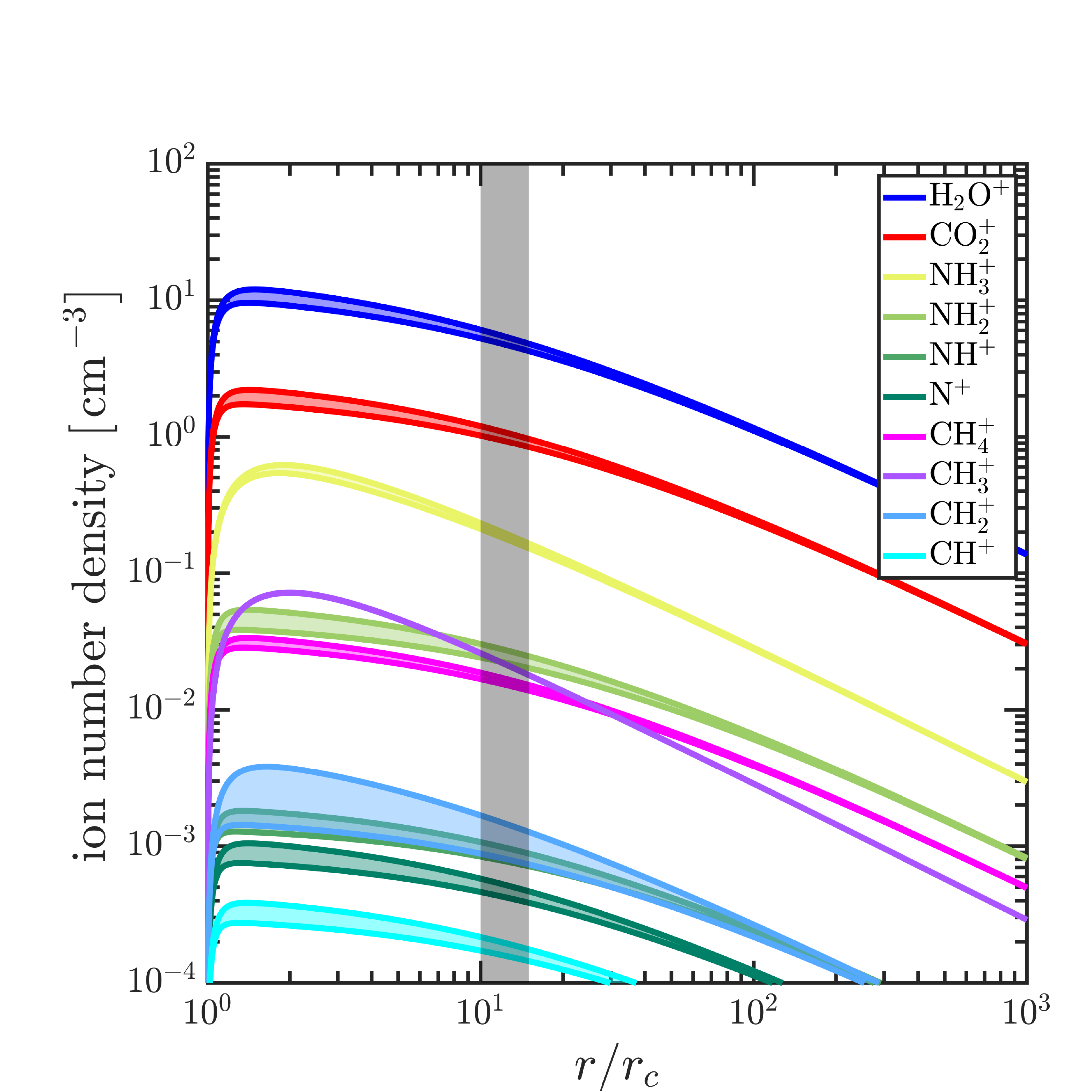}\\
	\includegraphics[width=.85\linewidth,clip,trim=0.5cm 1.9cm 0cm  4.1cm]{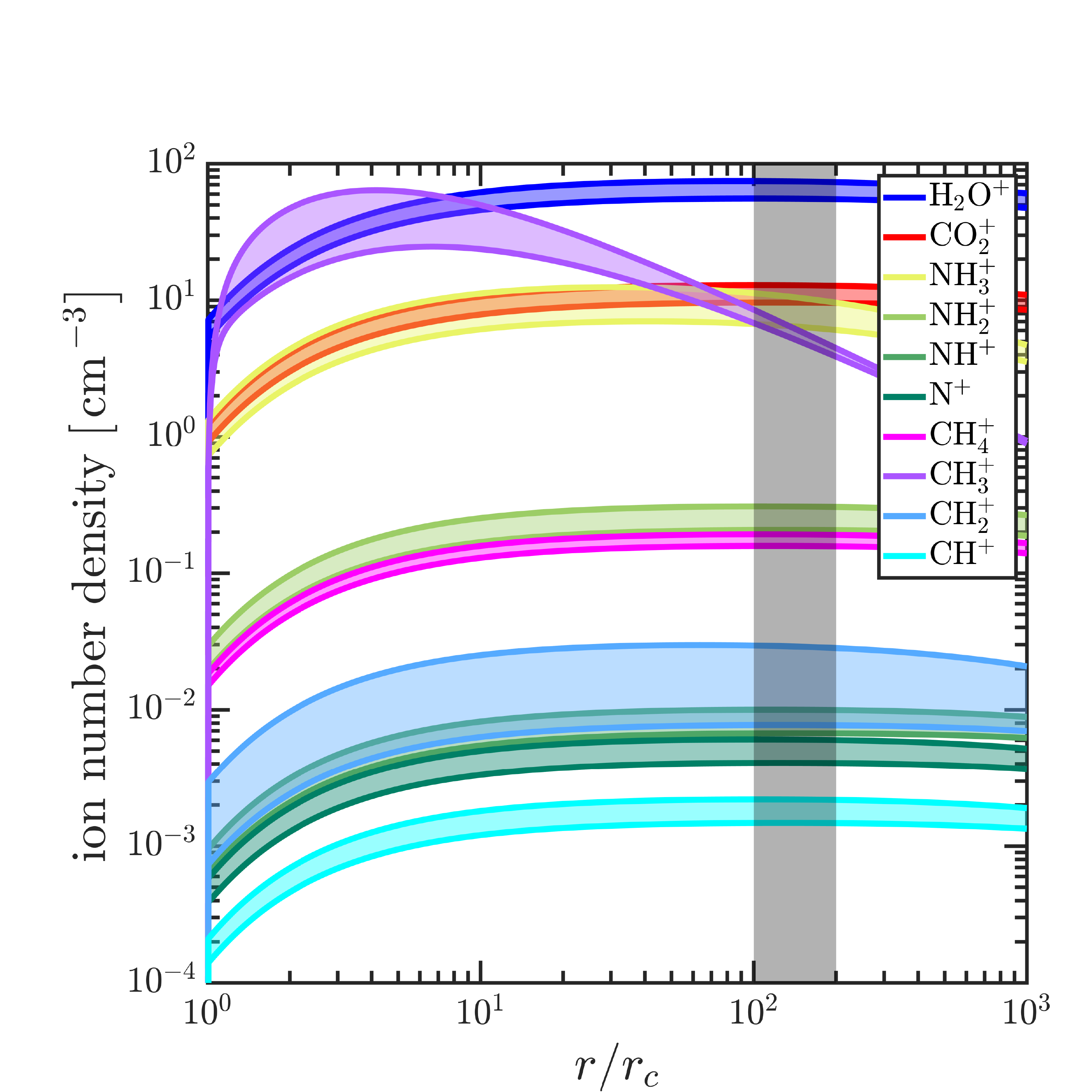}\\
	\includegraphics[width=.85\linewidth,clip,trim=0.5cm 0cm 0cm  4.1cm]{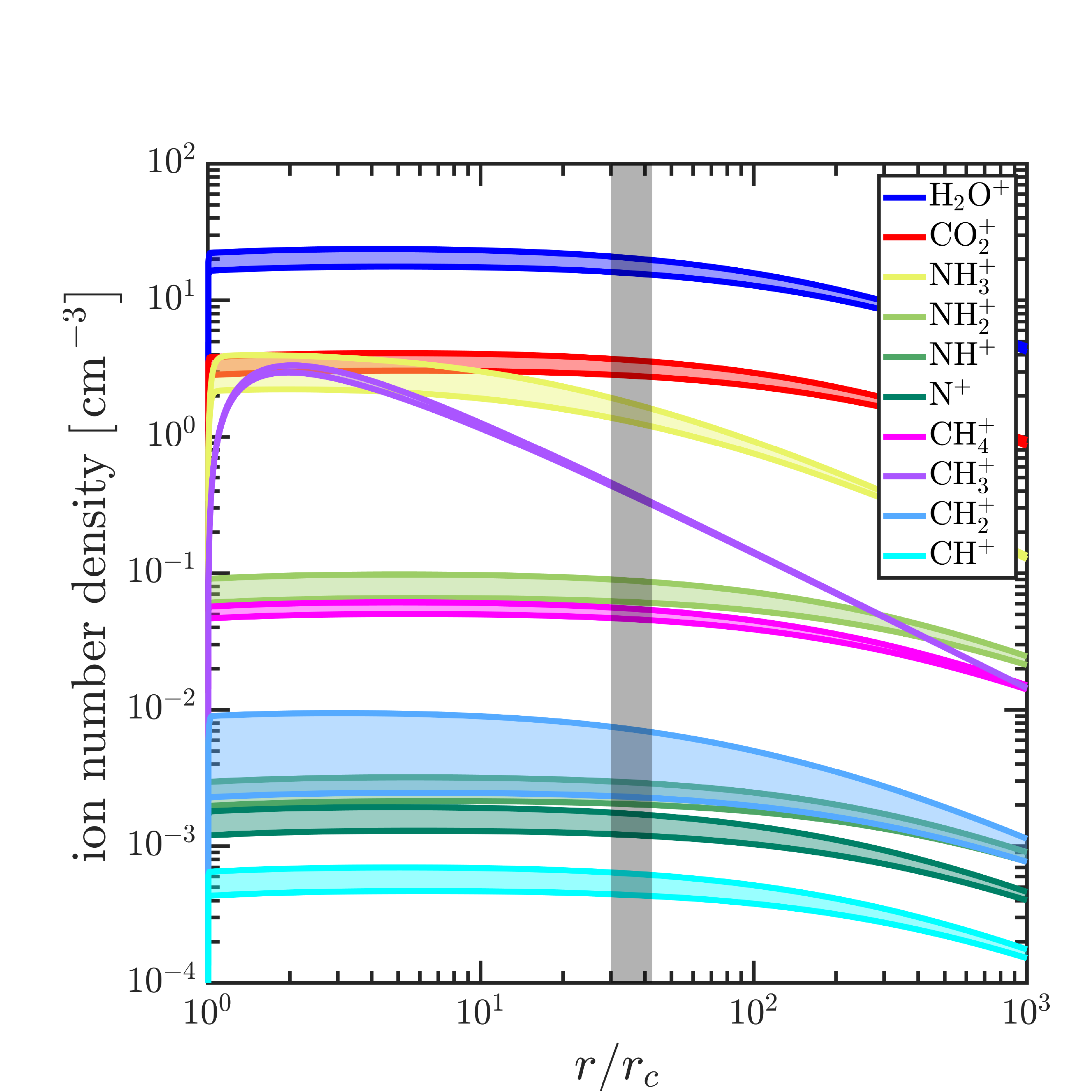}\\
	\caption{Ion density profiles of a few ions as a function of the cometocentric distance under different conditions: December 2014 ($Q=7.5\times 10^{25}$~s$^{-1}$, 2.75~au, top panel), end of August 2015 ($Q=5\times 10^{28}$~s$^{-1}$, 1.24~au, middle panel), end of January 2016 ($Q=2.4\times 10^{27}$~s$^{-1}$, 2.2~au, bottom panel). We have used the same neutral composition for each case: 88.7\% H$_2$O, 10\% CO$_2$, 1\% NH$_3$, 0.3\% CH$_4$. The photo-ionisation frequencies $\nu_p$ are taken from \citet{Huebner2015} at low solar activity at 1~au and scaled with respect to the heliocentric distance. The expanding speed of the gas is set to a constant $U=900$~m\textperiodcentered s$^{-1}$ and the nucleus' radius $r_c$ to 2~km. Ions are produced from photo-ionisation or photo-dissociative ionisation of the neutral molecules (see Section~\ref{AppB}). The gas temperature is assumed constant at $T=100$~K. The grey areas correspond to the cometocentric distance of Rosetta during these periods. The uncertainties from the kinetic rates on the ion number density are represented through coloured shades.\label{fig17}}
\end{figure}

Fig.~\ref{fig17} shows, for different cometary outgassing conditions, the number density profiles for ions with a number density lower than H$_3$O$^+$ between the nucleus and the location of Rosetta: H$_2$O$^+$, CO$_2^+$, CH$_n^+$ ($n=1-4$), and NH$_m^+$ ($m=0-3$). These ions are produced by (dissociative) ionisation of parent molecules (H$_2$O, CO$_2$, CH$_4$, and NH$_3$) and lost through transport or chemistry with mainly H$_2$O. H$_2$O$^+$ may also be produced by ion-neutral chemistry (charge transfer between cations and H$_2$O; e.g. H$_2$O+CO$_2^+$) but this process has been neglected here as it is significantly less efficient than photo-ionisation. As H$_2$O is the main EUV absorber in the coma near perihelion, we have used the same optical depth (i.e. 3) in Fig.~\ref{fig17} for all photo-ionisation rates. Results shown in Fig.~\ref{fig17} for three outgassing conditions provide a simple insight about the ions of main interest and their number density profiles, even if obtained using several simplifying assumptions. Firstly, only photo-ionisation has been considered, whereas \citet{Galand2016} and \citet{Heritier2018} have shown that electron-impact ionisation is usually the main ion source, against photo-ionisation, at large heliocentric distances. Solar-wind charge exchange has also been neglected although it may contribute to the ionisation of neutrals \citep{Simon2019c} in some cases at large heliocentric distances. Increasing the ionisation rate would shift profiles to higher number densities. Then, the ion dynamics and any significant acceleration of the cometary plasma has been neglected. \citet{Galand2016} and \citet{Heritier2018} have shown that considering the ion velocity to be that of neutral species is a good approximation to assess the plasma density at large heliocentric distances. Near perihelion, if the ion speed was significantly higher than that of neutrals as suggested by \citet{Odelstad2018}, $\alpha_{X^+}$ would be lower, preventing ion-neutral chemistry to happen and chemically-produced species, such as NH$_4^+$ and CH$_3$OH$_2^+$, from being present. Results shown in Fig.~\ref{fig17} are thus more reliable when one compares various ions originating from the same parent molecule such as, for example, CH$_3^+$ and CH$_4^+$. At large cometocentric distances (see Fig.~\ref{fig17}, top panel), transport is always dominating such that CH$_4^+$ number density is slightly higher than that of CH$_3^+$ (assuming both only produced from CH$_4$). At the location of Rosetta however, because CH$_4^+$ does react with H$_2$O (F2) and CH$_3^+$ does not (F1), CH$_3^+$ is similar to (top panel, $n_{\text{CH}_3^+}/n_{\text{CH}_4^+}\approx1$) or dominates over CH$_4^+$ (middle and bottom panels, $n_{\text{CH}_3^+}/n_{\text{CH}_4^+}>10$).

Figures~\ref{fig16} and~\ref{fig17} may help defining the most favourable conditions for detecting these ions. At low outgassing activity, when transport dominates the ion loss, the ion number density is $\propto\nu_pQ_p$, both parameters increase as the comet gets closer to the Sun. The ion number density profile peaks at $r=2r_c$, meaning that the best location for detection is close to the nucleus.
Under high outgassing activity (when photochemical equilibrium is achieved), the ion number density is approximately $\propto\nu_p$ , i.e. only the photo-ionisation rate and its variations with the heliocentric distances will control the ion number density. However, the location at which the ion density profile peaks also depends on the absorption of solar EUV radiation by neutral species in the coma, essentially H$_2$O. If photo-absorption is neglected, the ion density peaks closer to the nucleus' surface and has a plateau that may extend to distances ranging between a few tens to thousands of kilometres depending on the cometary activity (see Fig.~\ref{fig16}, left panel). Under optically thick conditions, the EUV radiation cannot penetrate deep enough into the coma to ionise neutrals such that the ion number density peaks farther away from the nucleus' surface (see Fig.~\ref{fig16} and compare profiles of similar colours).

Depending on the ion species and whether or not it reacts with H$_2$O, the peak of the ion number density is not located at the same cometocentric distance:
\begin{itemize}
	\item[(F1)] for ions which do not react with H$_2$O nor with any other major neutral species and are mainly produced by ionisation of parent molecules (e.g. CH$_3^+$), the maximum number density is reached around $r\approx2r_c$ and the ion density decreases asymptotically in $1/r$,
	\item[(F2)] for ions which do react with H$_2$O (e.g. HO$^+$ and CH$_4^+$), their number density peaks at different cometocentric distances, depending on the cometary activity (see Fig.~\ref{fig17}). For low activity, the ion number density peaks at $r\approx2r_c$. For high activity, these species start to reach photochemical equilibrium such that their number density profile exhibits a plateau from the nucleus' surface to tens or hundreds of cometary radii, and then decreases asymptotically in $1/r$,
	\item[(F3)] for high proton-affinity molecular ions (e.g. NH$_4^+$, H$_3$O$^+$, and CH$_3$OH$_2^+$), their number density peaks close to the comet's surface, but decreases asymptotically faster, in $\log(r)/r^2$ \citep{Beth2019}.
\end{itemize}
The aforementioned statements are true when photoabsorption is negligible. For outgassing rates higher than $10^{28}$-$10^{29}$~s$^{-1}$, photoabsorption matters \citep{Beth2019}: the ion number densities peak farther away from the comet's surface as EUV solar radiation cannot penetrate deep enough into the coma to ionise neutrals (see Fig.~\ref{fig16}, right panel).

\subsection{High proton affinity\label{sec42}}

The detection of protonated high-proton-affinity molecules other than H$_3$O$^+$ at 67P, such as NH$_4^+$, confirms the expectations from the models. Because of the insufficient resolution of the ion spectrometers on board Giotto, NH$_4^+$ was blended with H$_2$O$^+$ such that its contribution to the peak at 18~u\textperiodcentered q$^{-1}$ has only been assessed from a photochemical modelling based on the measured neutral composition. Prior to Rosetta's arrival at 67P, \citet{Vigren2013} attempted to assess the total contribution of these water ions inside the diamagnetic cavity. Though this cavity, detected around perihelion, was not as extended as expected \citep{Goetz2016b}, that does not prevent these ions from being produced close to the nucleus, then transported outwards, outside the diamagnetic cavity, and finally detected by the instrument. Indeed, during the escort phase and near perihelion, \citet{Beth2016} have unambiguously detected NH$_4^+$ in the coma of 67P, well separated from H$_2$O$^+$ thanks to the high resolution of the ROSINA-DFMS HR mode. Fig.~\ref{fig6} clearly shows that near perihelion, at the location of Rosetta, H$_2$O$^+$ was on the average more abundant compared with NH$_4^+$. In the present paper, we have confirmed the presence of three additional protonated molecules in the coma: HCNH$^+$, H$_2$COH$^{+}$, and CH$_3$OH$_2^{+}$,that were previously predicted from a photo-chemical modelling \citep{Heritier2017a}. Their detection attests the importance of ion-neutral chemistry and collisions in a high activity environment.

Except H$_3$S$^+$ (undetected in HR), HCNH$^+$, H$_2$COH$^{+}$, and CH$_3$OH$_2^{+}$ are strong candidates for the peaks observed at 28, 31, and 33~u\textperiodcentered q$^{-1}$ respectively in the LR mode. By means of a photochemical modelling, \citet{Heritier2017a} investigated the relative contribution at 28~u\textperiodcentered q$^{-1}$. They found that CO$^+$ should be lower than HCNH$^+$. However, their modelling did not include C$_2$H$_4^+$, another candidate at 28~u\textperiodcentered q$^{-1}$ in LR, and its associated chemistry, although this species is clearly detected in HR simultaneously with HCNH$^+$ on the same spectrum (see Fig.~\ref{fig11}, middle). The non-detection of H$_3$S$^+$ in HR is still not clearly understood. It has been pointed out that it may be related to the energy acceptance of the instrument which decreases at higher masses \citep{Heritier2017a}. It may also come from the origin and the source of H$_2$S. In the photo-chemical model, the source and background of the neutral species are supposed to exclusively come from the gas released following the ice sublimation at the surface, excluding extended sources. However, \citet{Calmonte2015} have suggested from DFMS neutral measurements that part of H$_2$S is associated with dust grains.

Finally, CH$_3^+$ might be considered to be part of (F3) as CH$_2$ has a higher proton affinity than HO and H$_2$O \citep{Altwegg1994}. However, Fig.~\ref{fig4} shows no evidence for higher counts of CH$_3^+$ at perihelion when proton transfer reactions are favoured, compared to large heliocentric distances.

\subsection{Water isotopologues \label{sec43}}

Figs.~\ref{fig7} and~\ref{fig8} attest the presence of water ion isotopologues, namely H$_2$$^{18}$O$^+$, H$_2$DO$^+$, and H$_3$$^{18}$O$^+$, near perihelion. As a consequence, DO$^+$ and H$^{18}$O$^+$ must also be present but the mass resolution of ROSINA-DFMS is not high enough to separate them from H$_2$O$^+$ and H$_3$O$^+$ signals, respectively (see Fig.~\ref{fig6}). In view of the isotopic ratios D/H \citep[$\sim5.3\times 10^{-4}$,][]{Altwegg2015} and $^{18}$O/$^{16}$O \citep[$\sim1.8\times 10^{-3}$,][]{Schroeder2019} derived from the neutral mode of DFMS, the detection and the count rates of H$_2$$^{18}$O$^+$, H$_2$DO$^+$, and H$_3$$^{18}$O$^+$ ($\lesssim 10$) are consistent with those of H$_2$O$^+$ and H$_3$O$^+$ (10$^3$-10$^4$) derived in HR near perihelion. In principle, D/H and $^{18}$O/$^{16}$O isotopic ratios might be determined from the LR ion observations \citep{Balsiger1995} using count rates at 19 and 20~u\textperiodcentered q$^{-1}$. However, as already noticed in Section~\ref{section32}, the quality of DFMS measurements at 19~u\textperiodcentered q$^{-1}$ close to the edge of the detector is not good enough to provide the necessary accuracy for the data processing. Moreover, ions at 19, 20, and 21~u\textperiodcentered q$^{-1}$ were not probed exactly at the exact same time (i.e. approximately 30 seconds between individual spectra) and plasma conditions (and thus the spacecraft potential) have been shown to vary on short time scales.

On the 16$^{th}$ of September 2015 from 08:59~UT, two successive HR spectra exhibit reliable peaks in both channels: 1) at 20~u\textperiodcentered q$^{-1}$ associated with H$_3$$^{18}$O$^{+}$ and the doublet H$_3$$^{17}$O$^+$ and H$_2$DO$^{+}$ (not separable) and 2) at 21~u\textperiodcentered q$^{-1}$ with a peak associated with H$_3$$^{18}$O$^+$. Peaks have been fitted with one Gaussian or double-Gaussian for the highly abundant ion H$_3$O$^+$ only while the total counts corresponding to the peaks of its isotopologues with very small amplitudes were simply obtained by summing the individual pixel counts. In this occasion, the isotopic ratio may be derived although with a limited accuracy. Firstly, (3D/H+$^{17}$O/$^{16}$O) may be inferred from (H$_2$DO$^+$+H$_3$$^{17}$O$^+$)/H$_3$O$^+$ \citep{Eberhardt1995}. We obtained $\sim10^{-3}$ which is compared with (2.05$\pm$0.3)$\times10^{-3}$ from neutral isotopic ratio obtained by DFMS observations in the neutral mode \citep{Altwegg2015,Schroeder2019}. Secondly, ($^{18}$O/$^{16}$O) may be inferred from H$_3$$^{18}$O$^+$/H$_3$O$^+$ \citep{Eberhardt1995}. We obtained $\sim10^{-3}$ which is compared with (2.25$\pm$0.18)$\times10^{-3}$ from the DFMS neutral mode \citep{Schroeder2019}.

As aforementioned, the isotopic measurements have only been possible on very few occasions when the count rates in the ion HR mode were sufficient. Moreover, the accuracy of their derivation is limited due to several factors such as the intrinsic smaller effective sensitivity of DFMS and the variable instrument energy acceptance controlled by the large and varying spacecraft potential. It is therefore out of scope to perform accurate isotopic measurements using HR ion data. However, our results show a relatively good agreement between both the ion-derived and neutral-derived isotopic abundances.

\subsection{Dications\label{sec44}}

As shown in Section~\ref{sec33}, ROSINA-DFMS provides the first unambiguous detection of the doubly-charged (or dication) CO$_2^{++}$ in a cometary ionosphere. Dications have been previously observed in dense planetary ionospheres throughout the Solar System: O$^{++}$ at Earth \citep{Hoffman1967}, Venus \citep{Taylor1980}, perhaps at Mars \citep{Dubinin2008} and at Io, along with S$^{++}$ \citep{Frank1996} and C$^{++}$ \citep{Sandel1979}. The recent (and only) review about ionospheric doubly-charged ions was published by \citet{Thissen2011}. Additional information on organic dications and their structures, aiming at presenting their chemical properties may be found in \citet{Lammertsma1989}. \citet{Thissen2011} reviewed stable dications stemming from the most abundant atoms and molecules found in planetary atmospheres, namely: C$^{++}$, N$^{++}$, O$^{++}$, CO$^{++}$, N$_2^{++}$, NO$^{++}$, O$_2^{++}$, Ar$^{++}$, CO$_2^{++}$. Notwithstanding a couple of ions indistinguishable by mass spectrometry (N$_2^{++}$ and N$^{+}$, O$_2^{+}$ and O$^+$), many of these dications have u\textperiodcentered q$^{-1}$ very close to monocations from different neutral atoms, molecules, and radicals (e.g. N$^{++}$ and Li$^{+}$, CO$^{++}$ and N$^{+}$, NO$^{++}$ and CH$_3^{+}$, Ar$^{++}$ and Ne$^{+}$) and their detection is difficult or impossible even with high performance mass spectrometers such as ROSINA-DFMS. Moreover, there were only a few attempts in assessing their potential role within an ionosphere and exosphere. For instance, \citet{Lilensten2013} showed that the presence of CO$_2^{++}$ and its associated chemistry in the upper atmosphere may play a non-negligible role in the ion escape. Furthermore, \citet{Falcinelli2016} showed that the Coulomb explosion of CO$_2^{++}$ produces CO$^+$ and O$^{+}$ at a few eVs, which is energetic enough to overcome the gravitational attraction of the planet.

ROSINA-DFMS has two advantages for the detection of these peculiar and scarce ions compared with the ion spectrometer on {\it Giotto}: a higher sensitivity and a wider mass coverage. By covering several integers in terms of u\textperiodcentered q$^{-1}$, DFMS may probe those stemming from double ionisation of odd-mass-number parent molecules.

\begin{figure}
	\centering
	\stackinset{r}{0.6cm}{b}{3.4cm}{\fbox{LR}}{%
	\includegraphics[width=\linewidth,clip,trim=0cm 1.9cm 0cm 5cm]{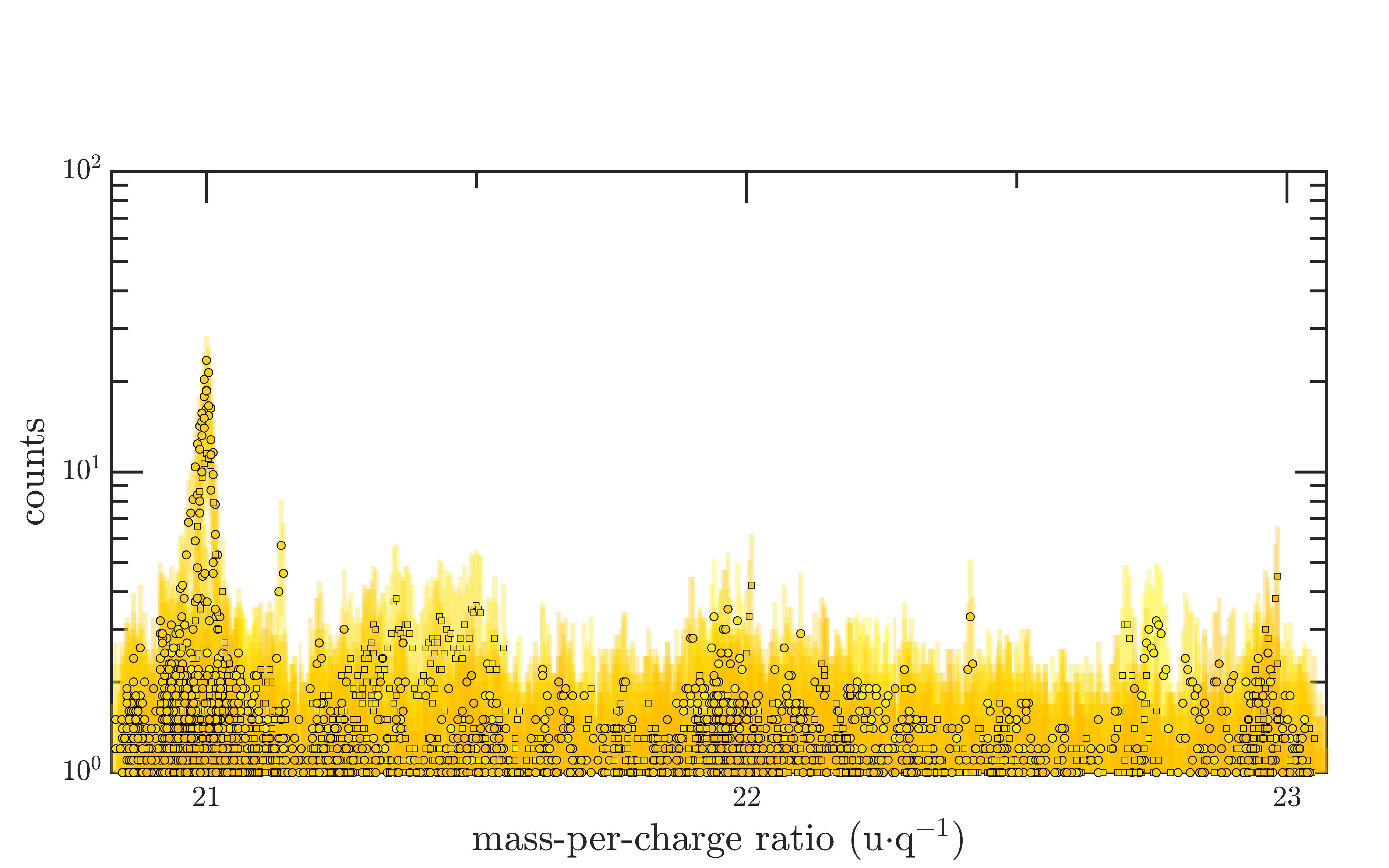}}\\
	\stackinset{r}{0.6cm}{b}{3.4cm}{\fbox{LR}}{%
	\includegraphics[width=\linewidth,clip,trim=0cm 1.9cm 0cm 5cm]{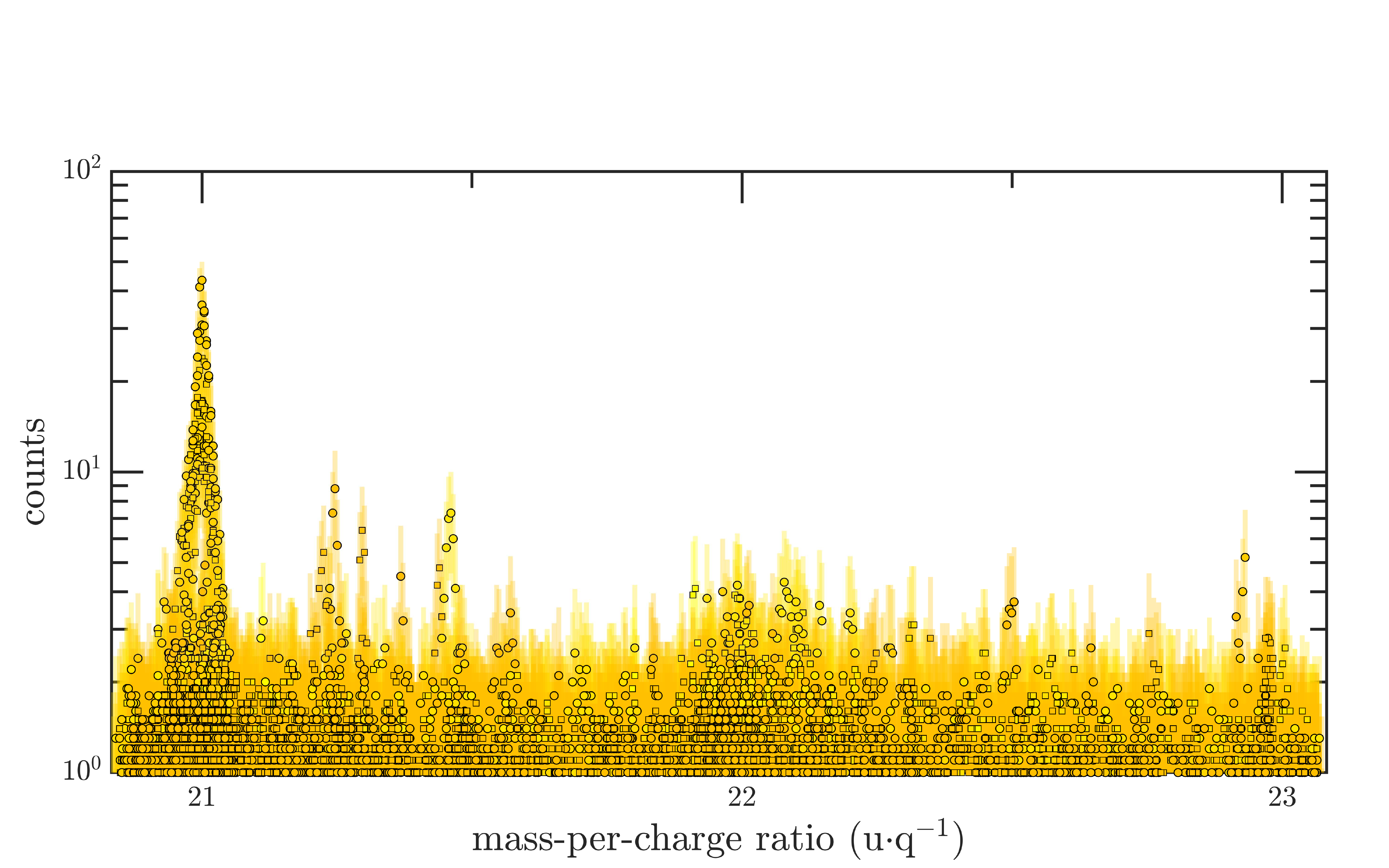}}\\
	\stackinset{r}{0.6cm}{b}{0.33667905cm+3.4cm}{\fbox{LR}}{%
	\includegraphics[width=\linewidth,clip,trim=0cm 0cm 0cm 5cm]{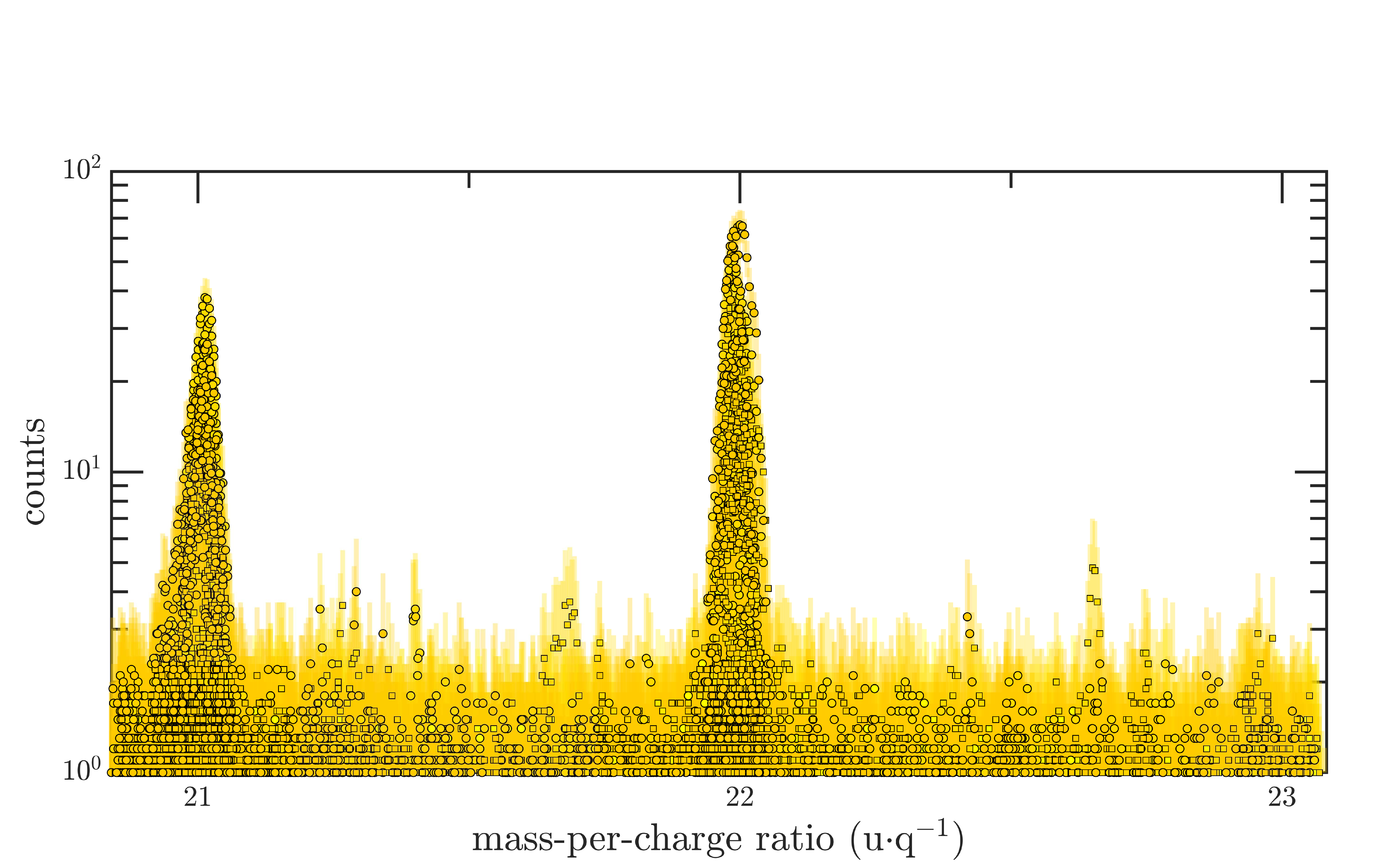}}
	\caption{Spectra in LR at 22~u\textperiodcentered q$^{-1}$ over three latitudinal regions prior to the 1$^{st}$ of March 2015: above +30$^\circ$ (upper panel, $\sim170$ stacked individual spectra), between -30$^\circ$ and +30$^\circ$ (middle panel, $\sim470$ stacked individual spectra), and below -30$^\circ$ (bottom panel, $\sim290$ stacked individual spectra). Due to a uneven latitudinal coverage when DFMS was operating in ion mode, more spectra were acquired above the Southern Hemisphere which might be a source of observational bias. \label{fig18}}
\end{figure}

It is likely that even-mass-number dications are also present but buried into monocations' signals (e.g. O$_2^{++}$ with O$^{+}$ or N$_2^{++}$ with N$^{+}$), but, fortunately, this is not the case for CO$_2^{++}$ at 22~u\textperiodcentered q$^{-1}$. In neutral mode, the neutrals are ionised by electron impact with energies of 45~eV. The recorded signal at 22~u\textperiodcentered q$^{-1}$ stems from CO$_2$ whose double ionisation threshold is $\sim 37$~eV. In the ion mode, ions have not been ionised within the instrument and originate from the cometary plasma. In order to firmly establish the nature of ions detected at 22~u\textperiodcentered q$^{-1}$, we have considered the following possibilities. We first check for candidate monocations: $^{22}$Ne$^{+}$, D$_2$$^{18}$O$^+$, HD$_2$$^{17}$O$^+$, and H$_2$D$^{18}$O$^+$ \citep{Balsiger1995}. Ne (Neon) has never been detected \citep{Rubin2018} and, regarding the isotopic composition \citep{Altwegg2015,Schroeder2019}, hydronium ion isotopologues have a too low abundance to be detected. In a second step, we check for candidate dications and consider other neutral candidates than CO$_2$ at 44~u: CS, CH$_2$ON, C$_2$H$_4$O, C$_2$H$_6$N, and C$_3$H$_8$ \citep{Altwegg2017}. Within this list, two arguments are in favour of the CO$_2^{++}$ dication: CO$_2$ is dominant at 44~u and the peak at 22~u\textperiodcentered q$^{-1}$ correlates with latitude, as CO$_2$ does. Fig.~\ref{fig18} shows LR spectra at 22~u\textperiodcentered q$^{-1}$ above both hemispheres. To exclude seasonal variations, we selected the time period from October 2014 to February 2015 (pre-equinox). ROSINA-DFMS probed frequent and strong signals at 22~u\textperiodcentered q$^{-1}$ above the Southern Hemisphere, where CO$_2$ is abundant \citep{Hassig2015}. Over the pre-equinox Northern Hemisphere where CO$_2$ is minor, no peak is present at 22~u\textperiodcentered q$^{-1}$. ROSINA-DFMS cannot simultaneously probe both ion and neutral compositions such that a `direct' correlation cannot be assessed. However, we find that, pre-equinox, both the neutral CO$_2$ abundance and the signal at 22~u\textperiodcentered q$^{-1}$ in ion mode correlate with the spacecraft latitude. Finally, to strengthen our argumentation, one should compare the pre-equinox count rates at 22~u\textperiodcentered q$^{-1}$ and at 44~u\textperiodcentered q$^{-1}$. At 22~u\textperiodcentered q$^{-1}$, the count rates were almost 100 while at 44~u\textperiodcentered q$^{-1}$, they were almost 1000. However, as aforementioned, the instrument sensitivity is a tenfold higher at 22~u\textperiodcentered q$^{-1}$ than at 44~u\textperiodcentered q$^{-1}$ such that the 44~u\textperiodcentered q$^{-1}$/22~u\textperiodcentered q$^{-1}$ ratio is $\approx100$, consistent with the ratio of photo-ionisation cross-sections of CO$_2$ leading to CO$_2^{+}$ and CO$_2^{++}$ \citep{Masuoka1994,Tian1998a}.

Unfortunately, no detection of dications except at 22~u\textperiodcentered q$^{-1}$ could be achieved from DFMS data, even if the design of ROSINA covers half-odd-integer masses. Indeed, in neutral mode, when neutrals are ionised in the gas chamber and then travel within the instrument, dications are observed at 13.5~u\textperiodcentered q$^{-1}$ for example. There are some hints to explain the lack of detection of stable doubly-charged ions other than CO$_2^{++}$. Dications are primarily produced from direct ionisation of neutrals. As ROSINA-DFMS detection starts at 13~u\textperiodcentered q$^{-1}$, only dications associated with neutral species above 26~u could be detected. Considering in first instance CO, H$_2$CO, CH$_3$OH, H$_2$S, and hydrocarbons (saturated or not), all of them have an even-integer mass number and the corresponding u\textperiodcentered q$^{-1}$ of the doubly-charged ion falls on a similar value of existing singly-charged ions except at 22~u\textperiodcentered q$^{-1}$ (see Table~\ref{table1}). The peak at 22~u\textperiodcentered q$^{-1}$ never exceeded 100 counts. Let's assume that double-ionisation of neutral species follows a similar pattern than that of CO$_2$, i.e. it occurs hundredfold less than a single ionisation \citep{Masuoka1994,Tian1998b}. A neutral parent species with an odd-integer mass number (which offers no overlap of the dication peak with other species) should exceed $\sim1$\% of the CO$_2$ volume mixing ratio in order for its daughter dication to generate 1 count, and even more to have the peak above the noise level. No cometary neutral species fulfil these two requirements.

The detection of stable dications also raises questions about their effects upon plasma dynamics and behaviour. Unlike terrestrial ionospheres, gravity is not at play at comets such that any neutral or ionised species will leave independently of their mass. The dynamics of an ion will only depend on its mass-per-charge ratio, as the dynamics is usually dominated by electromagnetic forces. In the case of CO$_2^{++}$, its mass-per-charge is not so different from the water-group ions and, therefore, its effect is marginal on the dynamics. However, it might play a role in the ion-chemistry and in the production of `energetic' CO$^+$ and O$^+$, of about a few eVs. The term energetic may mislead the reader though. Within a terrestrial ionosphere like those of Mars, Earth, or Titan, the ions are (almost) thermalised with the ambient neutral species such that their temperature is equal to the neutral temperature which never exceeds 1000 to 2000~K at most. In that case, an ion of 1~eV or more is classified as energetic compared with the neutral species. Within a cometary ionosphere, ions already have a significant amount of energy. Considering ions travelling between 1 (that of neutral species) and 8~km\textperiodcentered s$^{-1}$, as observed by and at Rosetta \citep{Odelstad2018}, the ion kinetic energy is between 0.09 and 6.0~eV for H$_2$O$^+$, between 0.15 and 9.3~eV for CO$^+$, and between 0.083 and 5.3~eV for O$^+$. In comparison, the kinetic energy released by the Coulomb explosion of CO$_2^{++}$ into CO$^+$ and O$^+$ lies within these values \citep{Falcinelli2016} such that this process is not a significant additional source of energy. Aside this energetic aspect, the double photoionisation of CO$_{2}$ might be a marginal source of CO$^+$ and O$^+$ ions. As suggested in Section~\ref{sec33}, DFMS likely detected the `stable' CO$_2^{++}$ which is produced in lower quantities than its metastable/unstable counterpart, quickly dissociating into CO$^+$+O$^+$ \citep{Masuoka1994,Slattery2015,Falcinelli2016}.

Lastly, our analysis shows that not only the low outgassing rate ($Q<10^{25}-10^{26}$~s$^{-1}$) but also the proximity of Rosetta to 67P (i.e. a few to tens of kilometres) played a key role in the detection of CO$_2^{++}$ because of the low lifetime of CO$_2^{++}$ \citep[$\leq4$~s,][]{Mathur1995}. Hence, we anticipate its detection is possible at a comet only if these conditions are met.

\subsection{Future missions}

Given the instrumental, operational, and orbital constraints, we showed that DFMS had a remarkable capability to assess the ion composition of the cometary plasma, though this was not its primary goal. Unlike {\it Giotto}, Rosetta was a non-spinning and quasi-stationary spacecraft with respect to 67P. DFMS was often operating in ion mode when it was pointing towards the comet, though manoeuvres were ongoing. These different characteristics may greatly influence the ion detection. In the case of a flyby for a future mission, the spacecraft will fly at speeds of tens of km\textperiodcentered s$^{-1}$ with respect to the target. There are two advantages of such a trajectory. Firstly, it gives a radial coverage of the plasma number density and composition over thousands of kilometres over a short period of time (typically a few hours) limiting time-dependent effects (e.g. changes in outgassing and neutral composition). Secondly, most of the cometary ions will be collimated in a limited region of velocity phase space in the spacecraft frame of reference (since the ion thermal speed is below or of the order of their mean speed) such that the instrument will capture the bulk of the cometary ions with a limited field of view. In addition, having a fast spacecraft minimises the troublesome effect of the spacecraft potential, very negative most of the time and varying for Rosetta. Indeed, considering a cometary ion, its energy, mainly being kinetic, will be $\tfrac{1}{2}m_i (\vec{v}_i-\vec{v}_\text{SC})^2\approx\tfrac{1}{2}m_i \vec{v}^2_\text{SC}$ with respect to the spacecraft. Typically, for H$_2$O$^+$, its kinetic energy in the spacecraft frame spans from 9~eV (${v}_\text{SC}=10$~km\textperiodcentered s$^{-1}$) to 330~eV (${v}_\text{SC}=60$~km\textperiodcentered s$^{-1}$) to be compared with the largest negative spacecraft potential of Rosetta, around -30~V.

In the case of an escorting spacecraft like Rosetta, being closer (a few tens of kilometres instead of hundreds of kilometres, or performing the dayside excursion) to the nucleus near perihelion would have improved the signal to noise and the detection of ions resulting from chemistry and dissociative ionisation of neutrals. However, being too close (typically $\sim10$~km) may not be the best location for very active comets and is not safe for the spacecraft. Indeed, the maximum in the plasma number density is not located at the same cometocentric distance depending on the outgassing activity (see Section~\ref{sec41}).

Although 67P has a low-to-intermediate outgassing activity, many cations have been detected even at $150-200$~km from the nucleus near perihelion. This result indicates that a very active comet is not a requirement regarding the current instrumental capabilities for mass spectrometers like DFMS. For weakly active comets (<10$^{27}$-10$^{28}$ s$^{-1}$), an escorting spacecraft, like Rosetta, is the best option as it allows to measure the ion composition over a long period, at different stages of its outgassing activity. However, additional aspects should be considered for future similar missions: limiting the manoeuvres during the spectrometer's scans, allocating operational time for the ion mode close to the nucleus, and a more uniform time coverage throughout the mission. As shown in Fig.~\ref{fig1}, DFMS ion dataset is relatively sparse, excluding safe mode and excursions. Regular ion scans would have helped to track the evolution of the ion composition through the mission. Running alternatively LR and HR modes should be also considered. For very active comets (>10$^{29}$~s$^{-1}$), a flyby is the best option from an ion composition perspective, though a different instrumentation is required such as a time-of-flight (TOF) mass spectrometer. A fly-by requires a high time resolution and spectrometers like DFMS do not suit (one DFMS spectrum at a given u\textperiodcentered q$^{-1}$ is acquired during 20~s every 10-15~min). TOF spectrometers have a much higher time resolution acquiring several u\textperiodcentered q$^{-1}$ all at once, at the expense of a lower sensitivity. Beside fostering the ion-neutral chemistry, very active comets exhibit several boundaries \citep{Mandt2019} and regions including the diamagnetic cavity \citep{Cravens1989}. Ion composition may differ inside and outside such a cavity \citep{Balsiger1986}. Although it was detected at 67P \citep{Goetz2016a,Goetz2016b}, diamagnetic crossings were on average less than 30 minutes, which corresponds to less than 3 scans for a specific mass-per-charge ratio by DFMS, limiting its ability to probe the composition inside the cavity. More time should be spent inside this region and that can be best achieved at very active comets. In addition, one flyby will allow to assess the composition inside and outside over a short time period.

\section{Conclusion\label{sec5}}

The mass spectrometer DFMS with its HR mode outperforms any in situ measurements made at comets so far. In particular, its HR ion mode has been extremely valuable for identifying ions present within a coma at close range. Although the time coverage in ion mode is more restricted than that of the neutral mode by far, DFMS has produced invaluable results, which are presented in this paper. Amongst all of them, a very new and interesting result is the first detection of the CO$_2^{++}$ dication. For future studies, to make the most out of the ion ROSINA-DFMS dataset, cross-analysis should be performed with the set of instruments from the RPC Consortium.

\begin{acknowledgements}
	 The authors very warmly thank Cyril Simon Wedlund for having provided invaluable feedback on the manuscript. The dataset analysed during the current study together with a user manual for data analysis are available in the ESA-PSA archive (ftp://psa.esac.esa.int/pub/mirror/INTERNATIONAL-ROSETTA-MISSION/ROSINA/) and on request to the authors. Rosetta is an ESA mission with contributions from its Member States and the U.S. National Aeronautics and Space Administration (NASA). Work at Imperial College London is supported by STFC of UK under grant ST/N000692/1 and ESA under contract No.4000119035/16/ES/JD. AB and MG thank the support from Imperial College London through the European Partners Fund. AB thanks also the support from the Swedish National Space Agency (grant 108/18). JDK was supported by the Belgian Science Policy Office via PRODEX/ROSINA PEA 90020. We would like to acknowledge ISSI for the great opportunity it offered us for very valuable discussions on this topic as part of the International Team ``Plasma Environment at 67P After Rosetta''.
\end{acknowledgements}

\newpage
\newpage

\bibliographystyle{aa} 
\bibliography{biblio}

\newpage

\begin{appendix}
\section{Correction for 13, 14, and 15~u\textperiodcentered q$^{-1}$\label{AppA}}
\begin{figure}[H]
		\stackinset{r}{0.6cm}{b}{3.4cm}{\fbox{HR}}{%
		\includegraphics[width=\linewidth,trim=0 1.9cm 0 2.5cm,clip]{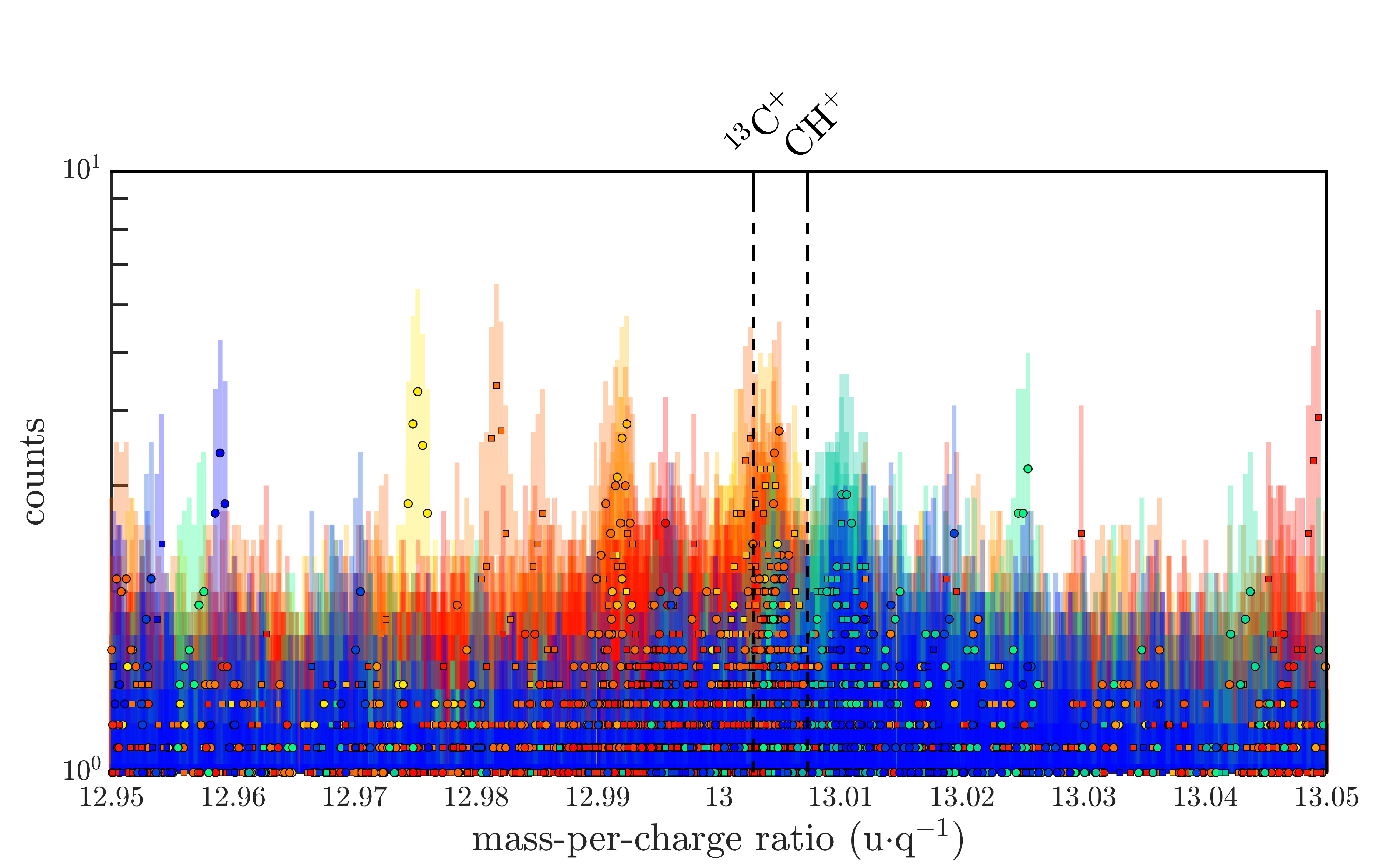}\\}
		\stackinset{r}{0.6cm}{b}{3.4cm}{\fbox{HR}}{%
		\includegraphics[width=\linewidth,trim=0 1.9cm 0 2.5cm,clip]{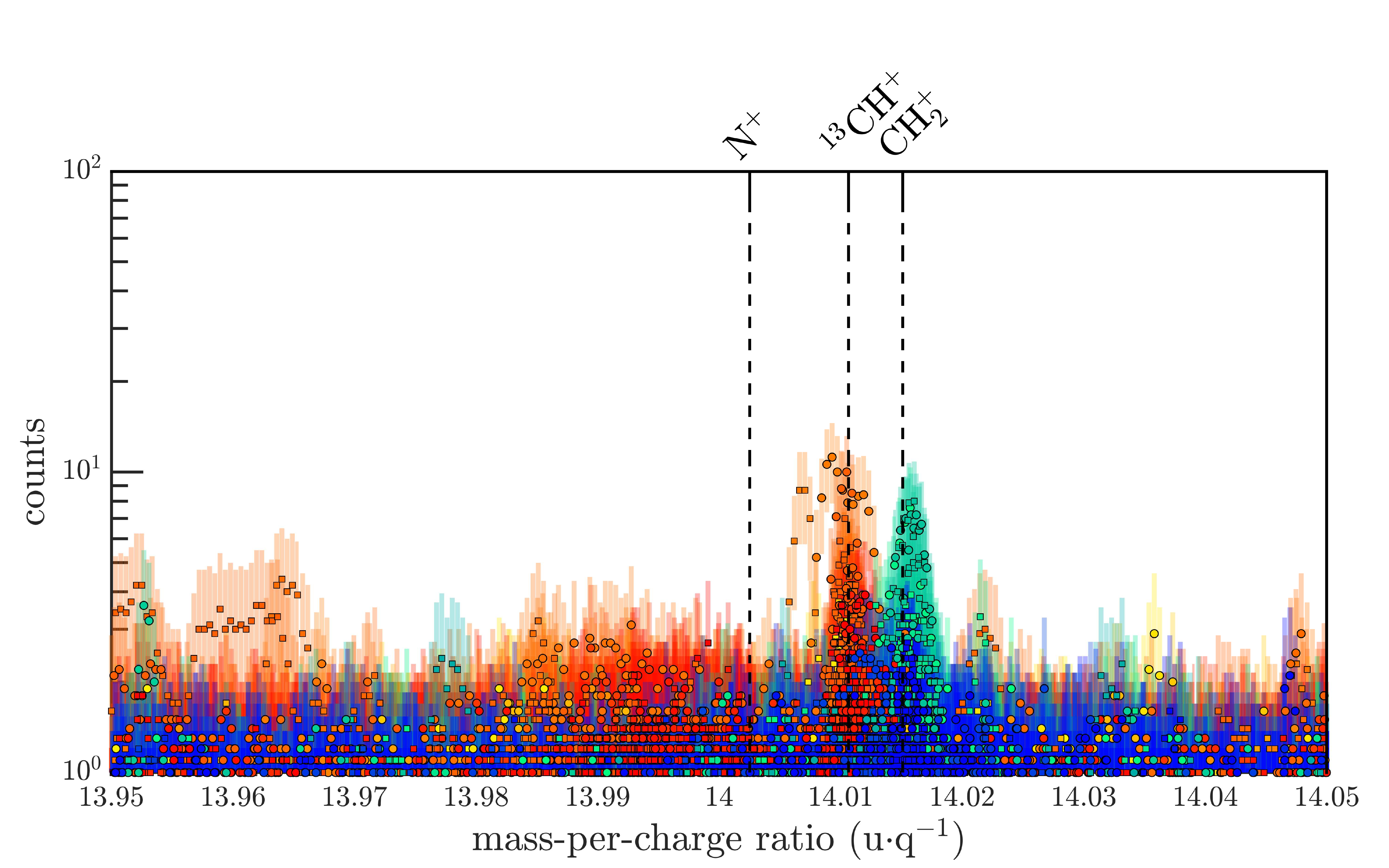}\\}
		\stackinset{r}{0.6cm}{b}{0.33667905cm+3.4cm}{\fbox{HR}}{%
		\includegraphics[width=\linewidth,trim=0 0cm 0 2.5cm,clip]{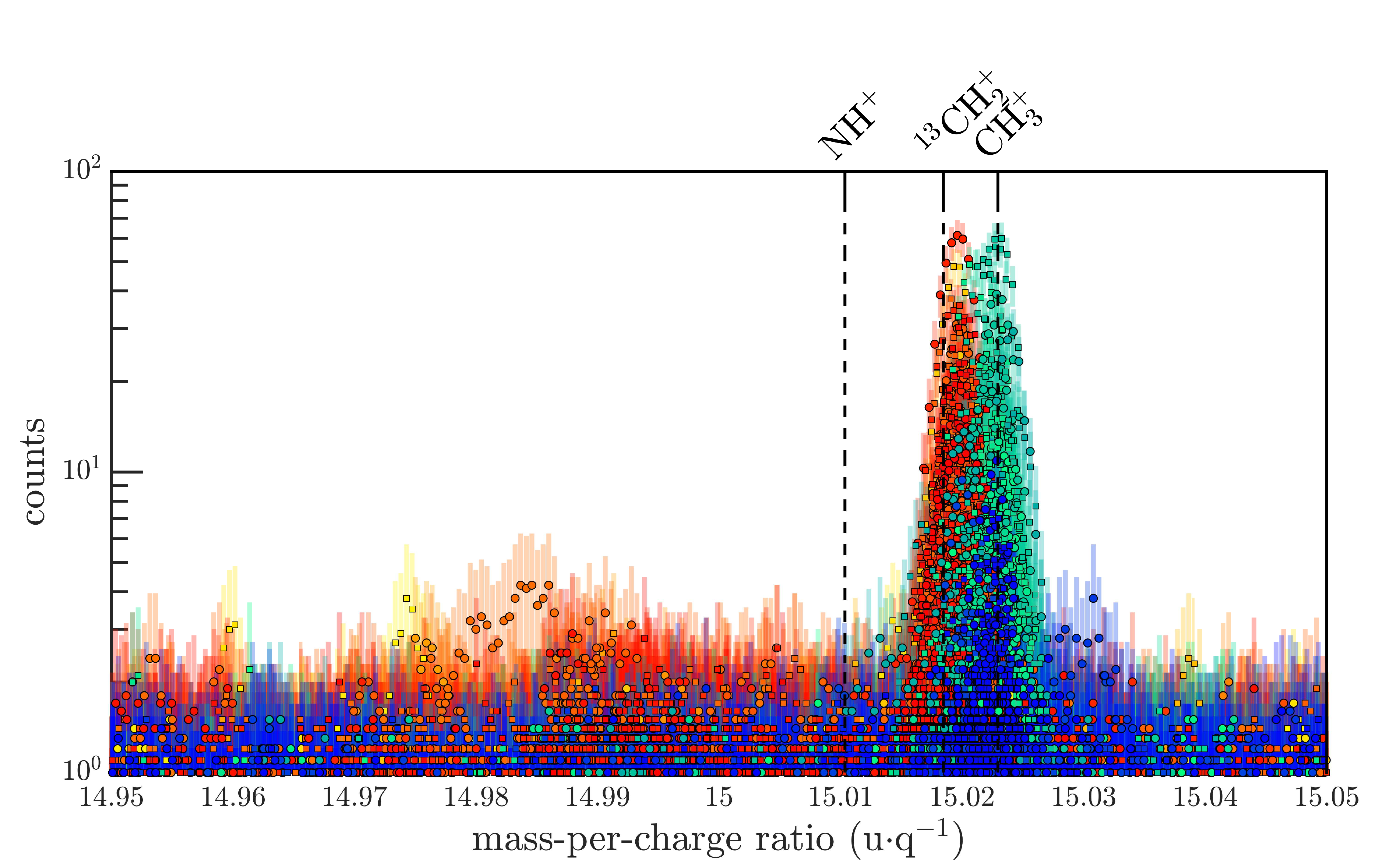}}
		\caption{Original spectra when applying the procedure described in Section~\ref{analysis} with a zoom factor of 6.4 instead of 5.5 as suggested. Two distinct peaks appear whose location depends on the time of acquisition (warm colour, $t<t_\text{shift}$; cold colours, $t>t_\text{shift}$) and therefore are attributed to the change of the post-acceleration. In particular, the red colour is right before the shift and the green colour right afterwards according to the colour scale. Even if we applied a different zoom factor, more likely to be 5.5 here and therefore lower than assumed for the figure, both peaks for each spectrum would move to the left and be closer to each other and $m_0$.\label{figa1}}
	\end{figure}

Additional corrections are required for the mass calibration at 13, 14, and 15~u\textperiodcentered q$^{-1}$. First, as mentioned in Section~\ref{analysis}, the zoom factor $z$ is slightly lower than at higher mass-per-charge ratios, at 5.5. Moreover, by applying the procedure described in Section~\ref{analysis}, Fig.~\ref{figa1} shows spectra with two distinct peaks, each of them associated with a different period: warm-coloured before the 27$^{th}$ of January 2016 and cold-coloured afterwards. On the 27$^{th}$ of January 2016, the location of the impinging beam was shifted such that the ions would hit the  detector on less-deteriorated parts \citep[see Fig.~A.6 in][]{Schroeder2019}.
	
To corroborate that further corrections at theses specific u\textperiodcentered q$^{-1}$ are required, both peaks are associated with the same species and the separation in pixel of the peaks from both periods is instrumental (i.e. pixels $p_{0,19}\neq p_{0,13} \neq p_{0,14} \neq p_{0,15}$): the separation of the warm- and cold-coloured peaks shrinks as the mass-per-charge ratio goes from 13 to 15~u\textperiodcentered q$^{-1}$ (see Fig.~\ref{figa1}), though the mass separation should be constant if both peaks from both periods are associated with two distinct ions:
$m_{\text{CH}^+}-m_{^{13}\text{C}^+}=m_{\text{CH}_2^+}-m_{^{13}\text{CH}^+}=m_{\text{CH}_3^+}-m_{^{13}\text{CH}_2^+}=4.48\times10^{-3}$ u. Moreover, low resolution spectra do not show a strong difference in terms of counts between both periods at these masses, consistent with similar counts of both peaks in HR such that they might correspond to the same ion. In conclusion, the peaks at 13, 14, and 15~u\textperiodcentered q$^{-1}$ are CH$^+$, CH$_2^+$, and CH$_3^+$. As a result, we have interpolated the behaviour of the instrument at 13, 14, and 15~u\textperiodcentered q$^{-1}$ as follows: $p_{0,\{13,14,15\}}<p_{0,19}$ before the 27$^{th}$ of January 2016, $p_{0,13}>p_{0,19}$ and $p_{0,14}=p_{0,15}=p_{0,19}$ after the 27$^{th}$ of January 2016.

\section{Photo-ionisation and kinetic rates\label{AppB}}

In Section~\ref{sec4}, photo-ionisation frequencies of some neutral species (i.e. H$_2$O, CO$_2$, CH$_4$, NH$_3$) and the kinetic rates of their ionised fragments with H$_2$O are used. We list here in Table~\ref{tableb1} and~\ref{tableb2} these values.

\begin{table}[H]
	\begin{tabularx}{\linewidth}{c@{\hskip 3pt}l@{ + }l@{\hskip 3pt}R}
		Neutral species&\multicolumn{2}{l}{Products}&Photo-ionisation frequency at 1~au (s$^{-1}$)\\
		\hline
		H$_2$O&H$_2$O$^+$&e$^-$& $3.31\times10^{-7}$\\
		CO$_2$&CO$_2^+$&e$^-$& $6.55\times10^{-7}$\\
		NH$_3$&NH$_3^+$&e$^-$& $6.10\times10^{-7}$\\
		"&NH$_2^+$&products& $1.77\times10^{-7}$\\
		"&NH$^+$&products& $6.92\times10^{-9}$\\
		CH$_4$&CH$_4^+$&e$^-$& $3.58\times10^{-7}$\\
		"&CH$_3^+$&products& $1.98\times10^{-7}$\\
		"&CH$_2^+$&products& $2.08\times10^{-8}$\\
		"&CH$^+$&products&\ $4.21\times10^{-9}$\
	\end{tabularx}
\caption{Photo-ionisation frequencies at 1~au from \citet{Huebner2015} with a quiet Sun and their associated photo-products.\label{tableb1}}
\end{table}

\begin{table*}
	\centering
	\begin{tabularx}{\linewidth}{l@{+ }l@{ $\longrightarrow$ }l@{+ }l@{\hskip 3pt}cc@{ $\times$ }lcl}
		\multicolumn{2}{l}{Reactants}&\multicolumn{2}{c}{Products}&&\multicolumn{2}{l}{Kinetic rate (cm$^{3}$\textperiodcentered s$^{-1}$)}&Uncertainties&References\\
		\hline
		H$_2$O$^+$&H$_2$O&H$_3$O$^+$&HO&&1.85&$10^{-9\phantom{0}}$ $(300/T)^{1/2}$&$\pm 15\%$&\citet{Anicich1993}\\
		CO$_2^+$&H$_2$O&H$_2$O$^+$&CO$_2$&\multirow{1}{*}{\ldelim \} {2}{1 mm}}&\multicolumn{2}{l}{\multirow{2}{*}{2.40 $\times$ $10^{-9\phantom{0}}$ $ (300/T)^{1/2}$}}&\multirow{2}{*}{$\pm 15\%$}&\multirow{2}{*}{"}\\
		\multicolumn{2}{c}{}&HCO$_2^+$&HO&&\multicolumn{2}{l}{}&&\\
		CH$_4^+$&H$_2$O&H$_3$O$^+$&CH$_3$&&2.50&$10^{-9\phantom{0}}$ $(300/T)^{1/2}$&$\pm10\%$&\\
		CH$_3^+$&H$_2$O&CH$_3$OH$_2^+$&$h\nu$&&$4.61$&$10^{-13}$ $(300/T)^{1.73}$*&$\pm100\%$&\citet{Bates1983,Herbst1985}\\
		CH$_2^+$&H$_2$O&H$_2$COH$^+$&H&&2.05&$10^{-9\phantom{0}}$ $(300/T)^{1/2}$&$\pm60\%$&\citet{Anicich1993}\\
		CH$^+$&H$_2$O&HCO$^+$&H$_2$O&\multirow{1}{*}{\ldelim \} {3}{1 mm}}&\multicolumn{2}{l}{\multirow{3}{*}{2.90 $\times$ $10^{-9\phantom{0}}$ $ (300/T)^{1/2}$}}&\multirow{3}{*}{$\pm20\%$}&\\
		\multicolumn{2}{c}{}&H$_3$O$^+$&C&&\multicolumn{2}{l}{}&&"\\
		\multicolumn{2}{c}{}&H$_2$CO$^+$&H&&\multicolumn{2}{l}{}&&\\
		NH$_3^+$&H$_2$O&NH$_4^+$&HO&&2.50&$10^{-10}$ $(300/T)^{1/2}$&$\pm30\%$&"\\
		NH$_2^+$&H$_2$O&H$_3$O$^+$&NH&\multirow{1}{*}{\ldelim \} {3}{1 mm}}&\multicolumn{2}{l}{\multirow{3}{*}{2.90 $\times$ $10^{-9\phantom{0}}$ $ (300/T)^{1/2}$}}&\multirow{3}{*}{$\pm20\%$}&\multirow{3}{*}{"}\\
		\multicolumn{2}{c}{}&NH$_4^+$&O&&\multicolumn{2}{l}{}&&\\
		\multicolumn{2}{c}{}&NH$_3^+$&HO&&\multicolumn{2}{l}{}&&\\
		NH$^+$&H$_2$O&HNO$^+$&H$_2$&\multirow{1}{*}{\ldelim \} {5}{1 mm}}&\multicolumn{2}{l}{\multirow{5}{*}{3.50 $\times$ $10^{-9\phantom{0}}$ $ (300/T)^{1/2}$}}&\multirow{5}{*}{$\pm20\%$}&\multirow{5}{*}{"}\\
		\multicolumn{2}{c}{}&H$_3$O$^+$&N&&\multicolumn{2}{l}{}&\\
		\multicolumn{2}{c}{}&H$_2$O$^+$&NH&&\multicolumn{2}{l}{}&\\
		\multicolumn{2}{c}{}&NH$_3^+$&O&&\multicolumn{2}{l}{}&\\
		\multicolumn{2}{c}{}&NH$_2^+$&HO&&\multicolumn{2}{l}{}&\\
		N$^+$&H$_2$O&H$_3$O$^+$&N&&2.70&$10^{-9\phantom{0}}$ $(300/T)^{1/2}$&$\pm20\%$&"\\
	\end{tabularx}
\caption{Kinetic loss rates used in Section~\ref{sec41}. Most of the rates are experimentally determined at 300~K. A modified Arrhenius law, with a $T^{-1/2}$ coefficient, has been employed for all except one which is a radiative association (*). In addition, for the latter which is theoretical, no uncertainties were provided so that we assumed $\pm100\%$. However, \citet{Luca2002} showed that this kinetic rate is most likely overestimated by almost one order of magnitude. Finally, different but theoretical kinetic rates are available in the literature \citep[e.g.][]{Woon2009}. \label{tableb2}} 
\end{table*}

\section{Exact mass-per-charge ratios\label{AppC}}

\begin{table*}
	\centering
	\setlength{\tabcolsep}{0.50em}
	\renewcommand{\arraystretch}{1.4}
	\begin{tabular}[t]{|lcr|}
			\cline{1-3}
		$m_0\ (\text{u\textperiodcentered q}^{-1})$&$\Delta m\ (10^{-3}\text{ u\textperiodcentered q}^{-1})$&ion\\
		\cline{1-3}	
		13&+02.81&$^{13}$C$^{+}$\\
		  &+07.28&CH$^{+}$\\\cline{1-3}		
		14&+02.53&N$^{+}$\\
		  &+10.63&$^{13}$CH$^{+}$\\
		  &+15.10&CH$_2^{+}$\\\cline{1-3}	
		15&+10.35&NH$^{+}$\\
		  &+18.46&$^{13}$CH$_2^{+}$\\
		  &+22.93&CH$_3^{+}$\\\cline{1-3}	
		16&--\! 05.63&O$^{+}$\\
		  &+18.18&NH$_2^{+}$\\
		  &+26.28&$^{13}$CH$_3^{+}$\\
		  &+30.75&CH$_4^{+}$\\\cline{1-3}	
		17&--\! 01.42&$^{17}$O$^{+}$\\
		  &+02.19&HO$^{+}$\\
		  &+26.00&NH$_3^{+}$\\
		  &+34.11&$^{13}$CH$_4^{+}$\\
		  &+38.58&CH$_5^{+}$\\\cline{1-3}	
		18&--\! 01.39&$^{18}$O$^{+}$\\
		  &+06.41&H$^{17}$O$^{+}$\\
		  &+08.47&DO$^{+}$\\
		  &+10.02&H$_2$O$^{+}$\\
		  &+33.83&NH$_4^{+}$\\
		  \hline
		\end{tabular}
		\begin{tabular}[t]{|lcr|}
				\cline{1-3}
			$m_0\ (\text{u\textperiodcentered q}^{-1})$&$\Delta m\ (10^{-3}\text{ u\textperiodcentered q}^{-1})$&ion\\
			\cline{1-3}	
		19&+06.44&H$^{18}$O$^{+}$\\
			&+14.23&H$_2$$^{17}$O$^{+}$\\
			&+16.29&HDO$^{+}$\\
			&+17.84&H$_3$O$^{+}$\\\cline{1-3}	
			20&+14.26&H$_2$$^{18}$O$^{+}$\\
			&+22.06&H$_3$$^{17}$O$^{+}$\\
			&+24.12&H$_2$DO$^{+}$\\\cline{1-3}	
		21&+22.09&H$_3$$^{18}$O$^{+}$\\\cline{1-3}
		22&--\! 05.63&CO$_2^{++}$\\\cline{1-3}
		23&--\! 10.78&Na$^{+}$\\\cline{1-3}	
		24&--\! 00.55&C$_2^{+}$\\\cline{1-3}
		25&+07.28&C$_2$H$^{+}$\\\cline{1-3}
		26&+02.53&CN$^{+}$\\
		  &+15.10&C$_2$H$_2^{+}$\\\cline{1-3}	
		27&+10.35&HCN$^{+}$\\
		  &+22.93&C$_2$H$_3^{+}$\\\cline{1-3}	
		28&--\! 23.62&Si$^{+}$\\
		  &--\! 05.63&CO$^{+}$\\
		  &+05.60&N$_2^{+}$\\
		  &+18.18&HCNH$^{+}$\\
		  &+30.75&C$_2$H$_4^{+}$\\
		  &&\\
		  \hline
	\end{tabular}
		\begin{tabular}[t]{|lcr|}
	\cline{1-3}
	$m_0\ (\text{u\textperiodcentered q}^{-1})$&$\Delta m\ (10^{-3}\text{ u\textperiodcentered q}^{-1})$&ion\\
	\cline{1-3}	
	29&--\! 15.80&SiH$^{+}$\\
	&+02.19&HCO$^{+}$\\
	&+13.42&N$_2$H$^{+}$\\
	&+38.58&C$_2$H$_5^{+}$\\\cline{1-3}	
	30&--\! 02.56&NO$^{+}$\\
	&+10.02&H$_2$CO$^{+}$\\
	&+33.83&CH$_2$NH$_2^{+}$\\
	&+46.40&C$_2$H$_6^{+}$\\\cline{1-3}
	31&--\! 26.79&P$^{+}$\\
	&+05.27&HNO$^{+}$\\
	&+17.84&H$_2$COH$^{+}$\\\cline{1-3}	
	32&--\! 28.48&S$^{+}$\\
	&--\! 10.72&O$_2^{+}$\\
	&+25.67&CH$_3$OH$^{+}$\\
	&+36.90&N$_2$H$_4^{+}$\\\cline{1-3}	
	33&--\! 20.65&HS$^{+}$\\
	&--\! 02.89&HO$_2^{+}$\\
	&+33.49&CH$_3$OH$_2^{+}$\\
	&+44.72&N$_2$H$_5^{+}$\\\cline{1-3}	
	39&--\! 36.84&K$^{+}$\\
	&+22.93&C$_3$H$_3^{+}$\\   
	&&\\
	\hline
\end{tabular}
\caption{List of the ions displayed in the different spectra with their exact mono-isotopic mass ($m_0+\Delta m$). Near the pixel $p_0$, the difference between two pixels corresponds to $\sim0.03\times 10^{-3}m_0$ u\textperiodcentered q$^{-1}$ in terms of mass-per-charge ratio in HR. As the HR resolution is $>3000$ at 1\% peak height, ion species should be separated by at least 10 pixels or $\sim0.33\times 10^{-3} m_0$ to be resolved if the counts do not exceed 100 times the noise level, the mass-per-charge separation has to be higher otherwise. First column: commanded mass-per-charge ratio $m_0$ to which ions belong. Second column: algebraic mass-per-charge shift with respect to $m_0$. Third column: the ion species. \label{tablec1}}
\end{table*}

\end{appendix}

\end{document}